\documentclass[a4paper,11pt]{article}
\pdfoutput=1 

\usepackage{jheppub} 
\usepackage{framed}
\usepackage{enumerate}
\usepackage{amsmath,amssymb}
\usepackage{float}
\usepackage{xcolor}
\usepackage[T1]{fontenc} 
\usepackage{epigraph}

\usepackage{amssymb}
\usepackage{amsmath}
\usepackage{graphicx}
\usepackage{subfig}
\usepackage{cancel}

\newcommand{\la}[1]{\label{#1}}
\newcommand{\eq}[1]{\eqref{#1}}

\def\[{\left[}
\def\]{\right]}
\def\({\left(}
\def\){\right)}
\def\d{\partial}
\newcommand{\beq}{\begin{equation}}
\newcommand{\eeq}{\end{equation}}
\newcommand\beqa{\begin{eqnarray}}
\newcommand\eeqa{\end{eqnarray}}
\newcommand{\nn}{\nonumber}

\definecolor{color1}{rgb}{0.471412, 0.108766, 0.527016}
\definecolor{color2}{rgb}{0.246296, 0.31595666666666666, 0.80044}
\definecolor{color3}{rgb}{0.324106, 0.6089696666666666, 0.7083413333333334}
\definecolor{color4}{rgb}{0.513417, 0.72992, 0.440682}
\definecolor{color5}{rgb}{0.764712, 0.7283023333333333, 0.27360833333333334}
\definecolor{color6}{rgb}{0.901627, 0.5398719999999999, 0.208366}
\definecolor{color7}{rgb}{0.857359, 0.131106, 0.132128}

\title{\boldmath Bootstrability in Defect CFT: Integrated Correlators and Sharper Bounds}

\author{Andrea Cavagli\`a$^a$}
\author{Nikolay Gromov$^{a}$}
\author{Julius Julius$^a$}
\author{Michelangelo Preti$^a$}%
\affiliation{%
 $^a$ Department of Mathematics, King's College London, Strand WC2R 2LS 
 \\
}%
 \emailAdd{andrea.cavaglia@kcl.ac.uk}
 \emailAdd{nikolay.gromov@kcl.ac.uk}
 \emailAdd{julius.julius@kcl.ac.uk}
 \emailAdd{michelangelo.preti@kcl.ac.uk}

\abstract{
We continue to develop {Bootstrability}
-- a method merging Integrability and Conformal Bootstrap to extract CFT data in 
 integrable conformal gauge theories such as $\mathcal{N}$=4 SYM. 
In this paper, we consider the 1D defect CFT defined on a $\frac{1}{2}$-BPS Wilson line in the theory, 
whose non-perturbative spectrum is governed by the Quantum Spectral Curve (QSC). In addition, we use that 
the deformed setup 
of a cusped Wilson line is also controlled by the QSC. In terms of the defect CFT, this translates into two nontrivial relations connecting integrated 4-point correlators to cusp spectral data, such as the Bremsstrahlung and Curvature functions -- known analytically from the QSC.
Combining these new constraints and the spectrum of the $10$ lowest-lying  states with the  Numerical Conformal Bootstrap, we obtain very sharp rigorous numerical bounds for the structure constant of the first non-protected state, giving this observable with seven digits precision for the 't Hooft coupling in the intermediate coupling region $\frac{\sqrt{\lambda}}{4\pi}\sim 1$, with the error  decreasing quickly at large 't Hooft coupling.
Furthermore, for the same structure constant we obtain a $4$-loop analytic result at weak coupling. We also present results for excited states.
}

\begin{document} 
\maketitle
\flushbottom
\section{Introduction}
\label{sec:intro}
Despite many efforts, solving an interacting gauge theory in 4D still remains an unsolved problem. At the same time there is a number of exact non-perturbative results available in the maximally supersymmetric gauge theory in 4D.
Even though this theory has a large number of symmetries, it is still a highly nontrivial interacting
theory, which generates physically significant observables interpolating   between a free 4D QFT at $\lambda=0$ and the dynamics of a classically integrable 2D string worldsheet at strong coupling. In fact, integrability was also noticed on the QFT side first in two different regimes, \cite{Lipatov:1993yb,Faddeev:1994zg} and \cite{Minahan:2002ve}, which are now understood to be related by a unified non-perturbative quantum integrable structure known as Quantum Spectral Curve (QSC) \cite{Gromov:2013pga,Gromov:2014caa}. 
It is believed that integrability  governs all observables at least in the planar limit, and could be used to solve the full theory. 

At the moment, the status of this program is the following. The QSC unlocks the  non-perturbative spectrum of anomalous dimensions of all single trace operators. In order to solve the planar theory, one would also need to understand how to compute all correlation functions at finite $\lambda$. At weak coupling, the Hexagon formalism~\cite{Basso:2015zoa,Fleury:2016ykk,Eden:2016xvg,Bargheer:2017nne} allows to effectively reformulate the computation of Feynman diagrams into a simpler diagrammatic description involving the scattering of ``magnons''. For generic correlators, the complexity of this method still grows exponentially with the order in perturbation theory. Nevertheless, in some regimes of infinitely heavy operators there was some progress in resumming the series recently, e.g.~\cite{Coronado:2018cxj,Kostov:2019stn,Bargheer:2019kxb,Bargheer:2019exp}.  
There are two other approaches trying to tackle the finite coupling region for general operators -- one based on the Separation of Variables method~(see \cite{Jiang:2015lda,Gromov:2016itr,Cavaglia:2018lxi,Giombi:2018hsx,Cavaglia:2019pow,Gromov:2019wmz,Cavaglia:2021mft,Gromov:2022waj} for recent progress), and  another based on a combination of  Integrability with the Numerical Conformal Bootstrap (NCB)~\cite{Rattazzi:2008pe,El-Showk:2012cjh}, started in our previous paper \cite{Cavaglia:2021bnz}\footnote{In different contexts unrelated to gauge theories, the integration of NCB with exact spectral data coming from integrability was also used, see e.g. \cite{Picco:2016ilr,He:2020rfk} and similar ideas in \cite{Nakayama:2016cim,Gliozzi:2015qsa,Gliozzi:2016cmg}. 
}.
In this work we continue developing this approach -- which we call ``Bootstrability''. It incorporates  both new exact analytical results and new numerical techniques. 

\begin{figure}
    \centering
    \includegraphics[width=0.82\linewidth]{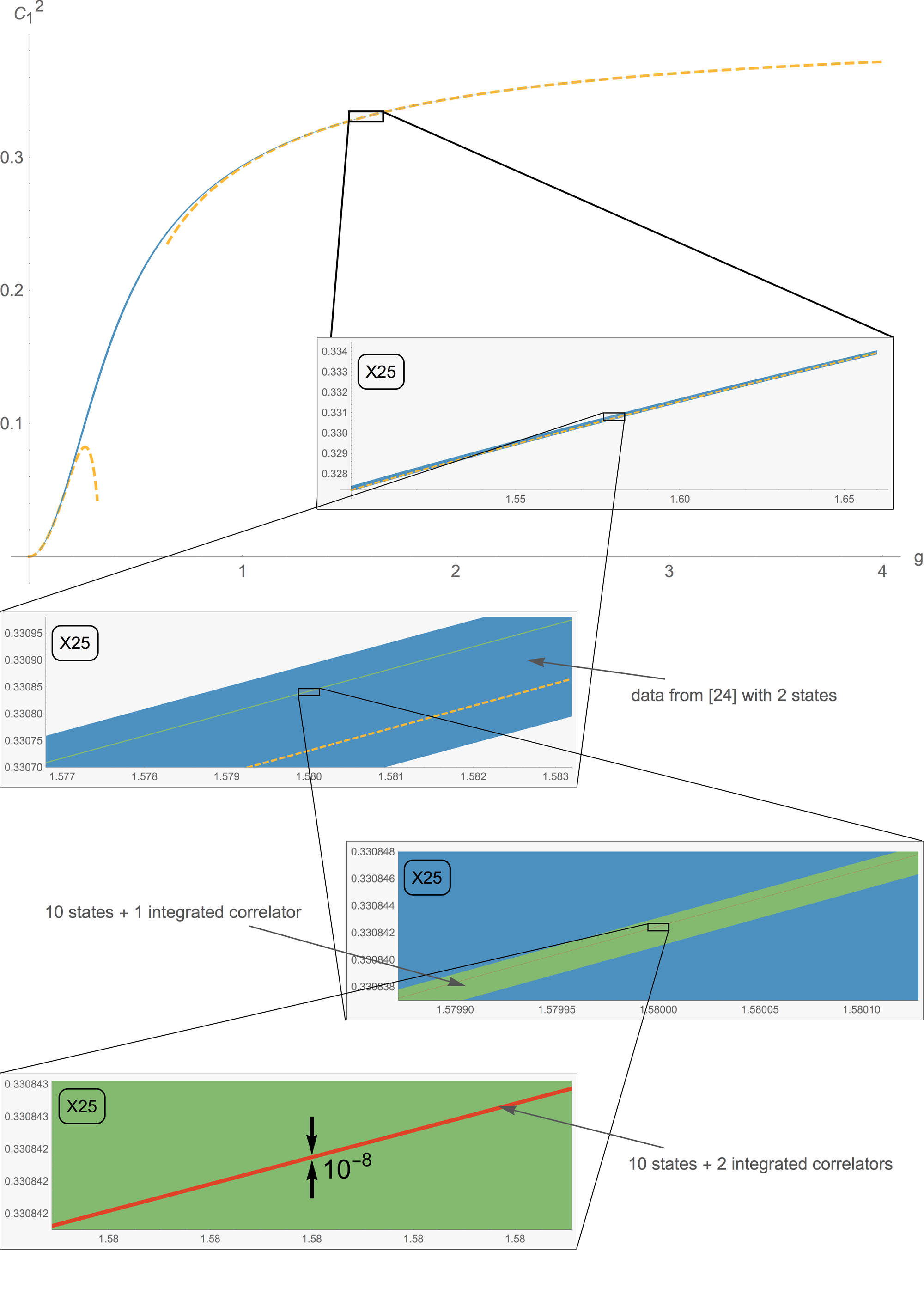}
    \caption{
    The square OPE coefficient $C_1^2$ of two protected line deformation operators $\Phi_{\perp}$ into the non-protected one $\Phi_{||}$, as a function of the coupling $g$. The
thickness of the solid lines gives the allowed regions for $C_1^2$ for different methods. The yellow dashed lines give the weak coupling \eqref{eq:C1finalresult} and strong coupling \cite{Ferrero:2021bsb} analytic expansion. The four tiles are magnification of the previous one of a factor of $25$. The blue domain is the previous result of~\cite{Cavaglia:2021bnz} where only two states were included. The green domain includes $10$ states and one integrated correlator while the red one includes both the constraints.}
    \label{fig:zoom}
\end{figure}

In this paper we study the same observables as in our previous paper \cite{Cavaglia:2021bnz} -- namely operator insertions on an infinite, straight $\frac{1}{2}$-BPS supersymmetric Wilson line  in $\mathcal{N}$=4 SYM. They define a set of  correlators respecting the properties of a 1D CFT~\cite{Drukker:2006xg}. These physical observables have been investigated intensively in the past few years with a wide variety of methods, from string and gauge theory computations to localisation, integrability  and the conformal bootstrap~\cite{Drukker:2006xg,Correa:2012hh,Drukker:2012de,Giombi:2017cqn,Kim:2017sju,Cooke:2017qgm,Liendo:2018ukf,Giombi:2018qox,Giombi:2018hsx,Giombi:2020amn,Grabner:2020nis,Giombi:2021zfb,Barrat:2021yvp,Ferrero:2021bsb,Barrat:2021tpn,Giombi:2022anm} (for less supersymmetric setups see e.g. \cite{Polchinski:2011im,Beccaria:2017rbe,Gimenez-Grau:2019hez}). Beside being interesting for the connection to  the study of general Wilson loops, this setting is also a  convenient laboratory to develop exact methods. For example, operators inserted on Wilson lines are simpler to study in the SoV approach~\cite{Cavaglia:2018lxi}, since they naturally have twisted boundary conditions. Moreover, the 1D defect CFT is a very nice setup to start experimenting with Bootstrability, since we can just restrict to the 't Hooft large N limit and it is still described by a consistent 1D CFT. 

A recent important advance was the adaptation of the QSC to describe the spectrum of the  defect CFT in the planar limit~\cite{Grabner:2020nis,Julius:2021uka,spec1DCFT}. 
The approach we started in \cite{Cavaglia:2021bnz} is to exploit
the knowledge of the exact spectrum to extract maximal information from the NCB. 
In particular, we showed how  the knowledge of only two non-protected states in the spectrum allows to deduce (with rigorous NCB methods) a very narrow numerical estimate for an OPE coefficient involving the simplest non-protected operator.\footnote{One could argue that we use some extra input, beyond the spectrum coming from integrability. In fact, we also use the knowledge of an OPE coefficient involving a supersymmetric operator (see section \ref{sec:bootstrapsetup}), which was first obtained using information from localisation in~\cite{Liendo:2018ukf}. However, we show in this paper that it is also possible to obtain the same result for $C_{BPS}$ using purely a comparison with integrability data, see appendix \ref{app:CBPS}.}

The main new ingredient of this paper are two new integrated correlator  constraints~\cite{upcomingAJMNderivation}, which greatly increase the precision of our estimates.
This new insight comes from the knowledge about the spectrum of the deformed observable with a defect -- such as cusp or a parameter change.
Conceptually, this is similar to what was observed for the bulk $\mathcal{N}$=4 SYM theory, where integrated correlator relations were deduced from localisation~\cite{Binder:2019jwn} in a deformed model, and proved to be very constraining in the bootstrap~\cite{Chester:2021aun}. 

We reserve the derivation of these integral relations to~\cite{upcomingAJMNderivation}, where we consider the deformation obtained by forming a \emph{cusp}  (in this case a discontinuity) in R-space on the  line.  This is associated to the cusp anomalous dimension~\cite{Polyakov:1980ca,Korchemsky:1987wg}, which can be studied with integrability in this context~\cite{Correa:2012hh,Drukker:2012de,Gromov:2015dfa}. It was showed in~\cite{Cooke:2017qgm} that deformations of the contour of a Wilson loop -- in physical or $R$-space -- are equivalent to summing over integrated correlators on the undeformed loop (see for instance \cite{Polyakov:2000ti,Semenoff:2004qr,Zarembo:2016bbk,Cooke:2017qgm}). In a certain limit, we obtain a nontrivial identity relating an integrated 4-point function in the 1D CFT to quantities in the cusp setup known from integrability.  
Considering similar arguments but a different deformation, we obtained a second independent relation, giving a total of two new constraints.

We  show that including the new constraints in the Bootstrability approach leads to a huge gain in precision, illustrated in Figure \ref{fig:zoom}. We also introduce new tricks on the numerical side of the analysis, allowing  us to incorporate knowledge from more states of the spectrum and to deduce new results for the  OPE coefficients involving excited states.

Furthermore, we also develop an analytic bootstrap approach at weak coupling (inspired by a similar strategy at strong coupling~\cite{Ferrero:2021bsb}, but now including input from integrability as well as the new integrated correlator constraints). 
With this method, we obtain a 4-loop prediction for the leading OPE coefficient, fusing two protected line deformation operators $\Phi_{\perp}^i$ into one non-protected operator $\Phi_{||}$, 
\beq\begin{split}\nonumber
C_1^2(g) = &2 g^2 - \left(24-\frac{4 \pi ^2}{3}\right)g^4+
\left(320-16 \pi
   ^2+48 \zeta_3-\frac{76 \pi ^4}{45}\right)g^6\\
   &-\left(4480-\frac{832 \pi ^2}{3}+256 \zeta_3-\frac{224 \pi ^4}{15}+880 \zeta_5-\frac{64 \pi ^6}{45}\right)g^8+
O(g^{10})\;.
\end{split}\eeq
We also give the analytic expression at 2 loops for a 4-point correlation function~\eqref{eqn:Gweakg4}.

The rest of this paper is organised as follows. In section \ref{sec:setup}, we describe the main setup, and the new integrated correlator constraints are presented in section \ref{sec:integrated}. Section \ref{sec:numerical} contains a detailed discussion of numerical Bootstrability illustrating our main results bounding three OPE coefficients, while section \ref{sec:analytical} develops the analytical approach at weak coupling. Finally, in section \ref{sec:discussion} we summarise the results and discuss   future directions. The Appendices contain technical details as well as data for the spectrum of the first 10 states and numerical bounds for three structure constants. Appendix \ref{app:CBPS} contains a new non-perturbative derivation of the form of an  OPE coefficient involving a supersymmetric operator, using a comparison with integrability data. 

\section{Setup}\label{sec:setup}
In this section we describe in detail the setup. First, we introduce the supersymmetric Wilson line, and its important deformation obtained by forming a cusp, defining two crucial quantities -- the Bremsstrahlung and Curvature functions. 
We then introduce the 1D defect CFT, 
 and present the conformal data for the spectrum from \cite{Cavaglia:2021bnz} to be used in the following. Finally, we describe the conformal bootstrap problem considered in the rest of the paper. 
 
\subsection{The line CFT and the cusp as its deformation}\label{sec:linecusp}
\paragraph{The line defect.}
The defect CFT we consider lives on an infinite supersymmetric Wilson line in four dimensional ${\cal N} = 4$ SYM. 
 This is the so-called  Maldacena--Wilson line (MWL), defined as
\begin{align}
\label{eqn:MWLdefine}
    {\cal W} = \operatorname{Tr}W_{-\infty}^{+\infty}, \qquad \text{ with }  W_{-\infty}^{+\infty}\equiv \operatorname{P}\exp\int_{-\infty}^{+\infty}dt\,(i\, A_{\mu}\dot{x}^\mu + \Phi_{||}|\dot{x}|)\;,
\end{align}
where $\operatorname{P}$ is the path-ordering and $x^\mu(t)$ is the parametrisation of a straight line.
Here $\Phi_{||}$ denotes one scalar field out of the six real scalars of the theory. The five remaining scalars, that do not couple to the line, are denoted as $\Phi_{\perp}^{i}$ with $i = 1,\cdots, 5$.
The MWL preserves the following symmetries of the full theory:
\begin{itemize}
    \item \texttt{Spacetime symmetries.} The 4D conformal symmetry is broken in presence of the MWL. However, there is an ${\rm SO}(3)$ symmetry of physical rotations about the line. Additionally, the 1D conformal group ${\rm SO}(1,2)$ is preserved along the line. 
    \item \texttt{$R$-symmetry.} Since the MWL couples to only one of the six scalars, the ${\rm SO}(6)$ $R$-symmetry of the parent theory is broken to ${\rm SO}(5)$ in the defect theory. This symmetry represents the flavour symmetry which allows one to rotate the five scalars $\Phi_{\perp}^{i}$ orthogonal to the line. 
    \item \texttt{Supersymmetry.} The MWL is a $1/2$-BPS observable since preserves half of the supersymmetries. This implies that its expectation value is trivial and it is given by $\langle \mathcal{W} \rangle = 1$~\cite{Drukker:1999zq,Erickson:2000af,Zarembo:2002an}. 
\end{itemize}
Altogether, the $1/2$-BPS MWL preserves a ${\rm OSp}(2,2|4)$ subgroup of the superconformal symmetry ${\rm PSU}(2,2|4)$ of the full theory.

\begin{figure}
    \centering
    \includegraphics[scale = 1]{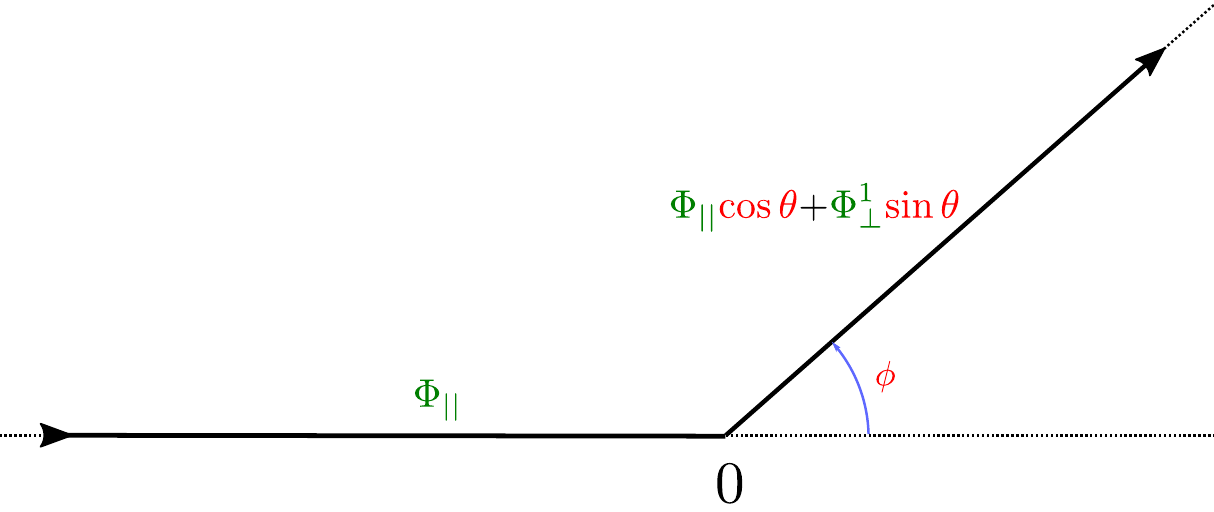}
    \caption{Two semi-infinite Wilson rays meet at the origin at a cusp. This is a cusp both in physical as well as $R$-space. The physical cusp is parametrised by the angle $\phi$ subtended by the second ray on the first. The ``internal'' angle $\theta$ represents a change in the scalar polarisation.}
    \label{fig:cusp}
\end{figure}

\paragraph{The cusp.}
An important integrable deformation of the $\frac{1}{2}$-BPS MWL can be constructed by introducing a cusp, see Fig. \ref{fig:cusp}, where two semi-infinite lines connect. In general this is defined as
\begin{align}
    {\cal W}_< \equiv \operatorname{Tr}\left[ W_{-\infty}^0(0,0)
    W_{0}^{+\infty}(\phi,\theta)\right]
    \;,
\end{align}
where the second infinite segment is rotated both in space-time with angle $\phi$, as well as in the space of scalar couplings with an internal angle $\theta$. Without loss of generality, choosing planes for these rotations we can write
\begin{align}\label{eqn:thetaphinotation}
    {W}_{t_1}^{t_2}(\phi,\theta) &= \operatorname{P}\exp\int_{t_1}^{t_2}dt\bigg[i\, A_\mu \dot{x}^{\mu}(t) + (\Phi_{||}\cos\theta + \Phi^1_\perp\sin\theta)\,|\dot{x}(t)|\bigg]\;,\\
    x(t) &= \big(t\cos\phi,t\sin\phi,0,0\big)\; .
\end{align}

The MWL with a cusp is no longer finite and scales with both UV and IR regulator. In other words it is associated to an anomalous dimension (the \emph{cusp anomalous dimension}) defined through its divergence,  

\begin{align}
   \langle\,{\cal W}_<\,\rangle\sim \left(\frac{\Lambda_{\rm IR}}{\Lambda_{\rm UV}}\right)^{- \Gamma(g,\phi,\theta)}\;,
\end{align}
where  $\Lambda_{\text{UV}}$, $\Lambda_{\text{IR}}$ are UV and IR cutoffs, respectively, screening the points $0$ and $\infty$.  $\Gamma(g,\phi,\theta)$ is called the \textit{generalised cusp anomalous dimension}. It was introduced and studied  at weak and strong coupling in~\cite{Drukker:2011za}. A set of TBA equations for it was introduced  in~\cite{Correa:2012hh,Drukker:2012de} and reformulated using the QSC in~\cite{Gromov:2015dfa}, which allowed for its non-perturbative analysis.
In the near-BPS limit $\phi\to\pm\theta$, the generalised cusp anomalous dimension at order $(\phi-\theta)^2$ is given by 
\beq
\label{dexpphi}
	\Gamma(g,\phi,\theta)=\frac{\cos\phi-\cos\theta}{\sin\phi}\Gamma^{(1)}(g,\phi)
	+\left(\frac{\cos\phi-\cos\theta}{\sin\phi}\right)^2\Gamma^{(2)}(g,\phi)+{O}((\cos\phi-\cos\theta)^3)\;.
\eeq
The first coefficient corresponds to the \textit{generalised Bremsstrahlung function} originally computed at any coupling in \cite{Correa:2012at,Fiol:2012sg} by relating this observable to expectation values of $\frac{1}{4}$-BPS Wilson loops  accessible with  localisation techniques~\cite{Erickson:2000af,Drukker:2000rr,Drukker:2006ga,Pestun:2009nn}.  The result was later reproduced and extended to an infinite family of observables from integrability in  \cite{Gromov:2012eu,Gromov:2013qga} and  then checked at strong and weak coupling in \cite{Sizov:2013joa,Bonini:2015fng}.
It reads
\beq\label{eq:Gamma1}
	\Gamma^{(1)}(g,\phi)\equiv \frac{2\phi}{1-\frac{\phi^2}{\pi^2}} \mathbb{B}_\phi(g)\ =
	\frac{2\phi g}{\sqrt{\pi^2-\phi^2}}
	\frac{I_2\(4 g\sqrt{\pi^2-\phi^2}\)}{I_1\(4 g\sqrt{\pi^2-{\phi^2}}\)} \;,
\eeq
where $I_n(x)$ is the modified Bessel function of the first kind. 

The second coefficient in \eqref{dexpphi} was computed analytically in \cite{Gromov:2015dfa} using the QSC formalism. It can be written in terms of the following double contour integral 
\begin{equation}\begin{split}
\label{Gamma2res}
	\Gamma^{(2)}(g,\phi)&=\!-\frac{1}{2}\!\oint\!\frac{du_x}{2\pi i}\!\oint\!\frac{du_y}{2\pi i}
	u_x u_y
{\Big [ }D_+(x,y)K_{+\phi}(u_x-u_y)
\!+\!D_0(x,y)K_{0}(u_x-u_y)
\\
&\qquad \qquad\qquad \qquad\qquad \qquad \qquad  \qquad \qquad \qquad \!+\!D_-(x,y) K_{-\phi}(u_x-u_y){\Big ] }\\
\equiv &\phi^2 \mathbb{C}_\phi(g) \;,
\end{split}\end{equation}
where both integrals run clockwise around the cut $[-2g,2g]$, and we are using the Zhukovsky parametrisation $u_x = g ( x + 1/x)$, and similarly for $u_y$ and $y$. The kernels $D$ and $K$ are given in Appendix A. In the following we will refer to this observable as \textit{generalised Curvature function} $\mathbb{C}_\phi(g)$.

In this paper we will focus on the case in which the euclidean angle $\phi$ is set to zero and $\theta$ is small. This corresponds to an expansion close to 1/2 BPS Wilson line. Considering that the coefficients appearing in \eqref{dexpphi} scale at small $\phi$ as follows,
\begin{equation}
    \Gamma^{(1)}\sim 2\phi\mathbb{B}(g)+O(\phi^3)\quad\text{and}\quad
    \Gamma^{(2)}\sim \phi^2\mathbb{C}(g)+O(\phi^4) \; ,
\end{equation}
we have
\begin{equation}
    \Gamma(g,\phi=0,\theta\rightarrow 0)=
     \mathbb{B}(g)\,\sin^2\theta
    +\frac{1}{4}\left(\mathbb{B}(g)+\mathbb{C}(g)\right)\,\sin^4\theta +O(\sin^6\theta)\;,
\end{equation}
where $\mathbb{B}$, simply called the Bremsstrahlung function, is given by
\beq\label{eq:B0}
\mathbb{B}(g) = \frac{g}{\pi}\frac{I_2(4 \pi g)}{I_1(4 \pi g)}\;,
\eeq
and $\mathbb{C}$ is the Curvature function\footnote{It is denoted by $f_2(g)$ in \cite{Gromov:2015dfa}.}~\cite{Gromov:2015dfa} 
\beq\label{curvaturedef}
\mathbb{C}(g) = -4\,\mathbb{B}^2(g)
	-\frac{1}{2}\oint\frac{du_x}{2\pi i}\oint\frac{d
 u_y}{2\pi i}
    K_0(u_x-u_y)F[x, y]\;.
\eeq
This result is obtained by  carefully taking the $\phi\rightarrow 0$ limit of the generalised expression  \eqref{Gamma2res}, with the integrands  defined in Appendix \ref{app:curvature}. 
Solving the integral \eqref{curvaturedef}, one can compute the perturbative expansion of the curvature function at weak coupling, 
\beq\begin{split}\label{eq:weakC0}
\mathbb{C}&=
4 g^4-\left(24 \zeta_3+\frac{16 \pi ^2}{3}\right)g^6+ \left(\frac{64 \pi ^2 \zeta_3}{3}+360 \zeta_5+\frac{64 \pi ^4}{9}\right)g^8\\
&- \left(\frac{112 \pi ^4 \zeta_3}{5}+272 \pi ^2 \zeta_5+4816 \zeta_7+\frac{416 \pi ^6}{45}\right)g^{10}\\
&+\left(\frac{3488 \pi ^6 \zeta_3}{135}+\frac{2192 \pi ^4 \zeta_5}{9}+\frac{9184 \pi ^2 \zeta_7}{3}+63504 \zeta_9+\frac{176 \pi
  ^8}{15}\right)g^{12} +O\left(g^{14}\right)
\end{split}\; , \eeq
and at strong coupling
\beq\begin{split}
\mathbb{C}&=
\frac{\left(2 \pi ^2-3\right) g}{6 \pi ^3}+\frac{-24 \zeta_3+5-4 \pi
   ^2}{32 \pi ^4}+\frac{11+2 \pi ^2}{256 \pi ^5 g}+\frac{96 \zeta_3+75+8 \pi ^2}{4096 \pi ^6 g^2}\\
   &+\frac{3 \left(408 \zeta_3-240
   \zeta_5+213+14 \pi ^2\right)}{65536 \pi ^7 g^3}+\frac{3 \left(315
   \zeta_3-240 \zeta_5+149+6 \pi ^2\right)}{65536 \pi ^8
   g^4}+O\left(\frac{1}{g^5}\right)\label{eq:strongC}
\end{split}\;,\eeq
matching the first two perturbative orders obtained by direct field theory and string theory computations in \cite{Drukker:2011za}.
The expression at strong coupling is a new result of this paper. In order to compute it, we solved the integral \eqref{curvaturedef} numerically with very high precision and then we converted the result in a linear combination of Riemann Zetas.

\subsection{1D CFT on the line}

The focus of our study is the 1D CFT that lives on the $1/2$-BPS MWL~\cite{Drukker:2006xg}.  Its correlation functions are defined by the expectation values of \emph{operator insertions on the line}
\begin{equation}
    \left\langle\left\langle O_{1}\left(t_{1}\right) O_{2}\left(t_{2}\right) \cdots O_{n}\left(t_{n}\right)\right\rangle\right\rangle
    \equiv \langle
    \operatorname{Tr}
   \operatorname{P}{ O}_1(t_1)
 \,{ O}_{2}(t_{2})
 \,   \ldots 
\, O_n(t_n)\, W_{-\infty}^{+\infty}
    \rangle/\langle  W_{-\infty}^{+\infty} \rangle
    \; ,
\end{equation}
where $O_i$ are composite fields transforming in the adjoint representation of the gauge group. 
 It descends from the embedding in $\mathcal{N}$=4 SYM that such correlators satisfy the properties of $n$-point functions of a 1D conformal field  theory~\cite{Drukker:2006xg}.  
 
  The states of the  CFT live in unitary representations of the ${\rm OSp}(2,2|4)$ symmetry left unbroken by the defect. Thus, they can be organised in superconformal multiplets labelled by four quantum numbers --- the scaling dimension $\Delta$ associated with the 1D conformal group, the ${\rm sp}(4)\cong {\rm so}(5)$ Dynkin lables $\[a,b\]$ associated with the $R$-symmetry, and the spin $s$ associated with ${\rm su}(2) \cong {\rm so}(3)$ symmetry of rotations about the line. These representations were classified in \cite{Gunaydin:1990ag,Liendo:2016ymz}.

In particular, the CFT admits $\frac{1}{2}$-BPS multiplets denoted as ${\cal B}_{k}$, whose superconformal primaries have the quantum numbers
\begin{align}
    \{\Delta, \[a,b\], s\} = \{k,\[0,k\],0\}\;, \quad k\in \mathbb{Z}\; .
\end{align}
The conformal dimension of these operators is protected by supersymmetry. Two of such multiplets play an important role in our
setup, which is the same considered in \cite{Giombi:2017cqn,Liendo:2016ymz,Liendo:2018ukf,Ferrero:2021bsb}. 

First is the multiplet $\mathcal{B}_1$. This contains the simplest superconformal primary operators of the theory, corresponding to the orthogonal scalars $\Phi^i_\perp$, $i = 1,\dots, 5$, and is also known as the displacement multiplet. In fact, it contains the components of the field-strength corresponding to the displacement operator associated to translations perpendicular to the Wilson line. Furthermore, the primary operators $\Phi_{\perp}^i$ can be seen as displacement operators corresponding to broken symmetries in R-space.

The other important multiplet for our analysis is $\mathcal{B}_2$, which contains the superconformal primary operators $\Phi_{\perp}^{\left\{i  \right.} \Phi_{\perp}^{\left.\;j\right\}} - \frac{1}{5}\delta^{ij} (\Phi_{\perp}\cdot  \Phi_{\perp} )$ of protected dimension $2$.

In addition there are the long multiplets which in principle preserve no supercharges. They are denoted by ${\cal L}^{\Delta}_{s,\[a,b\]}$ with the indices corresponding to the quantum numbers of the primary. For long operators, $\Delta$ is a non-trivial function of the coupling $g$. 

In \cite{Liendo:2018ukf}, several selection rules for the OPE in the CFT were deduced. In particular, the following will be relevant for us: at finite coupling,
\beq\label{eq:OPEfusion}
\underbrace{\mathcal{B}_1 \times \mathcal{B}_1 }_{\text{OPE}} = \mathcal{I} + \mathcal{B}_2 + \sum_{\Delta>1} \mathcal{L}_{[0,0]}^{\Delta} \;,
\eeq
where $\mathcal{I}$ denotes the identity multiplet, and ${\cal L}_{0,\[0,0\]}^\Delta$ are the non-protected multiplets transforming as singlets under the global ${\rm SO}(5)\times{\rm SO}(3)$ symmetry. There are infinitely many multiplets with such quantum numbers, all with unprotected scaling dimensions. As shown in (\ref{eq:OPEfusion}), the whole infinite set of such multiplets can appear in the fusion of two operators in the displacement multiplet, and this is the basis of the bootstrap problem discussed in section \ref{sec:bootstrapsetup}.

\subsubsection{Spectrum}

The QSC method was shown to be applicable to compute the spectrum of such neutral operators in~\cite{Grabner:2020nis}, where the non-perturbative scaling dimension of the lightest state was obtained. 
The QSC equations relevant to this case descend from the ones written down for the cusped Wilson line in~\cite{Gromov:2015dfa}. In fact, as first noticed in a special limit in \cite{Cavaglia:2018lxi}, the cusp QSC equations admit infinitely many solutions corresponding to operator insertions at the cusp with the same quantum numbers as the vacuum. Then, in~\cite{Grabner:2020nis}  it was found how to take the non-trivial straight-line limit to describe states of the defect CFT. 

It is currently understood that solutions of the QSC equations in \cite{Grabner:2020nis} are in one-to-one correspondence with the multiplets $\mathcal{L}_{0,[0,0]}^{\Delta}$. 
A systematic way to find solutions with $\Delta > 1$, by first solving the equations at weak coupling, was developed in~\cite{Julius:2021uka}, giving access to an infinite family of states, and a generalisation of these techniques to the entire singlet sector of the defect CFT was developed shortly after~\cite{spec1DCFT,Cavaglia:2021bnz}. A plot of 35 states in the spectrum in the singlet sector was presented in~\cite{Cavaglia:2021bnz}. Details of  the setup and of the nontrivial techniques to find excited state solutions of the QSC will be provided elsewhere, along with a generalisation to long operators which carry a non-zero $R$-charge~\cite{spec1DCFT}.

In the analysis of this paper, we will need the values of the lowest lying 10 states in the singlet sector~{\it cf.}~figure~\ref{fig:spec}. Perturbative data for these states are given in appendix~\ref{apd:anylSpec}. Their numerical values with at least 12 digits precision are listed for several values of the coupling constant  in appendix~\ref{apd:numSpec}. 
Both the perturbative and numerical data are also shared in a~\texttt{Mathematica} notebook attached to this paper.
Inclusion of the other states does not seem to lead to a significant improvement of the bounds obtained in this paper.

\begin{figure}
    \centering
    \includegraphics[scale=1.5]{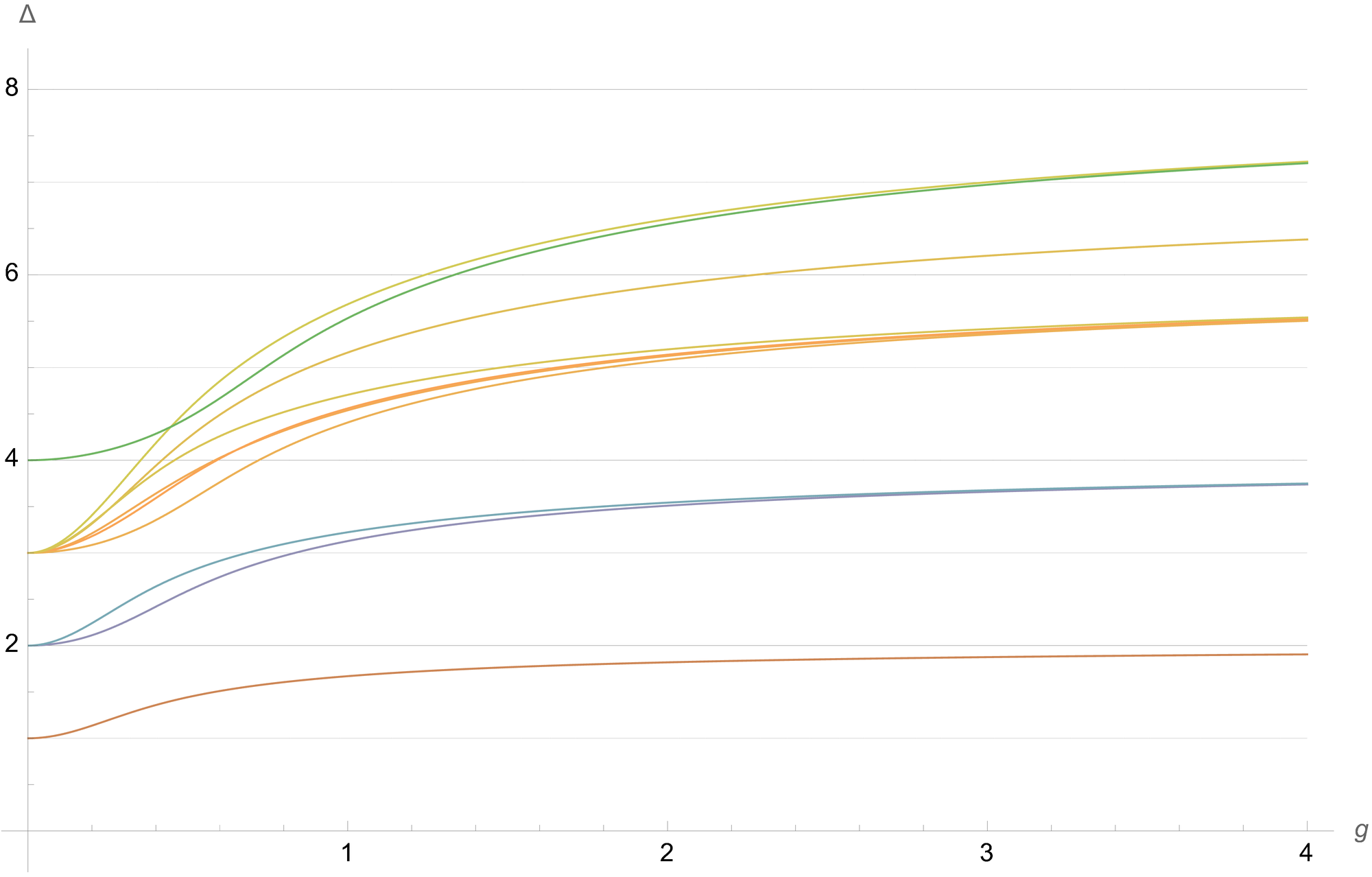}
    \caption{The first 10 lowest-lying states of the spectrum computed with the QSC. These levels will be the input in the bootstrap algorithms of this paper. For a plot including 35 states, see \cite{Cavaglia:2021bnz}. In this paper, we label these states, for any given value of the coupling, as $\left\{\Delta_n\right\}$,  ordered as $\Delta_{n}<\Delta_{n+1}$. }
    \label{fig:spec}
\end{figure}

\subsubsection{Conformal bootstrap setup}\label{sec:bootstrapsetup}
Following \cite{Giombi:2017cqn,Liendo:2018ukf}, we  study the 4-point function of four identical scalars polarised in the same direction, which we take for definiteness to be $\Phi_{\perp}^1$. Due to conformal symmetry, their correlator can be written in terms of a function of a single variable\footnote{See appendix \ref{app:covariant} for a covariant expression in the R-symmetry indices.}
\beq
\langle \langle \Phi_{\perp}^1(x_1) \Phi_{\perp}^1(x_2) \Phi_{\perp}^1(x_3) \Phi_{\perp}^1(x_4)\rangle\rangle = {G}(x)\; \left( \langle \langle \Phi_{\perp}^1(x_1) \Phi_{\perp}^1(x_2) \rangle \rangle\, \langle \langle \Phi_{\perp}^1(x_3) \Phi_{\perp}^1(x_4) \rangle \rangle \right)\;\;, 
\eeq
where we normalised by the 2-point functions $\langle \langle \Phi_{\perp}^i(x_i) \Phi_{\perp}^j(x_j) \rangle \rangle \propto x_{ij}^{-2}\,\delta_{ij}$,  and introduced the cross ratio:
\beq
 x\equiv \frac{x_{12} x_{34}}{x_{13} x_{24}}, \;\;\;\;x_{ij}\equiv x_i - x_j .
\eeq
The invariance under the cyclic relabeling $(1234)\rightarrow (2341)$ gives the crossing equation:
\beq
x^2 {G}(1 - x) - (1-x)^2 {G}(x) = 0.
\eeq
To setup a bootstrap problem we decompose the correlator using the OPE. To take into account supersymmetry, it is best to use superconformal blocks. We follow the results of \cite{Liendo:2018ukf}. 
To write down the superconformal OPE, it is convenient to parametrise the 4-point function as
\beq\la{pt4}
{G}(x) = \mathbb{F} \;x^2 +  (2 x^{-1} - 1)f(x) -\left(x^2 - x +1\right)f'(x)\; ,
\eeq
where the function $f(x)$ satisfies crossing in the form
\beq\label{eq:fcrossing}
x^2 f(1 - x) + (1-x)^2 f(x) = 0\; .
\eeq
In the following it is convenient to notice that the crossing-invariant combination $G(x)/x^2$ is a total derivative:
\beq\label{eq:totalder}
\frac{G(x)}{x^2} = \partial_{x}\left( \mathbb{F} x - \( 1 - \frac{1}{x} +\frac{1}{x^2} \) f(x) \right) \;.
\eeq
In this paper, we use the following notation for the spectrum of non-protected operators $\left\{\Delta_n \right\}_{n=1}^{\infty}$ -- each level denoting the lowest scaling dimension in a multiplet $\mathcal{L}_{0,[0,0]}^{\Delta_n}$. We will order the states according to $\Delta_n < \Delta_{n+1}$.\footnote{Notice that there are some level crossings in the spectrum, see Figure \ref{fig:spec}. We keep our naming convention  separately for each value of $g$.} We denote the corresponding OPE coefficients as \beq
C_n \equiv C_{\Phi_{\perp}^1 ,\; \Phi_{\perp}^1 ,\;\mathcal{L}_{0,[0,0]}^{\Delta_n}}.\eeq
The OPE decomposition is:
\begin{equation}\label{eq:OPEf}
f(x) = F_{\mathcal{I}}(x) + { {C^2_{\rm BPS} \,  {F}_{\mathcal{B}_2}(x)}}  + \sum_{n } { {C^2_{n} \,  {F}_{{\Delta_n}}(x)}} \; ,
\end{equation}
where the superconformal blocks are 
\beqa\la{superblocks}
F_{\mathcal{I}}(x) &=& x\;,\\
F_{\mathcal{B}_2}(x) &=& x - x\, _2F_1(1,2,4;x ) \;,\\
F_{{\Delta}}(x) &=& \frac{x^{\Delta+1}}{1-\Delta}\, _2F_1(\Delta+1,\Delta+2,2 \Delta+4;x )\; ,
\eeqa
and $C_n$ are OPE coefficients for the non-protected states, which are nontrivial functions of the coupling and the main objective of our work. Finally, the constant $\mathbb{F}$  and the OPE coefficients corresponding to the $\mathcal{B}_2$ block are related as $\mathbb{F}(g) = 1 + C^2_{BPS}(g)$, and this OPE coefficient was fixed by comparison with a topological observable computable with localisation~\cite{Giombi:2017cqn,Liendo:2018ukf}. The same relation can also be derived using integrability as shown in Appendix \ref{app:CBPS}. The result reads
\beq
\mathbb{F}(g) = 1 + C^2_{BPS}(g) = \frac{3 I_1(4 g \pi ) \left(\left(2 \pi ^2 g^2+1\right) I_1(4 g \pi )-2 g \pi  I_0(4 g \pi )\right)}{2 g^2 \pi ^2 I_2(4 g \pi ){}^2} \;\;, \eeq
which can also be recast in terms of the  Bremsstrahlung function:
\beq\label{eq:F0}
\mathbb{F}(g) =  \frac{3 (g^2 -\mathbb{B}(g))}{\pi^2 (\mathbb{B}(g))^2}\;\;.
\eeq
The conformal bootstrap constraint, arising from the compatibility of the OPE decomposition with crossing, takes the form:
\begin{framed}
\beq\label{eq:bootstrapeq}
\sum_{n} C^2_n \;\mathcal{G}_{\Delta_n}(x) + \mathcal{G}_{\text{simple}}(g, x) = 0\;,
\eeq
\end{framed}
\noindent
where we introduced the \emph{crossed superconformal blocks}
\beq\label{eq:defGD}
\mathcal{G}_{\bullet}(x) \equiv (1 - x)^2 F_{\bullet}(x) +  x^2 F_{\bullet}(1 - x) , \qquad \bullet \in \left\{ \mathcal{I},\mathcal{B}_2, \Delta\right\}\;\;,
\eeq
and $\mathcal{G}_{\text{simple}}(g, x)$ is an explicitly known function:
\beq\label{eq:defGsimple}
\mathcal{G}_{\text{simple}}(g, x) \equiv \mathcal{G}_{\mathcal{I}}(x) + C^2_{BPS}(g)\; \mathcal{G}_{\mathcal{B}_2}(x)\;.
\eeq
The bootstrap constraint (\ref{eq:bootstrapeq}) contains two sets of nontrivial quantities: the scaling dimensions and OPE coefficients $\left\{C_n\right\}$ for the nontrivial operators. In our approach, we take advantage of the fact that integrability effectively solves the problem of computing the spectrum, and focus on determining the OPE coefficients. In the following section we present two new exact relations involving this correlation function. 

\section{Integrated
correlators}\label{sec:integrated}
The main new ingredient of this paper is the inclusion of the constraints on the 4-point function $G(x)$ arising from  integrable deformations of the straight line (in addition to the spectral data coming from the QSC which were already used  in~\cite{Cavaglia:2021bnz}). 
In this section we describe in detail these constraints, which take the form of integrals over the cross ratio for the amplitude $G(x)$ or, equivalently, $f(x)$. 
\subsection{New integral constraints}
We claim that the 4-point correlator introduced in section \ref{sec:bootstrapsetup} satisfies two integral identities involving the Bremsstrahlung and Curvature functions~\cite{upcomingAJMNderivation}:
\begin{framed}
\begin{align}
  &\text{ Constraint 1:  }&&  \int_0^1
    \delta G(x)\frac{1+\log x}{x^2}dx=\frac{3\mathbb{C}-\mathbb{B}}{8 \;\mathbb{B}^2}\; , \la{eq:constr1}\\
   & \text{ Constraint 2:  }&& \int_{0}^1 dx \frac{\delta f(x)}{x} = 
        \frac{\mathbb{C}}{4\;\mathbb{B}^2} + \mathbb{F}-3 \; ,\label{eq:constr2}
\end{align}
\end{framed}
\noindent
where $
\delta G(x) \equiv G(x) - G_{\text{weak}}^{(0)}(x)$, $ \delta f(x) \equiv f(x) - f_{\text{weak}}^{(0)}(x)$, and $G_{\text{weak}}^{(0)}$, $f_{\text{weak}}^{(0)}$ are the zero-coupling values:
\beq\label{fandGtree}
f_{\text{weak}}^{(0)}(x)=2 x+\frac{x}{x-1}\;, \;\;\; G_{\text{weak}}^{(0)}(x) = \frac{2 (x-1) x+1}{(x-1)^2}
\; ,
\eeq
which can be easily deduced from free field theory, i.e., 
\beq
\frac{G_{\text{weak}}^{(0)}(x)}{x_{12}^2 x_{34}^2 } = \frac{1}{x_{12}^2 x_{34}^2 } + \frac{1}{x_{14}^2 x_{23}^2 }\;.  
\eeq
We notice that both integrals are convergent at finite $g$, as \beqa\la{limits}
\delta f(x)\simeq \frac{(3-\mathbb{F} )}{2} \, x^2 
\;\;&,&\;\;
\delta G(x)\simeq \frac{(4 \,\mathbb{F}-1 )}{5} \, x^2   
\;\;,\;\;x\to 0\; 
,\\
\delta f(x)\simeq \frac{(\mathbb{F}-3)}{2} 
\;\;&,&\;\;
\delta G(x)\simeq \frac{3(\mathbb{F}-1 )}{2}
\;\;,\;\;x\to 1 \; ,\label{eq:limits2}
\eeqa
where corrections are $O(x^{\Delta_1+1})$ for $x\sim0$, $O((1-x)^{\Delta_1+1})$ for $x\sim 1$, scaling with the power $\Delta_1+1>2$, as follows from the OPE \eq{eq:OPEf} and crossing symmetry.
 Using (\ref{eq:totalder}) and integrating by parts,  (\ref{eq:constr1}) can also be rewritten in terms of $f(x)$ as 
\beqa\la{eq:constr1b}
    \int_{\delta_x}^1
     \delta f(x)\( 
    \frac{1}{x}+\frac{1}{x^3}
    \)dx
    -\frac{1}{2} (\mathbb{F}-3) \log {\delta_x}-\mathbb{F}+3
    &=&\frac{3\mathbb{C}-\mathbb{B}}{8 \;\mathbb{B}^2}\;,
\eeqa
in the limit of $\delta_x\rightarrow 0^+$. 
We see that the divergence in \eq{eq:constr1b} cancels as a consequence of \eq{limits}.

\paragraph{Heuristic explanation of the integral relations.} 
While a detailed proof will be presented in \cite{upcomingAJMNderivation}, let us discuss here the main steps. The existence of the constraints is based on a general principle~\cite{Cooke:2017qgm}:

 \emph{Every deformation of a Wilson line can be parametrised in terms of integrated correlators on the original line.}
 
This follows from the fact that operator insertions represent infinitesimal, localised deformations. Any general  deformation (both in space-time as in R-space) can be approached through a series of integrated correlators (see \cite{Cooke:2017qgm} for a systematic treatment of a spacetime deformation in perturbation theory). An example of this principle in action and an anticipation of the idea outlined below is presented in Appendix \ref{app:CBPS}. 

To deduce the constraints (\ref{eq:constr1}) and (\ref{eq:constr2}), 
we analyse two special types of deformations in R-space, which, at leading order, are related to  integrated values of $\langle\langle \Phi_{\perp}^i \Phi_{\perp}^i \rangle\rangle$. This was shown in \cite{Correa:2012at}, leading to the first determination of the Bremsstrahlung function $\mathbb{B}$. The constraints (\ref{eq:constr1}) and (\ref{eq:constr2}) arise from the extension of this analysis to the next order in the deformation parameters, as we describe in more detail below. 

First, we consider the deformation obtained by forming a cusp in R-space: i.e., we switch on the internal angle $\theta$ on half of the line, as discussed in section \ref{sec:linecusp}, while keeping $\phi = 0$. As shown in \cite{Correa:2012at}, at order ${\cal O}(\theta^2)$ this connects the integrated values  of $\langle\langle \Phi_{\perp}^i \Phi_{\perp}^i \rangle\rangle$ to $\mathbb{B}$ arising from the expansion of the cusp -- fixing in this way the normalisation the 2-point function. 
 We extended the analysis to the next order ${\cal O}(\theta^4)$, finding a relation between the Curvature function \eq{curvaturedef} and integrated 4-point functions of $\Phi_{\perp}^i$. 
 It will be shown in  \cite{upcomingAJMNderivation}  how this leads to a linear combination of the two constraints  (\ref{eq:constr1}) and (\ref{eq:constr2}). 

A second important deformation of the line is the  $\frac{1}{4}$-BPS deformation  defined in \cite{Drukker:2006ga}. This  is defined by an explicit parameter $\vartheta$ in the gauge connection. On a circular contour, the expectation value is known explicitly for any value of $g$ and of the deformation parameter~\cite{Drukker:2006ga,Pestun:2009nn}: it is equivalent to the vev on a $\frac{1}{2}$-BPS MWL, with a redefinition of the coupling $g\rightarrow g' = g \cos\vartheta$. The perturbative expansion in $O(\vartheta)$ generates a  series of identities for  correlators of $\Phi_{\perp}^i$, now integrated on the circular $\frac{1}{2}$-BPS line. 
At order $O(\vartheta^2)$, one again finds integrated 2-point functions, and this observation allowed the authors of  \cite{Correa:2012at} to compute $ \mathbb{B}$.  Extending the analysis to  the next order in $\vartheta$ leads to a second independent linear combination of the constraints (\ref{eq:constr1}), (\ref{eq:constr2}). 

While the above discussion sketches the main physical intuition, the full derivation of the constraints at finite coupling requires careful treatment. In fact, integrated correlators generated in these expansions produce UV divergences, which should be removed through a consistent regularisation scheme while preserving the symmetries of the setup. The derivation at finite coupling requires the use of next-to-leading order conformal perturbation theory, and will be presented in \cite{upcomingAJMNderivation}.

\subsection{Tests of the relations at strong coupling}
Strong coupling results for the 4-point function were recently obtained using the functional analytic bootstrap in \cite{Ferrero:2021bsb}, where results for $f(x)$ and $G(x)$ to the first five orders at large $g$, where these functions expand as 
\beq\label{fstrongexp}
G(x) = \sum_{\ell=0}^{\infty}\frac{G_{\text{strong}}^{(\ell)}(x)}{(4\pi g)^\ell}
\qquad
\text{and}
\qquad
f(x) = \sum_{\ell=0}^{\infty}\frac{f_{\text{strong}}^{(\ell)}(x)}{(4\pi g)^\ell}
\qquad g\rightarrow\infty\;.
\eeq
The first terms for the four-point function $G$ are
\beq\begin{split}
G_{\text{strong}}^{(0)}(x)&=\frac{((x-1) x+1)^2}{(x-1)^2}\;,    \\
G_{\text{strong}}^{(1)}(x)&=\frac{x^2 ((x-1) x (x (2 x-5)+4)+2) \log (x)-2 (x-1) ((x-1) x+1)^2}{(x-1)^3} \\
&\quad+\frac{\left(-2 x^4+x^3+x-2\right) \log (1-x)}{x}\;,
\end{split}\eeq
and for the reduced correlator $f$ are
\beqa\begin{split}\label{fstron01}
f_{\text{strong}}^{(0)}(x)&\equiv \frac{x ^2}{x - 1 } + x \;,\\ f_{\text{strong}}^{(1)}(x) &\equiv \frac{2 (x +1) \log (1-x ) (x -1)^3-2x  (2 x -1) (x -1)-2 (x -2) x ^3 \log (x )}{ (x -1)^2} \;.
\end{split}\eeqa
The remaining orders up to $\ell=4$ can be found in the supplementary material of \cite{Ferrero:2021bsb}. Plugging these results into the l.h.s. of the constraints and performing the integrations, one finds, from  (\ref{eq:constr1}),  
\beq
-\frac{3}{16 \pi  g}-\frac{3 (24 \zeta_3+7)}{256 \pi ^2 g^2}-\frac{3 (144 \zeta_3+7)}{2048 \pi ^3 g^3}-\frac{9 (352 \zeta_3-49)}{32768 \pi ^4 g^4}+\dots
= \frac{3\mathbb{C}-\mathbb{B}}{8 \;\mathbb{B}^2} \;,
\eeq
while (\ref{eq:constr2}) yields
\beq
\frac{2 \pi ^2-21}{24 \pi g}+\frac{4 \pi ^2-7-24 \zeta_3}{128 \pi ^2 g^2}+\frac{10 \pi ^2+83-144 \zeta_3}{1024 \pi ^3 g^3}+\frac{40 \pi ^2+867-1056 \zeta_3}{16384 \pi ^4 g^4}+\dots=  \frac{\mathbb{C}}{4\;\mathbb{B}^2} + \mathbb{F}-3 \;.
\eeq
These relations can be easily verified using the strong coupling expansions of  (\ref{eq:B0}),(\ref{eq:F0}) and (\ref{eq:strongC}).

\subsection{Tests at weak coupling}\label{sec:weakcouplingintegrals}
The functions $G(x)$ and $f(x)$
 also have a regular expansion at weak coupling,
\beq\label{fandG}
G(x) = \sum_{\ell=0}^\infty G^{(\ell)}_{\text{weak}}(x)\,g^{2\ell}\qquad\text{and}\qquad
f(x) = \sum_{\ell=0}^\infty
f^{(\ell)}_{\text{weak}}(x)\,g^{2\ell},
\qquad g\rightarrow 0 \;, 
\eeq
which furthermore should have a finite radius of convergence. Indeed, the general expectation for observables in planar $\mathcal{N}$=4 SYM is that the radius of convergence is $|g|<\frac{1}{4}$, see \cite{Marboe:2014gma,Volin:2008kd,ShotaTalk}. The leading order of this expansion is given in (\ref{fandGtree}), while the next-to-leading order is~\cite{Kiryu:2018phb}
\beqa\label{f1weak0}
f_{\text{weak}}^{(1)}(x) &=&
\frac{2 x}{3 (1-x)}  \left(6 \text{Li}_2(x ) +3 \log (1-x ) \log (x
   )-\pi ^2 x\right)\;,\\
  G_{\text{weak}}^{(1)}(x) &=& \left(-2 x-\frac{2}{x-1}\right) \log (1-x)+\left(\frac{2 x ((x-1) x+1)}{(x-1)^2}+\frac{(2-4 x) \log (1-x)}{(x-1)^2}\right) \log (x)\nn\\
  &&+\frac{2 \left(\pi ^2 x^2+(6-12 x) \text{Li}_2(x)\right)}{3 (x-1)^2} \;. \label{eq:G1weak}
\eeqa
Plugging (\ref{fandGtree}), (\ref{f1weak0}) into the second integral relation  (\ref{eq:constr2}) and computing the integrals, we get a perfect match with the expansion of the r.h.s., 
\beq
 0 \times g^0+ \left(\frac{2 \pi ^2}{3}-6 \zeta_3\right)g^2+ O(g^4)\simeq\frac{\mathbb{C}}{4\;\mathbb{B}^2} + \mathbb{F}-3 \;, \qquad \quad g\rightarrow 0.
\eeq
Verifying the first integral relation (\ref{eq:constr1}) in this regime, however, is more subtle. 
 In fact, plugging the weak coupling expansion of $G(x)$ into the l.h.s. of (\ref{eq:constr1})  leads, order by order for $\ell>0$,  to integrals of the form
 \beq\label{eq:pertintegr}
 \int_0^1  G_{\text{weak}}^{(\ell)}(x) \frac{1+\log x}{x^2} dx \;,
 \eeq
 which are log-divergent. 
 The reason is that at weak coupling $\Delta_1\to 1$, which creates additional divergences at each given order in $g\to 0$ due to the correction terms in (\ref{limits}).
 On the other hand, the integral $\int_0^1 \delta G(x) \frac{1+\log x}{x^2} dx$ is perfectly convergent at finite coupling, and -- \emph{after computing the integral at the non-perturbative level} -- the result can be expanded giving rise to a well-behaved weak coupling expansion (which, however, contains also one negative power of the coupling). We will explain this point in detail in section~\ref{sec:numerology}, where we will also use the integral relations for the analytic evaluation of the structure constants.

\section{Numerical Bootstrability}\label{sec:numerical}

The bootstrap constraint  (\ref{eq:bootstrapeq}) presents us with a functional equation linear in the OPE coefficients, depending parametrically on the spectrum. 

The equation is linear in the OPE coefficients -- which are our only unknowns since we know the spectrum from the QSC. Nevertheless, we are still faced with two challenges.  First, while integrability allows us \emph{in principle} to compute the scaling dimension of any state, in practice we can only focus on a finite number of them.\footnote{It is an interesting challenge to understand how to use the QSC formalism to deduce global properties of a spectrum (e.g., the asymptotic density of states or expansions around large quantum numbers).} This means we need a controlled way to truncate the bootstrap constraint to a finite number of levels.
Secondly, we need to deal with the functional nature of the equation. The Numerical Conformal Bootstrap (NCB) approach allows us to solve these challenges and obtain rigorous bounds. Here we describe a number of possible NCB algorithms which incorporate knowledge of the spectrum, in increasing degree of complexity. We will then describe how to include the new integral relations in this setup, which will lead us to the main numerical results of the paper. 
\subsection{Basics}\label{sec:algorithms0}
Let us start by reviewing well known aspects of the modern  NCB. Excellent reviews are  availalble~\cite{Rattazzi:2008pe,Chester:2019wfx,Poland:2018epd}, so we keep the discussion short.
\paragraph{The linear functionals approach. }
The 1D CFT we are dealing with is unitary (which descends from the ambient theory $\mathcal{N}$=4 SYM), which implies that $C_n^2\geq 0$. In the NCB approach for unitary theories, one exploits this fact to turn the bootstrap equation  into a set of inequalities, which bound the conformal data. 

A convenient way to deduce such constraints is to act on the bootstrap equation (\ref{eq:bootstrapeq}) with a \emph{linear functional}, which transforms  functions of the cross ratio $x$ into numbers. One typically considers functionals obtained as linear  combinations of derivatives acting on a specific point, 
for example\footnote{We consider only even derivatives, since we will act with these functionals on crossed conformal blocks $\mathcal{G}_{\Delta}$ at $x=\frac{1}{2}$. For an odd number of derivatives the action is trivial: $\partial_x^{2 n+1} \left.\mathcal{G}_{\Delta}(x) \right|_{x=\frac{1}{2}}$ = 0. }
\beq\label{eq:derspace}
\alpha \left[ F(x) \right]\equiv \sum_{n = 0}^{N_{\text{der}/2} } {A}_n \;\partial_x^{2 n} \left. F(x) \right|_{x=\frac{1}{2}}\;,
\eeq
for some coefficients $\vec{A}$, where the point $x=\frac{1}{2}$ is chosen because it guarantees maximal convergence of the OPE. The value of $N_{\text{der}}\in 2 \mathbb{N}$ gives a truncation on the space of functionals and will be a parameter in the approach. 

Acting with $\alpha$ on the bootstrap constraint (\ref{eq:bootstrapeq}), we get a linear equation for the OPE coefficients 
\beq\label{eq:bootstraplinear}
\sum_{n} C^2_n \; \alpha\left[ \mathcal{G}_{\Delta_n}\right] +\alpha\left[ \mathcal{G}_{\text{simple}} \right]= 0 \; ,
\eeq
where the cross-ratio dependence is removed,
and $\mathcal{G}_{\Delta}$, $\mathcal{G}_{\text{simple}}$ are defined in (\ref{eq:defGD}),(\ref{eq:defGsimple}). Given a functional of the form (\ref{eq:derspace}), this concretely is rewritten as
\beq\label{eq:vectorform}
\sum_{n} C^2_n \; \left(\vec{A}\cdot \vec{V}_{\Delta_n}\right) + \left(\vec{A}\cdot \vec{V}_{\text{simple}}\right)= 0 \;,
\eeq
which is true for arbitrary coefficients $\vec{A} = (A_0,A_1,\dots A_{N_{\text{der}}/2})$ , and where
\beqa
\vec{V}_{\Delta}&\equiv& \left. \left(  \mathcal{G}_{\Delta}(x)\,,\, \partial_x^2\mathcal{G}_{\Delta}(x), \dots, \partial_x^{2 n}\mathcal{G}_{\Delta}(x)\,,\, \dots\,,\, \partial_x^{N_{\text{der}}}\mathcal{G}_{\Delta}(x) \right)\right|_{x = \frac{1}{2}}\;,\label{eq:VDelta}\\
\vec{V}_{\text{simple}} &\equiv& \left. \left(  \mathcal{G}_{\text{simple}}(x)\,,\, \partial_x^2\mathcal{G}_{\text{simple}}(x)\,,\, \dots\,, \, \partial_x^{2 n}\mathcal{G}_{\text{simple}}(x)\,, \,\dots\,,\, \partial_x^{N_{\text{der}}}\mathcal{G}_{\text{simple}}(x) \right)\right|_{x = \frac{1}{2}} \;.\label{eq:Vsimple}
\eeqa
Starting from section \ref{sec:incorporating}, we will get other systems of equations of the form (\ref{eq:vectorform}), but for more more general definitions of $\vec{V}_{\Delta}$, $\vec{V}_{\text{simple}}$. 

The main principle of the NCB is finding appropriate functionals  allowing to extract maximal information from the equation. In particular, we will look for $\alpha$ which is positive semi-definite above a threshold $\Delta_{\ast}$:
\beq\label{eq:posi}
\texttt{Positivity condition: } \;\;\; \vec{A}\cdot \vec{V}_{\Delta}  \geq 0 , \;\;\;\; \forall \Delta \geq \Delta_{\ast} \;, 
\eeq
which allows to deduce an inequality:
\beq\label{eq:inequa}
\sum_{\left\{\Delta_n\right\},| \Delta_n < \Delta_{\ast} } \left(\vec{A}\cdot \vec{V}_{\Delta_n}\right) \; C^2_n + \left(\vec{A}\cdot \vec{V}_{\text{simple}}\right) \leq 0 \;.
\eeq
This relation involves the sum over a finite number of states in the spectrum and gives a rigorous truncation of the bootstrap.  

\paragraph{OPE bounds algorithm. } 
By further specifying the properties of the functional, one can deduce constraints on the OPE coefficients. Here we start by revising the algorithm used in our previous paper  \cite{Cavaglia:2021bnz}. It is an adaptation of a well-known NCB algorithm (see e.g.\cite{ Chester:2019wfx}) to our situation, where we know the spectrum exactly.
\begin{framed}
\texttt{Algorithm 1 (Upper Bound). }
We start from the bootstrap constraint written in the form (\ref{eq:vectorform}). 
We will scan among the functionals satisfying the \emph{positivity condition} (\ref{eq:posi}), with the threshold coinciding with the second state in the spectrum,  $\Delta_{\ast} \equiv \Delta_2$. Among such functionals, we search for $\vec{A}^{\text{up}}$ such that:
\begin{enumerate}[1)]
\item $(\vec{A}^{\text{up}}\cdot \vec{V}_{\Delta_1}) = 1 $,
\item $(\vec{A}^{\text{up}}\cdot \vec{V}_{\text{simple}}) $ is maximal.
\end{enumerate}
Then, (\ref{eq:inequa}) gives the optimal bound
\beq
C_1^2 \leq - (\vec{A}^{\text{up}}\cdot \vec{V}_{\text{simple}})\;.
\eeq
\end{framed}
\noindent
A small tweaking of the algorithm gives us a lower bound. 
\begin{framed}
\texttt{Algorithm 1 (Lower Bound). }
Again, choose the positivity threshold as the second state in the spectrum,  $\Delta_{\ast} \equiv \Delta_2$. Among the functionals satisfying the \emph{Positivity Condition}  (\ref{eq:posi}), search for $\vec{A}^{\text{low}} $ such that:
\begin{enumerate}[1)]
\item $ (\vec{A}^{\text{low}}  \cdot \vec{V}_{\Delta_1} ) = -1 $,
\item $(\vec{A}^{\text{low}}  \cdot \vec{V}_{\text{simple}} )$ is maximal.
\end{enumerate}
Then, (\ref{eq:inequa}) gives the optimal bound
\beq
C_1^2 \geq  (\vec{A}^{\text{low}}  \cdot \vec{V}_{\text{simple}} ) \;.
\eeq
\end{framed}
\noindent
In short, we can obtain bounds on the OPE coefficient of the lowest lying state, 
\beq
 (\vec{A}^{\text{low}}  \cdot \vec{V}_{\text{simple}} )\leq C_1^2\leq -   (\vec{A}^{\text{up}}  \cdot \vec{V}_{\text{simple}} )\;,
\eeq
using as input only the values of the scaling dimensions of the first two states, $\Delta_1$ and $\Delta_2$. 

This method was employed in \cite{Cavaglia:2021bnz} to obtain  a very narrow allowed region for $C_1$ as a function of the coupling. In the following sections we will see how this result can be dramatically improved by including the new integral relations, and how bounds for excited states can be obtained. Before moving to these new results, we discuss how the mathematical optimisation problem defined above can be solved efficiently using \texttt{SDPB}~\cite{Simmons-Duffin:2015qma}.

\paragraph{Implementation using SDPB. }
The space of  functionals satisfying positivity conditions such as (\ref{eq:posi}) can be navigated efficiently using semi-definite programming algorithms. These methods were introduced in the NCB context in \cite{Simmons-Duffin:2015qma} and implemented numerically in the software package \texttt{SDPB}~\cite{Simmons-Duffin:2015qma,Landry:2019qug}, which we used in our work. A full description of the internal algorithms can be found in the above references. Here, we briefly describe the main approximations involved and how they impact the results. 

First, in order to apply linear programming algorithms to impose the positivity conditions, we need to employ an approximation for the conformal blocks in terms of polynomials in $\Delta$. Similar to the cases  in higher dimensions~\cite{Chester:2019wfx}, this is  obtained using a truncated expansion of the form:
 \beq\label{eq:expandblocks}
f_{\Delta}(x) \sim \sum_{k = 0}^{\text{N}_{\text{blocks}}} a_k(\Delta) \; \left( r(x)\right)^k , \;\;\;\; r(x)\equiv \frac{x}{\left(\sqrt{1-x}+1\right)^2} \; .
 \eeq
Under this approximation, the action of the elementary functionals on the blocks takes the form:
\beq\label{eq:polyapprox}
\left. \frac{\partial^{2 n} }{\partial x^{2 n} } \mathcal{G}_{\Delta}(x)\right|_{x=\frac{1}{2}} \equiv \texttt{Pos}(\Delta) \times \texttt{Poly}_n(\Delta) \;,
\eeq
where $\texttt{Pos}(\Delta)$ is a strictly positive function\footnote{We  take $\texttt{Pos}(\Delta) = \left(4 r(\frac{1}{2}) \right)^{\Delta} \;\left((\Delta^2 - 1)\Delta (\Delta - 1) \prod_{n=1}^{\frac{N_{\text{blocks}}}{2} - 1}(\Delta + \frac{(3 + 2 n)}{2} )\right)^{-1}$.  The same positive prefactor allows to write a polynomial approximation both for the action of derivatives, and for the action of integral operators introduced in section \ref{sec:incorporating}. Notice that this is different from the case of the integrated correlator constraints in  \cite{Chester:2021aun}, which required to use different methods (linear programming rather than semidefinite programming). } of $\Delta$ and $\texttt{Poly}_n(\Delta)$ are polynomial in $\Delta$. This  polynomial approximation is used to  impose positivity conditions such as (\ref{eq:posi}) by treating $\Delta$ as a continuous parameter. 

In our work, we typically take $N_{\text{blocks}}\sim  30$. While this approximation introduces an error, in practice its impact is invisible on the scale of our bounds. We have verified  on selected points that $N_{\text{blocks}} = 50$ and  $N_{\text{blocks}} = 100$  give the same  bounds to the relevant digits. 

The most important parameter of the method is the integer  $N_{\text{der}}$ which characterises the dimension of the vector $\vec{A}$ -- namely, it restricts the space of functionals we explore. The choice of $N_{\text{der}}$ does affect significantly the bounds. However, importantly the bounds associated to the Numerical Conformal Bootstrap are rigorous, in the sense that they are true for any value of $N_{\text{der}}$. Moreover, they can only get sharper and sharper by taking this value larger and larger. 

While in \cite{Cavaglia:2021bnz} we presented an extrapolation to $N_{\text{der}}\rightarrow \infty$, here we simply take this number to be $N_{\text{der}} =  140$ for our main results (see section \ref{sec:results}). Even at this fixed value, thanks to the new integral relations, we will improve  significantly our previous results. 

\paragraph{Precision considerations related to the spectrum. } There is still one potential source of errors we have not discussed, namely the finite precision on the spectral data coming from the numerical solution of the QSC, which we are using as input. We checked that all methods discussed in this paper are quite stable with respect to this error:  an error on the spectrum propagates into a shift roughly of the same magnitude for the best estimate for the OPE coefficients. 
This means that, in order not to contaminate the bounds for $C_i^2$, it is sufficient to use spectral data with an error a couple of orders smaller than the width of the bounds.

We also noticed that the numerical bootstrap algorithms are quite sensitive to the injection of wrong spectral data, and often they cease to converge  if one inputs a spectrum which deviates too much from the QSC answer. For instance, at $g=3$, using the method discussed in section \ref{sec:incorporating} with spectral input from 10 states, it is enough to introduce an error on the spectrum on a scale of $5\times10^{-7}$ and the method to bound $C_1^2$ would no longer converge. 

To generate the results published in this paper, we used spectral data with at least 12 digits precision (we expect that the precision is actually  higher for most points, and exceeds 20 digits at weak coupling).
We estimate that, with such precision, errors on the  spectrum do not have a significant effect on the bounds for OPE coefficients, on the full range of the coupling we consider.

\subsection{Including more information on the spectrum}\label{sec:algorithms1}
The method of \texttt{Algorithm 1} uses as input only the scaling dimensions for the first two low-lying states. A natural way to improve the bounds is to include more information on the exact spectrum. 

\begin{figure}[t]
\centering
  \includegraphics[width=0.85\linewidth]{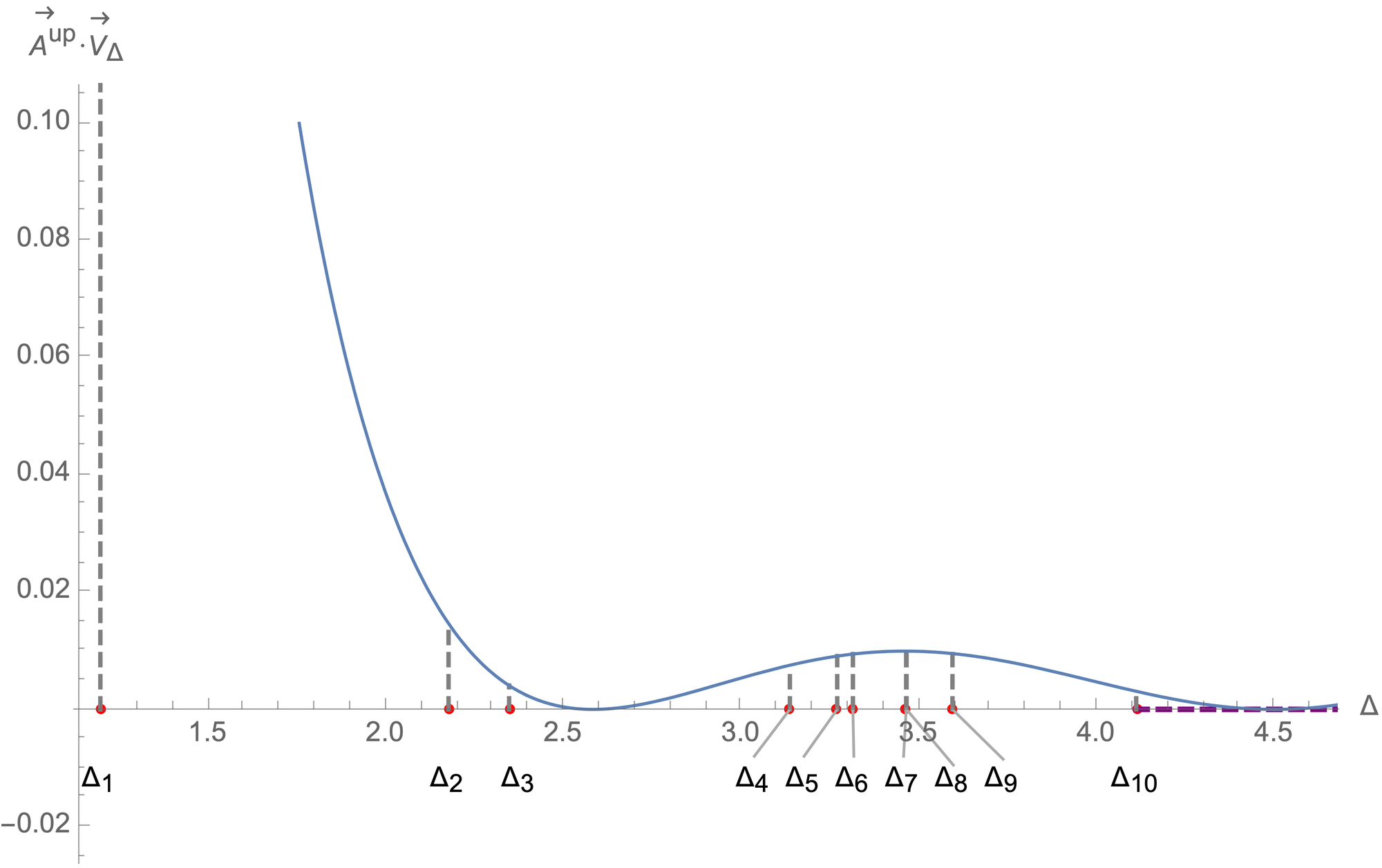}
  \captionof{figure}{A depiction of  $\vec{A}^{\text{up}} \cdot \vec{V}_{\Delta}$ as a function of $\Delta$, where $\vec{A}^{\text{up}}$ is calculated using \texttt{Algorithm 1} with $N_{\text{der}} = 60$, for $g = \frac{1}{4}$. The red points mark the position of the first 10 spectral levels (the levels $\Delta_7$ and $\Delta_8$ are very close and not distinguishable by eye). The value of the functional is positive for all levels and takes the value $\vec{A}^{\text{up}} \cdot \vec{V}_{\Delta_1} = 1$ on the ground state (outside the scale of the figure). }
  \label{fig:figfunc1}
\centering
  \includegraphics[width=0.85\linewidth]{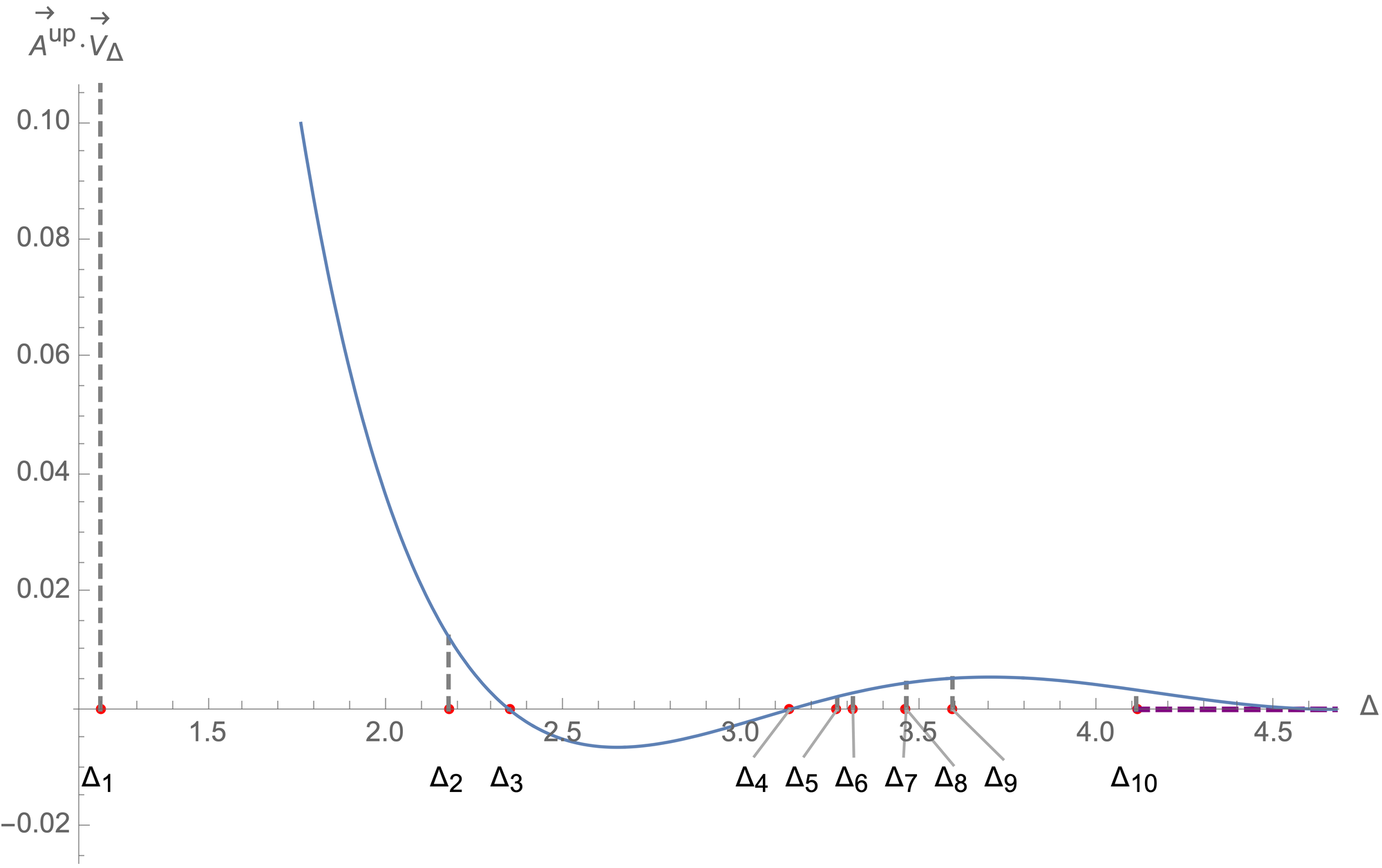}
  \captionof{figure}{Here, we depict again  $\vec{A}^{\text{up}} \cdot \vec{V}_{\Delta}$ as a function of $\Delta$ but obtained with the more general \texttt{Algorithm 2} which allows for it to take  negative values  between the exact values of the first 10 levels. As the optimal functional is drawn from a larger set, this leads to a stronger bound. As in Fig. \ref{fig:figfunc1}, here $N_{\text{der}} = 60$ and $g = \frac{1}{4}$.  }
  \label{fig:figfunc2}
\end{figure}

\begin{figure}
\centering
  \includegraphics[width=0.78\linewidth]{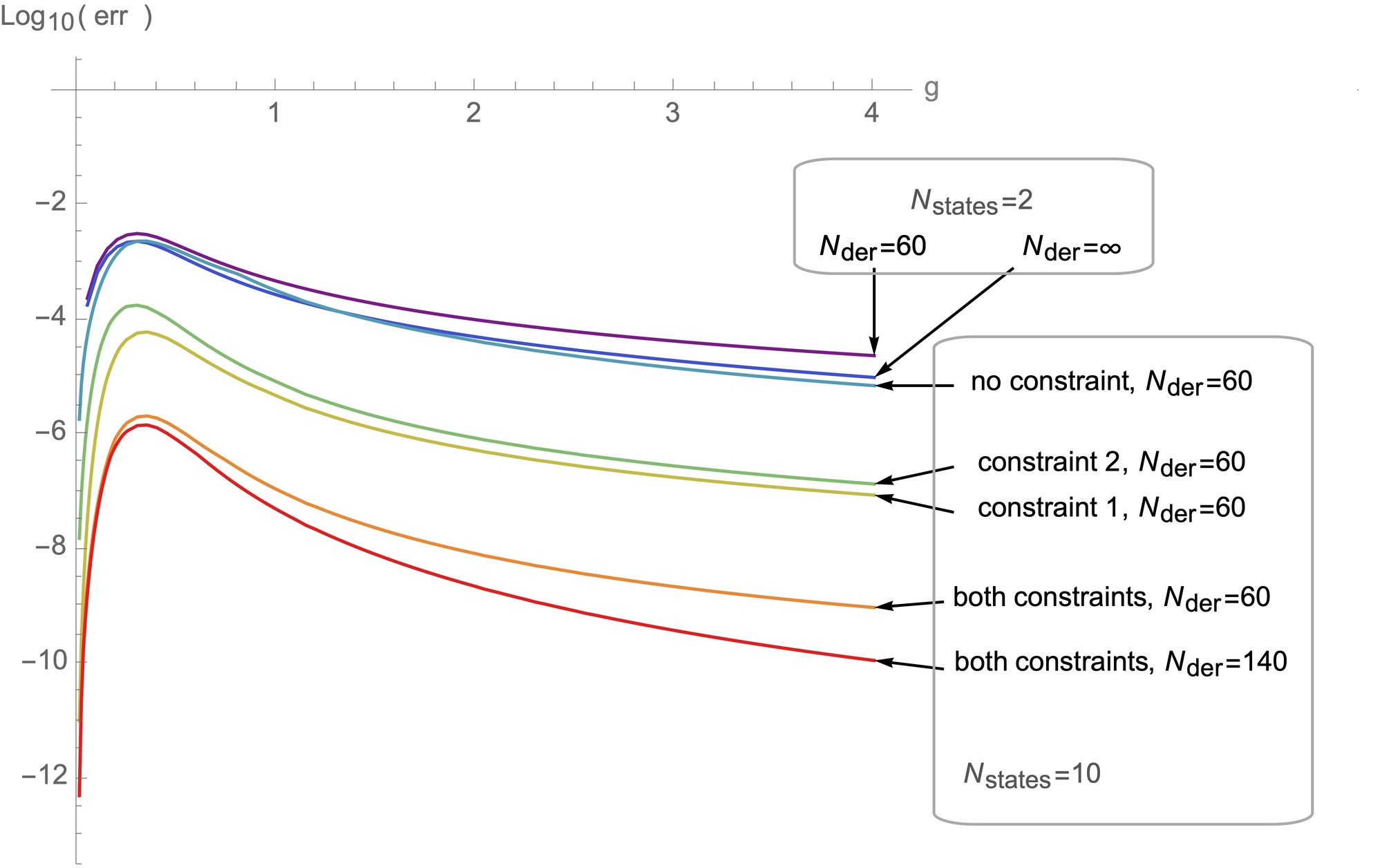}
  \captionof{figure}{The value of the error for the first OPE coefficient in logarithmic scale, $\log_{10}\left( \frac{1}{2} (C_{1,+}^2 - C_{1,-}^2 ) \right)$, as function of the coupling, resulting from various methods. We show two results from \cite{Cavaglia:2021bnz} (obtained with input from two states): the bounds at $N_{\text{der}}= 60$  (\fcolorbox{black}{color1}{\rule{0pt}{3pt}\rule{3pt}{0pt}}),  and the best result of \cite{Cavaglia:2021bnz}, obtained with the extrapolation $N_{\text{der}} \rightarrow \infty$ (\fcolorbox{black}{color2}{\rule{0pt}{3pt}\rule{3pt}{0pt}}). 
  These are compared to the new results, at $N_{\text{der}}= 60$, 
  obtained without using integral relations (\fcolorbox{black}{color3}{\rule{0pt}{3pt}\rule{3pt}{0pt}}), using either the second (\fcolorbox{black}{color4}{\rule{0pt}{3pt}\rule{3pt}{0pt}}) or the first integral relation (\fcolorbox{black}{color5}{\rule{0pt}{3pt}\rule{3pt}{0pt}}), or using both of them (\fcolorbox{black}{color6}{\rule{0pt}{3pt}\rule{3pt}{0pt}}). Our best result is obtained using both integral relations and $N_{\text{der}} = 140$ (\fcolorbox{black}{color7}{\rule{0pt}{3pt}\rule{3pt}{0pt}}). In all these new results we use \texttt{Algorithm 2 } with input from $N = 10$ states.  }
  \label{fig:ErrorLogScale}
\centering
  \includegraphics[width=0.78\linewidth]{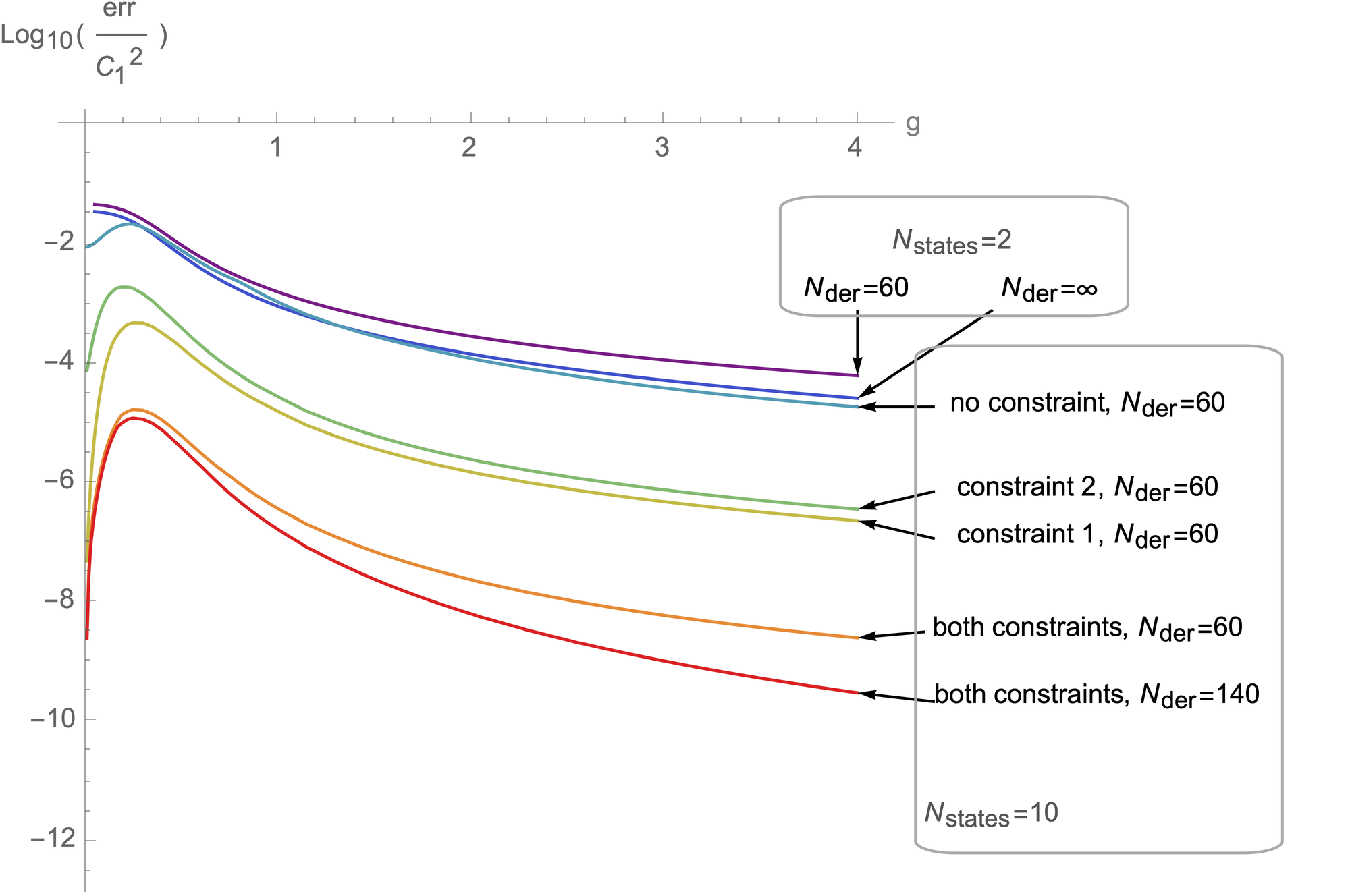}
  \captionof{figure}{With the same colour scheme as in Figure \ref{fig:ErrorLogScale}, we compare various methods for the value of the relative error on a logarithmic scale,  plotting $\log_{10}\left( \frac{ C_{1,+}^2 - C_{1,-}^2 }{ C_{1,+}^2 + C_{1,-}^2 } \right)$. }
  \label{fig:RelativeErrorLogScale}
\end{figure}

For instance, if we know the values of the next few states $\Delta_3,\Delta_4,\dots, \Delta_{N}$, we can impose a relaxed positivity condition, where the functional should satisfy
\beq\label{eq:posigap}
(\vec{A}\cdot \vec{V}_{\Delta})\geq 0 \,, \;\;\;\forall \Delta \geq \Delta_{N}\; ,
\eeq
together with a discrete set of additional constraints
\beq\label{eq:discreteposi}
(\vec{A}\cdot \vec{V}_{\Delta_n})\geq 0 , \;\;\; n = 2,3,\dots, N  -1 \;.
\eeq
Such generalised conditions can be easily implemented in \texttt{SDPB}\footnote{We are grateful to Petr Kravchuk for discussing this point.}. 
The functionals satisfying these relations now span a larger space, since they are allowed to assume negative values when $\Delta$ lies in the gaps between the first $N$ states. This is illustrated in figures \ref{fig:figfunc1} and \ref{fig:figfunc2}. If we now run again the previous algorithms, the bounds will become sharper. This is simply because the bounds come from maximising some quantities -- and we are now looking in a larger space for the functional that maximises it. 
 Moreover, this extension of the method also allows us to easily generate bounds for OPE coefficients for states other than the ground state (in fact, for any of the first $N-1$ states). We summarise concisely how this works below.
\begin{framed}
\texttt{Algorithm 2 -- Bounds including more states. } 
We now use the knowledge of the first $N$ states. We describe how to obtain bounds for $C_m^2$, with $m \leq N - 1$. 
We restrict to functionals satisfying the \emph{positivity conditions with gaps} 
\beqa
&&(\vec{A}\cdot \vec{V}_{\Delta} )\geq 0,\;\;\;\forall \Delta \geq \Delta_{N},\\  
&& (\vec{A}\cdot \vec{V}_{\Delta_n} ) \geq 0,\;\;\; n \in \left\{1,2,\dots,N-1\right\} , \; n\neq m \; .
\eeqa
Under these conditions, we search for  $\vec{A}^{\text{up},m}$ and $\vec{A}^{\text{low},m}$  such that:
\begin{enumerate}[1)]
\item $(\vec{A}^{\text{up},m}\cdot \vec{V}_{\Delta_m} ) = 1 $,
\item $(\vec{A}^{\text{up},m}\cdot \vec{V}_{\text{simple}} ) $ is maximal,
\item$(\vec{A}^{\text{low},m}\cdot \vec{V}_{\Delta_m} )= -1 $,
\item $(\vec{A}^{\text{low},m}\cdot \vec{V}_{\text{simple}} )  $ is maximal.
\end{enumerate}
Then, (\ref{eq:inequa}) give the  bounds
\beq
(\vec{A}^{\text{low},m}\cdot \vec{V}_{\text{simple}} )\leq C_m^2 \leq - (\vec{A}^{\text{up},m}\cdot \vec{V}_{\text{simple}} )\; .
\eeq
\end{framed}
For illustration, let us compare the bounds for $C_1$ obtained with \texttt{Algorithm 1} (i.e., including only $\Delta_1$ and $\Delta_2$ as input)  or with \texttt{Algorithm 2} with the input of the first $N=10$ states. The functional corresponding to the upper bound is shown in Figs. \ref{fig:figfunc1} and \ref{fig:figfunc2} for a specific choice of parameters. In the latter case, the functional is allowed to become negative in between the states. Notice that it assumes negative values between $\Delta_3$ and $\Delta_4$, while being essentially zero at these two points. In general, in the case illustrated in Fig. \ref{fig:figfunc2} the functional takes \emph{smaller} values for all levels $\Delta_n$, $2 \leq n \leq 10$. The consequence is that the truncation of the bootstrap equation is more efficient and the resulting upper bound becomes stronger. In the case illustrated, we used $N_{\text{der}} = 60$ and $g = \frac{1}{4}$. While the lower and upper bounds obtained with the simpler algorithm are $C_1^2 \in [0.0908377,0.0967516]$, with \texttt{Algorithm 2} the bounds are $C_1^2 \in [0.0908945,0.0949486 ]$, and the width of the bound reduces by 31 percent. Over the full range of values for the coupling, the gain in precision is at least $\sim$ 16 percent up to $g \sim 0.8$, and becomes significant at strong  coupling, e.g. for $g>3$ the error decreases by more than 75 percent of its original value. A comparison of the error with different methods will be presented in Figures \ref{fig:ErrorLogScale}, \ref{fig:RelativeErrorLogScale} at the end of the section.

\begin{figure}[t]
\centering
  \includegraphics[width=0.8\linewidth]{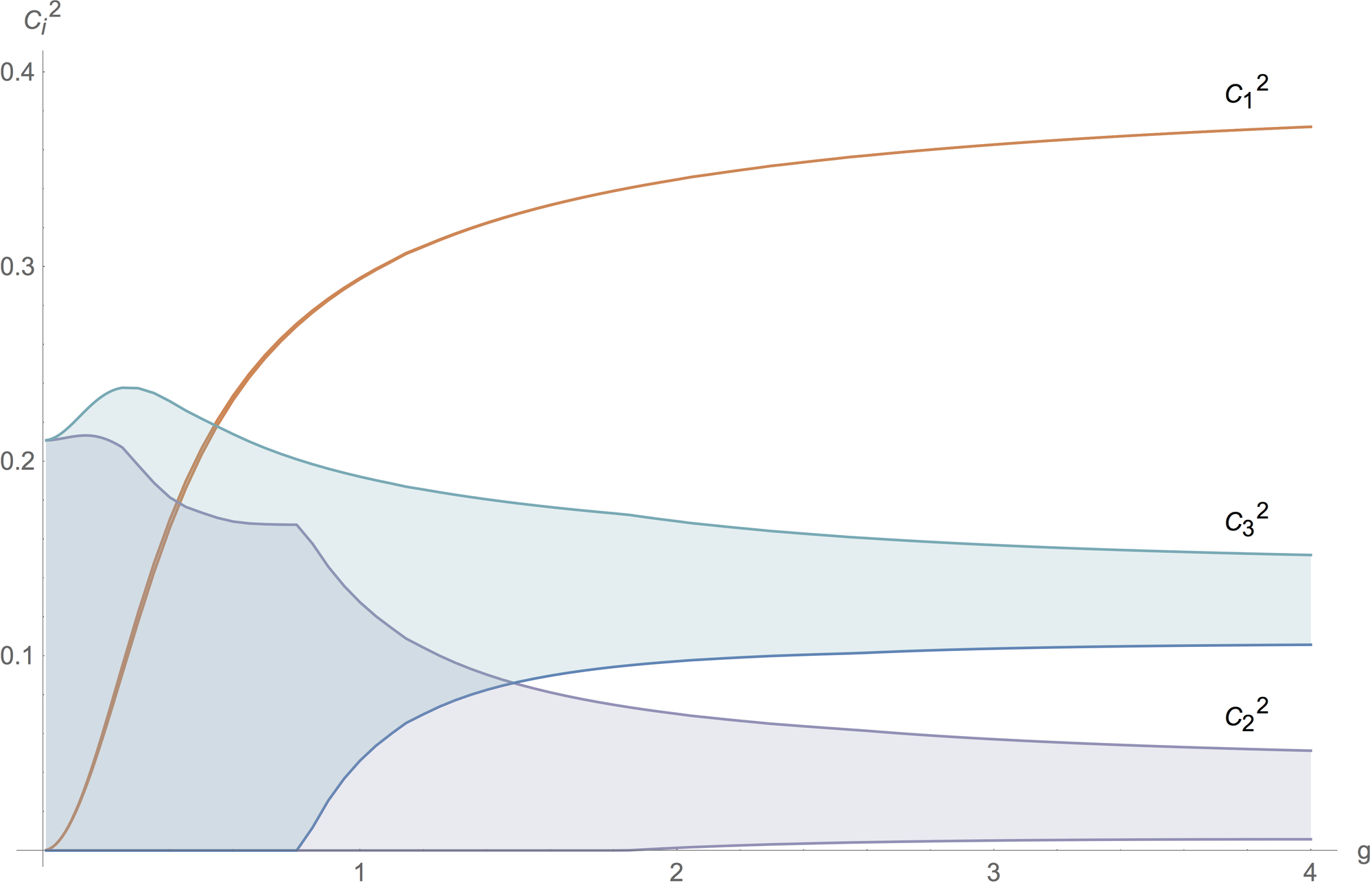}
  \captionof{figure}{Bounds for the first three OPE coefficients squared, 
  obtained with \texttt{Algorithm 2} with input of the first 10 states, and with $N_{\text{der}} = 60$. The allowed region is very narrow for the ground state (lower and upper bounds form a thin region which almost looks like a line), but the precision for excited states is much less with this method. The gain in precision achieved by including the new integral relations can be seen in Figures \ref{fig:plotintcompare}, \ref{fig:plotintboth2}.}
  \label{fig:bigplot1}
\end{figure}

Bounds for the OPE coefficients of the three lowest states obtained with \texttt{Algorithm 2} are shown in Fig. \ref{fig:bigplot1}, with the inclusion of 10 states as input and $N_{\text{der}} = 60$.  One immediately sees that the bounds for excited states are much less precise than the one for the ground state. At low values of the coupling, in particular, the algorithm produces negative values for the lower bound for $C_2^2$ and $C_3^2$, which are worse than the obvious estimate $C_i^2>0$, so they are not shown. In section \ref{sec:incorporating}, we will obtain significant improvements of these bounds by including the new integral relations. 

\begin{figure}[t]
\centering
  \includegraphics[width=0.9\linewidth]{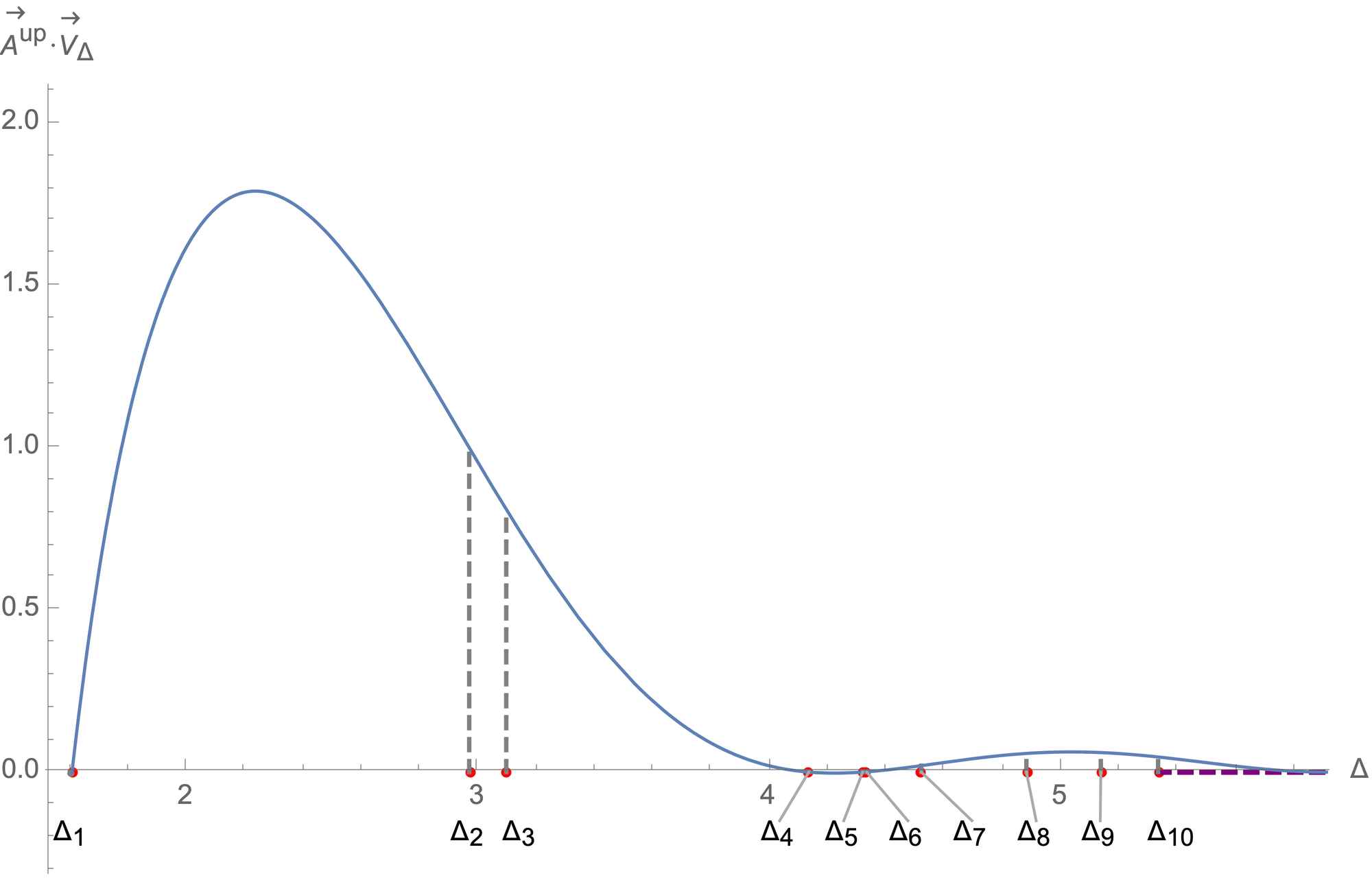}
  \captionof{figure}{Values of $(\vec{A}^{\text{up}} \cdot \vec{V}_{\Delta} )$, as a function of $\Delta$, for the optimal functional giving the upper bound for $C_2^2$, using \texttt{Algorithm 2} with $N_{\text{der}} = 60$ and $N_{\text{states}} = 10$. Here, $g = 4/5$. Notice the functional is positive between $\Delta_3$ and $\Delta_4$. }
  \label{fig:plotphase1}
\centering
  \includegraphics[width=0.9\linewidth]{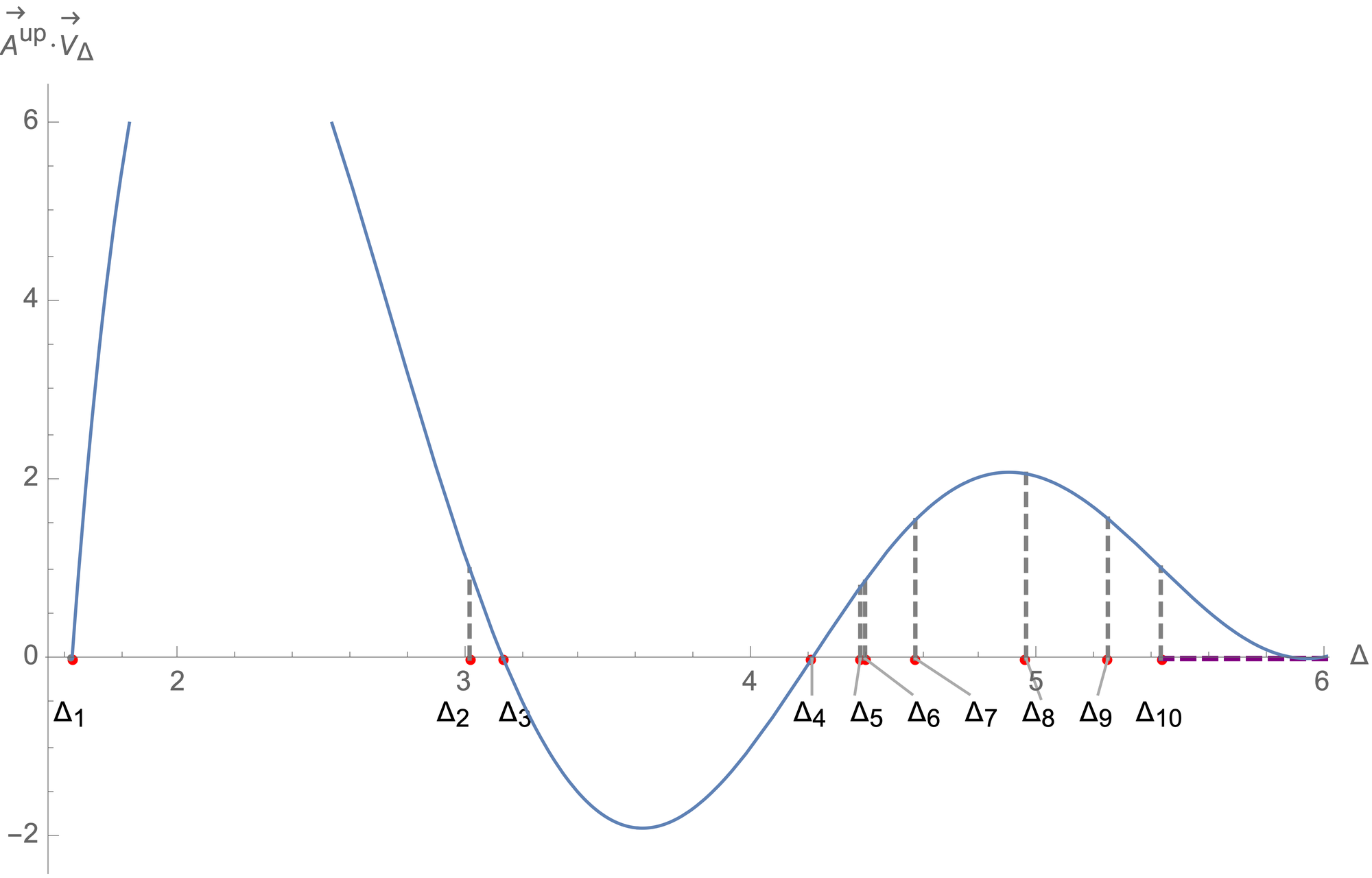}
  \captionof{figure}{The optimal functional giving the upper bound for $C_2^2$, for the same parameters as in \ref{fig:plotphase1}, but with $g = 17/20$. Notice the abrupt change in the shape of the functional, with the opening up of the gap between $\Delta_3$ and $\Delta_4$, associated to the phase transition. }
  \label{fig:plotphase2}
\end{figure}

\paragraph{Phase transitions in the optimal functionals. }
An interesting feature of Fig. \ref{fig:bigplot1} is the corner in the upper bound for $C_2^2$. 
Such point of non-analyticity is associated to a ``phase transition'' in the shape of the optimal functional corresponding to the bound. We plot the functional for $g = 4/5$ and $g = 17/20$  in Figs. \ref{fig:plotphase1} and \ref{fig:plotphase2}, respectively. These two points are very close to each other,  on different sides of the phase transition. Correspondingly, the shape of the functional changes sharply. In particular, in the case of Figure \ref{fig:plotphase2} the functional starts exploiting the gap between $\Delta_3$ and $\Delta_4$, and this leads to an abrupt improvement of the bound. 

We have observed that such type of phase transitions occur not only across different values of $g$, but also as $N_{\text{der}}$ is increased. Recall that this parameter controls the dimensionality of the linear space to which the functionals belong. As this  value is increased, the linear functionals acquire more degrees of freedom allowing them to penetrate the gaps between the states in the spectrum. We observed that the opening of a new gap is associated to a jump in the error.  

One may wonder if the error would shrink to zero provided we used some kind of limiting procedure, by incorporating more and more levels, and consistently using more and more derivatives. We suspect this is however not the case, and that a single correlator does not contain all information on the OPE coefficients, even provided we were able to input all the spectrum. The shrinking of the bounds to zero very likely requires the use of multiple correlators.

In any case, since some pairs of levels in the spectrum are very close to each other, it is in practice not possible to exploit all the gaps within computationally reasonable values for $N_{\text{der}}$. The presence of phase transitions as $N_{\text{der}}$ is increased also  means that the dependence of the bounds on this cutoff cannot be considered smooth. This prevents us from performing an extrapolation to $N_{\text{der}}\rightarrow \infty$, as we did in the analysis of \cite{Cavaglia:2021bnz}.\footnote{ In that case, we believe the procedure to be justified, since we were using \texttt{Algorithm 1}. We were not allowing for the functional to be negative between excited states, and therefore we expect no phase transitions. } Here, therefore, we will content ourselves with working at fixed large values for $N_{\text{der}}$.  

In the rest of this section we discuss  some further generalisations of the algorithms. These, however, were not used to obtain our main results, and at this point the reader can safely choose to jump to section \ref{sec:incorporating}, where we discuss the inclusion of the integrated correlator constraints.  

\paragraph{Incorporating existing bounds in the algorithm.} 
Suppose we know some bounds for the first OPE coefficient, $C^2_1\in\left[ C_{1,-}^2 , C_{1,+}^2  \right]$ (for instance obtained by running one of the previous algorithms), and we know the first $N$ states in the spectrum. Then, we can split $C_1^2 = C_{1,-}^2 + \delta_+ C^2_1$, where  $\delta_+ C^2_1$ should be a positive quantity, and rewrite the  bootstrap constraint (\ref{eq:vectorform}) as:
\beqa
0&=&\sum_{n\geq 1} C^2_n \; \left(\vec{A}\cdot \vec{V}_{\Delta_n}\right) + \left(\vec{A}\cdot \vec{{V}}_{\text{simple}}\right) \\
&=&\delta_+ C^2_1 \left(\vec{A}\cdot \vec{V}_{\Delta_1}\right) + \sum_{n\geq 2} C^2_n \; \left(\vec{A}\cdot \vec{V}_{\Delta_n}\right) + \left(\vec{A}\cdot \vec{\widetilde{V}}_{\text{simple}}\right) \;,
\eeqa
where we redefined
\beq \label{eq:simpletilde}
\vec{\widetilde{V}}_{\text{simple}} \equiv \vec{V}_{\text{simple}} + C_{1,-}^2\;\vec{V}_{\Delta_1} \;.
\eeq
Now, when computing and upper or lower bound for $C_m^2$ with $1<m < N$, we use a simple update of \texttt{Algorithm 2}: we search for a functional $\vec{A}$ which satisfies the conditions (\ref{eq:posigap}), (\ref{eq:discreteposi}) of the previous algorithm, but which maximises its value on the vector $\vec{\widetilde{V}}_{\text{simple}}$ defined in (\ref{eq:simpletilde}), 
 rather than $\vec{{V}}_{\text{simple}}$ if (\ref{eq:Vsimple}).  We can also clearly repeat the argument but using the knowledge of the upper bound. In this case, we would have
 \beq\label{eq:upperreformulation}
\delta_- C^2_1 \left(\vec{A}\cdot \vec{V}_{\Delta_1}\right) + \sum_{n\geq 2} C^2_n \; \left(\vec{A}\cdot \vec{V}_{\Delta_n}\right) + \left(\vec{A}\cdot \vec{\widetilde{V}}_{\text{simple}}\right) = 0 \;,
\eeq
where 
\beq \label{eq:upperreformulation2}
\vec{\widetilde{V}}_{\text{simple}} \equiv \vec{V}_{\text{simple}} + C_{1,+}^2\;\vec{V}_{\Delta_1} \; , 
\eeq
and where $\delta_- C^2_1 \equiv C^2_1 - C_{1,+}^2$ is now a negative quantity. In this case, to find bounds for $C_m^2$, $1<m<N$, one has to scan the space of functionals such that $\vec{A}\cdot \vec{V}_{\Delta_1} < 0$. 

We summarise  these considerations in the following iteration of the algorithm:
\begin{framed}
\texttt{Algorithm 3 -- Bounds including several states and previous bounds. }
We use the input of the first $N$ states,  and we assume knowledge of some bounds for the OPE coefficients, which we denote as:
\beq
C_k^2 \in \left[ C_{k, -}^2 \,,\; C_{k, +}^2  \right] , \;\;\; k = 1,\dots, N-1\;.
\eeq
These could come from previous iterations of the algorithms, or from other considerations. 

\noindent We now focus on obtaining a new bound for the $m$-th state, $1\leq m \leq N-1$. 

To do this, we choose a combination of \emph{signs}:
\beq
\left\{ \sigma_k \right\} , \;\;\; k \in \left\{ 1,\dots, N-1\right\}, k\neq m \;,
\eeq
where $\sigma_i \in \left\{-1,1\right\}$. Then the bootstrap equation rewrites as
\beq\label{eq:bootstraprewrite2}
 C^2_m \left(\vec{A}\cdot \vec{V}_{\Delta_m}\right) + \sum_{n\neq m, \; n\leq N-1} \delta_{\sigma_n} C^2_n \; \left(\vec{A}\cdot \vec{V}_{\Delta_n}\right) + \left(\vec{A}\cdot \vec{\widetilde{V}}_{\text{simple}}\right) + \sum_{n\geq N} C^2_n \; \left(\vec{A}\cdot \vec{V}_{\Delta_n}\right) = 0 \; ,
\eeq
where we define $\delta_{+1} C_n^2 \equiv C_n^2 - C_{k,-}^2 > 0$, and $\delta_{-1} C_n^2 \equiv C_n^2 - C_{k,+}^2  < 0$, and
\beq
\vec{\widetilde{V}}_{\text{simple}} \equiv \vec{V}_{\text{simple}} + \sum_{n\neq m, \; n\leq N}  C_{n, \; - \sigma_n}^2 \;\vec{V}_{\Delta_n} \; .
\eeq
\noindent We work in the space of functionals satisfying the conditions:
\beqa
&&(\vec{A}\cdot \vec{V}_{\Delta} )\geq 0,\;\;\;\forall \Delta \geq \Delta_{N},\\  
&& \sigma_n\; (\vec{A}\cdot \vec{V}_{\Delta_n} ) \geq 0,\;\;\; n \in \left\{1,2,\dots,N-1\right\} , \; n\neq m\; ,
\eeqa
so that (\ref{eq:bootstraprewrite2}) becomes
\beq\label{eq:bootstraprewrite3}
 C^2_m \left(\vec{A}\cdot \vec{V}_{\Delta_m}\right) + \left(\vec{A}\cdot \vec{\widetilde{V}}_{\text{simple}}\right) \leq 0 \; .
\eeq
Proceeding as in the previous cases, we now search for  $\vec{A}^{\text{up},m}$ and $\vec{A}^{\text{low},m}$  such that:
\begin{enumerate}[1)]
\item $(\vec{A}^{\text{up},m}\cdot \vec{V}_{\Delta_m} ) = 1 $,
\item $(\vec{A}^{\text{up},m}\cdot \vec{\widetilde{V}}_{\text{simple}}) $ is maximal,
\item$(\vec{A}^{\text{low},m}\cdot \vec{V}_{\Delta_m} )= -1 $,
\item $(\vec{A}^{\text{low},m}\cdot \vec{\widetilde{V}}_{\text{simple}})  $ is maximal.
\end{enumerate}
Then, (\ref{eq:bootstraprewrite2}) gives the  bounds
\beq
(\vec{A}^{\text{low},m}\cdot \vec{\widetilde{V}}_{\text{simple}} )\leq C_m^2 \leq - (\vec{A}^{\text{up},m}\cdot \vec{\widetilde{V}}_{\text{simple}} )\;.
\eeq
\end{framed}
We found that in some cases these tricks are effective. In particular, one can sometimes improve significantly the bounds for the excited states, provided a rather precise bound for $C_1^2$ is exploited as input. To make an example, for $g = 1/4$, using \texttt{Algorithm 2} with  $N_{\text{states}}= 10$, $N_{\text{der}} = 60$, one finds  $C_3^2\in[0,0.23779]$. 
 Incorporating the bounds for $C_1^2 \in [0.09313678, 0.09313996]$ as in \texttt{Algorithm 3}, one can recalculate the allowed region  for $C_3^2$, finding now $C_3^2 \in [0.0214, 0.1845]$, i.e. the allowed interval shrinks by 31 percent. The magnitude of this improvement is however very sensitive to the precision of the injected bounds for $C_1^2$. In this example in particular we used as input a bound for $C_1^2$  which is more precise than what could be achieved by the algorithms presented so far, and was obtained by including the integral relations into the game. 

While \texttt{Algorithm 3} could be useful in some contexts, we found that, once we include  uniformly the new integral constraints as explained in the next section, the precision is not significantly affected by using this  upgrade of the method. Our main results in section \ref{sec:results} were obtained simply with  \texttt{Algorithm 2}. 

\paragraph{Other approaches. }
Although we do not explore them in this paper, it is worth mentioning that there are other techniques to study OPE coefficients within the Numerical Bootstrap (for some modern developments see e.g. \cite{Chester:2019ifh}).  

One possible approach, for which preliminary results were presented in \cite{Cavaglia:2021bnz}, is to treat some of the OPE coefficients as variational parameters in order to find their allowed region. For example, in the case presented in \cite{Cavaglia:2021bnz}, we rewrote the bootstrap equation as
\beq
 C^2_1 \left(\vec{A}\cdot \vec{V}_{\Delta_1}\right) + \sum_{n\geq 4} C^2_n \; \left(\vec{A}\cdot \vec{V}_{\Delta_n}\right) + \left(\vec{A}\cdot \vec{\widetilde{\widetilde{V}}}_{\text{simple}}\right) = 0 \;,
\eeq
with 
$\vec{\widetilde{\widetilde{V}}}_{\text{simple}} \equiv \vec{V}_{\text{simple}} + C^2_2 \left(\vec{A}\cdot \vec{V}_{\Delta_2}\right) + C^2_3 \left(\vec{A}\cdot \vec{V}_{\Delta_3}\right)$. One can then treat $C_2^2$ and $C_3^2$ as parameters, and use one of the algorithms presented above to obtain bounds on $C_1^2$, $C_1^2 \in \left[ C_{1,-}^2(C_2^2, C_3^2) , \; C_{1,+}^2(C_2^2, C_3^2)\right] $, which depend parametrically on $C_2^2$ and $C_3^2 $. Such bounds indirectly define an allowed region for $C_2^2$ and $C_3^2$, which is given by the condition that  $C_{1,-}^2(C_2^2, C_3^2) < C_{1,+}^2(C_2^2, C_3^2)$. 
One can also use other types of bootstrap algorithm, where one does not impose optimisation conditions, but rather looks for a functional  with positivity conditions that exclude a certain set of conformal data.

It is definitely worth investigating if these techniques can lead to an improvement of our results. Exploring the higher dimensional parametric space of OPE coefficients, while computationally expensive, might reveal a finer structure than treating them individually as we do in this paper (for instance, in the case of \cite{Cavaglia:2021bnz} we observed that there is a linear combination of $C_2$ and $C_3$ for which the bound is much narrower than for each of them individually).

\subsection{Incorporating the integral relations}\label{sec:incorporating}
Here we explain how the new integral relations can be embedded in the NCB  framework, similarly to what was done in  \cite{Chester:2021aun} in 4D.

First of all, we rewrite the constraints using the OPE decomposition of the 4-point function, as new linear relations for the OPE coefficients. As shown in appendix  \ref{app:rewriteconstr}, using crossing symmetry and the OPE the two relations (\ref{eq:constr1}), (\ref{eq:constr2}) can be rewritten as
\beqa
\text{Constraint 1:  }&&\;\; \sum_{\Delta_n} C^2_n \; \texttt{Int}_1\left[\, f_{\Delta_n} \,\right] + \texttt{RHS}_1 = 0 \;, \label{eq:newconstr1}\\
\text{Constraint 2:  }&&\;\; \sum_{\Delta_n} C^2_n \; \texttt{Int}_2\left[\, f_{\Delta_n} \,\right] + \texttt{RHS}_2 = 0 \;,\label{eq:newconstr2}
\eeqa
where we introduced the integral operators
\beqa\la{ints}
&&\texttt{Int}_1\left[ F(x) \right] \equiv - \int_0^{\frac{1}{2}} (x-1-x^2)\frac{F(x)}{x^2} \partial_x\log\left( x (1-x)\right)\, dx \;,\\
&&\texttt{Int}_2\left[ F(x)\right]\equiv  \int_{0}^{\frac{1}{2}} dx \frac{ F(x) \; ( 2 x - 1) }{x^2} \;,
\eeqa
and the explicit functions of the coupling constant:
\beqa
\texttt{RHS}_1&=&\frac{\mathbb{B}-3
   \mathbb{C}}{8\mathbb{B}^2}+\left(7 \log(2) -\frac{41}{8}\right) (\mathbb{F}-1)+  \log
   (2)\; , \\
   \texttt{RHS}_2 &=& \frac{1-\mathbb{F}}{6}+(2-\mathbb{F}
   ) \log (2)+1 -\frac{\mathbb{C}}{4\;\mathbb{B}^2}\; .\label{eq:RHS2}
\eeqa
In order to obtain \eq{ints}, we used the crossing equation \eq{eq:fcrossing} to rewrite the integral over the half range $x\in [0,1/2]$. Within this interval the OPE expansion \eq{eq:OPEf}
converges rapidly, which also implies that the action of the integral operators on the blocks $f_{\Delta}$ are rapidly decreasing with $\Delta$. Further details on the derivation of (\ref{ints})-(\ref{eq:RHS2}) are contained in Appendix \ref{app:rewriteconstr}.

 To incorporate these equations in the bootstrap setup, we consider a generic linear combination of  derivatives acting on the bootstrap equation with the new constraints (\ref{eq:newconstr1}),(\ref{eq:newconstr2}):
 \beqa\label{eq:newbootstrap}
&& \sum_{n} C^2_n \; \left[ \sum_{k = 0}^{N_{\text{der}}/2} b_k \left. \partial_x^{2 k} \mathcal{G}_{\Delta_n}  \right|_{x = \frac{1}{2}} + b_{-1} \texttt{Int}_1[f_{\Delta_n}] + b_{-2} \texttt{Int}_2[f_{\Delta_n}]\right]\\ &&+\sum_{k = 0}^{N_{\text{der}}/2} b_k \left. \partial_x^{2 k} \mathcal{G}_{\text{simple}}  \right|_{x = \frac{1}{2}}+ b_{-1}  \texttt{RHS}_1 + b_{-2}  \texttt{RHS}_2= 0 \;.
\eeqa
This equation is true for any choice of the coefficients  $\left\{b_{-1},b_{-2}, b_0, \dots, b_{N_{\text{der}}/2}\right\}$.  It is now apparent that this more general equation takes the same form as (\ref{eq:vectorform}), which was the starting points of our bootstrap algorithms, where we redefine $\vec{A}$, $\vec{V}_{\text{simple}}$, $\vec{V}_{\Delta}$ as
\beqa
\vec{A} &\equiv& ( b_{-1},b_{-2})\oplus\left( b_0,b_1, \dots, b_{N_{\text{der}}/2}\right), \label{eq:speacialvectors1}\\
\vec{V}_{\Delta} &\equiv& (\texttt{Int}_1[f_{\Delta}], \texttt{Int}_2[ f_{\Delta}] )\oplus \left. \left(  \mathcal{G}_{\Delta}(x),\, \partial_x^2\mathcal{G}_{\Delta}(x),  \dots, \partial_x^{N_{\text{der}}}\mathcal{G}_{\Delta}(x) \right)\right|_{x = \frac{1}{2}} \; ,\\
\vec{V}_{\text{simple}} &\equiv & (\texttt{RHS}_1, \texttt{RHS}_2 )\oplus \left. \left(  \mathcal{G}_{\text{simple}}(x),\, \partial_x^2\mathcal{G}_{\text{simple}}(x),  \dots, \partial_x^{N_{\text{der}}}\mathcal{G}_{\text{simple}}(x) \right)\right|_{x = \frac{1}{2}} \; . \label{eq:speacialvectors3}
\eeqa
After these redefinitions, we can run the same algorithms described sections \ref{sec:algorithms0} and \ref{sec:algorithms1}.

At the level of implementation, using the expansion (\ref{eq:expandblocks}) we can approximate our integral operators as
\beq
\texttt{Int}_1[ f_{\Delta}] = \texttt{Pos}(\Delta) \times \texttt{Poly}_{(1)}(\Delta) \;, \;\;\;\; \texttt{Int}_2[ f_{\Delta}] =\texttt{Pos}(\Delta) \times \texttt{Poly}_{(2)}(\Delta) \; ,
\eeq
with polynomials $\texttt{Poly}_{(i)}(\Delta)$ and, crucially, the same positive prefactor as in (\ref{eq:polyapprox}). This is a difference to what was observed in \cite{Chester:2021aun}, and allows us to  include the integral constraints into the  \texttt{SDPB} setup. 

\begin{figure}
\centering
  \includegraphics[width=0.8\linewidth]{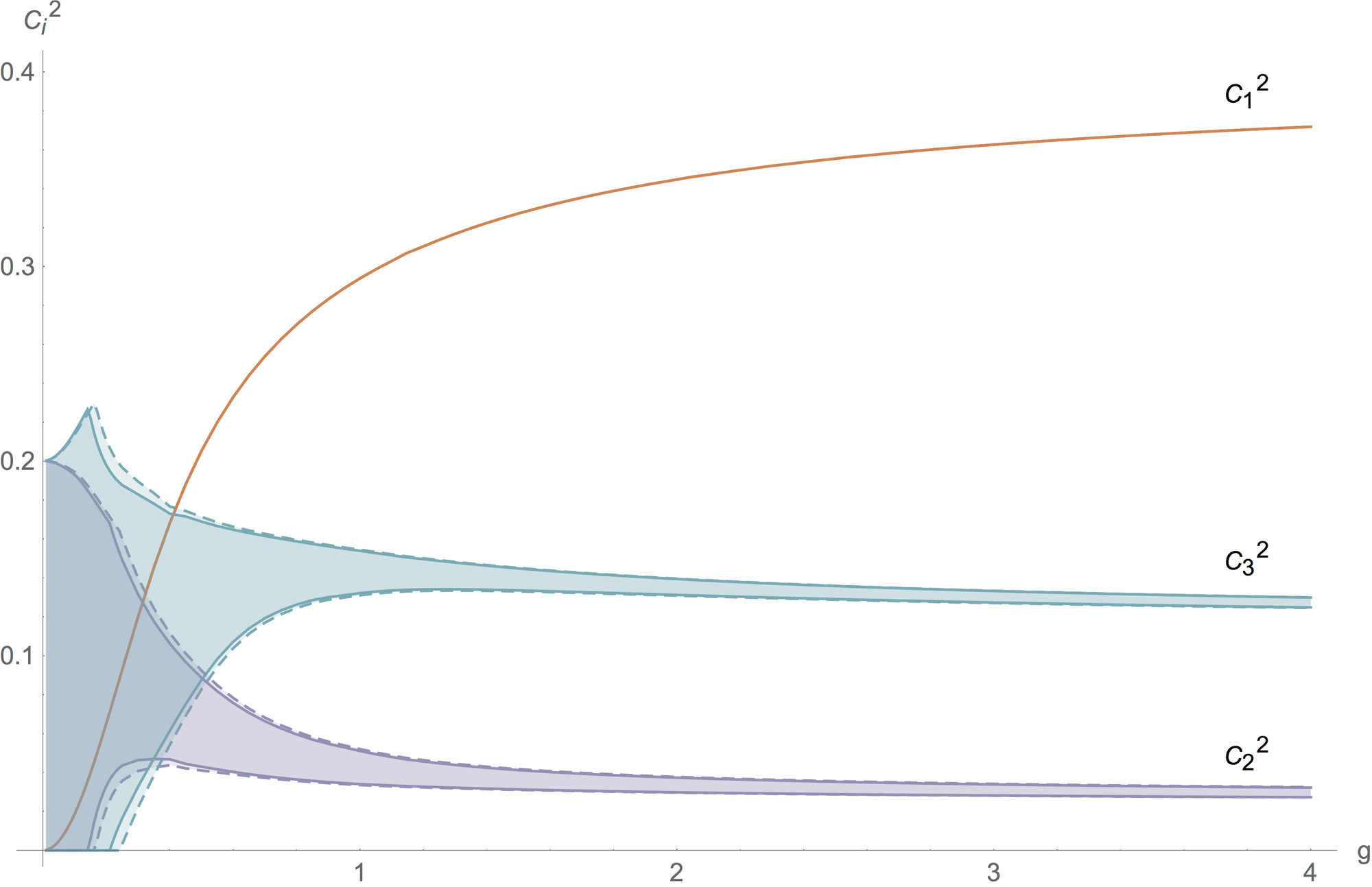}
  \captionof{figure}{Bounds for the first three OPE coefficients squared. We use the same  parameters as in Fig.  \ref{fig:bigplot1}, but including either the first or the second  integral relation (solid vs dashed lines). 
  The use of the constraints lead to a visible  improvement. 
  The two relations lead to comparable results, with the first integral relation being slightly more effective.}
  \label{fig:plotintcompare}
\end{figure}

\begin{figure}
\centering
  \includegraphics[width=0.8\linewidth]{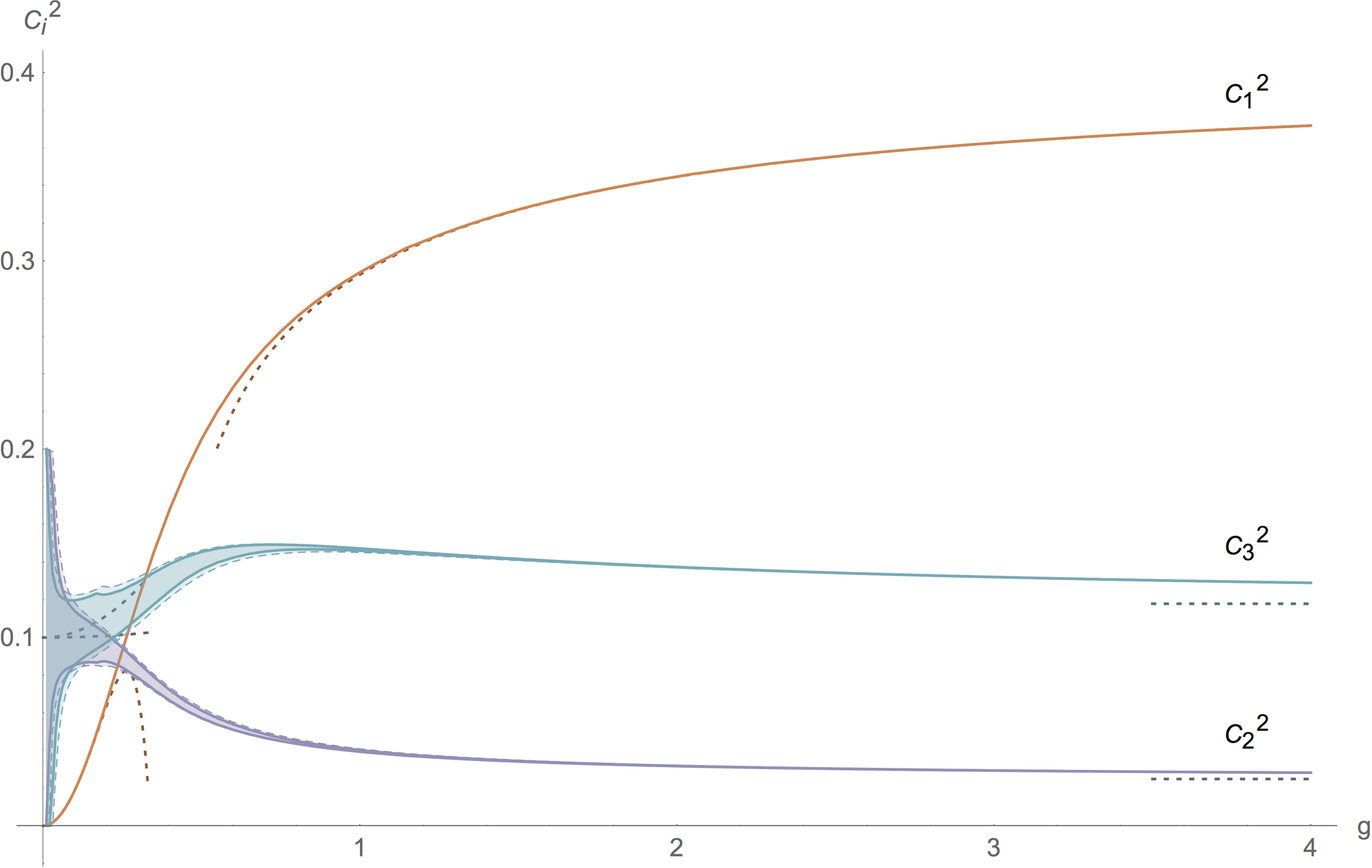}
  \captionof{figure}{Bounds for the first three OPE coefficients obtained using two integral relations simultaneously, with $N_{\text{der}} = 60$ and $N_{\text{der}}=140$ (darker). This is our main result. The data for the bounds are reported in Appendix \ref{app:boundsC}. Shown in the figures are also weak and strong coupling predictions (dotted lines), obtained in the next section \ref{sec:analytical}. }
  \label{fig:plotintboth2}
\end{figure}

\subsection{Results}\label{sec:results}
Including the integral relations leads to a dramatic improvement of the bounds. 
To quantify this effect, let us first describe some experiments where we add one relation at a time. This can be easily done by just dropping one component from (\ref{eq:speacialvectors1})-(\ref{eq:speacialvectors3}). 
 For instance, with the same parameters as in Figure \ref{fig:bigplot1}, but now including the second integral constraint (\ref{eq:constr2}), we find the width of the bound for $C_1^2$ decreasing by at least a factor of $10$ over all range in the coupling (for $g>1.5$, the gain is a factor of 50). Adding the first integral constraint  (\ref{eq:constr1}) on its own has an even stronger effect, with the bound decreasing by at least factor $30$, which becomes a factor $80$ for $g>1.9$. The improvement is also marked at weak coupling. A comparison of the error with various methods can be found in Figures \ref{fig:ErrorLogScale} and \ref{fig:RelativeErrorLogScale}. 
The gain in precision for the excited states is clearly visible in Figure \ref{fig:plotintcompare}. With either integral relation the bounds shrink approximately by a factor $2$ starting from $g \sim 0.3$, and by a factor $\sim 9$ at strong coupling. 
As can be seen in the figure, the two integral relations, separately, lead to very similar new bounds for the excited states coefficients, with the first relation (\ref{eq:constr1}) being slightly more constraining. One might even be suspicious that the two relations are not independent,  but one easily sees that this is not the case, as combining them reduces the error much further. 

 Our best results, obtained using both integral relations together in the algorithm, are shown in Figure \ref{fig:plotintboth2}. One can immediately see a significant improvement for the excited states, with the upper and lower bounds indistinguishable by eye  for a wide range of values of the coupling. Keeping fixed the value of $N_{\text{der}} = 60$, remarkably for $C_1^2$ the bound shrinks by at least a factor $10^3$ for all values of the coupling. For $C_2^2$ and $C_3^2$, the bound reduces monotonically with the coupling -- for $g = 0.3$  by at least  factor  $10^1$, which becomes $10^2$ for $g\sim 1.5$. At strong coupling, the gain is almost a factor of 200.  
We run the algorithm with $N_{\text{der}} = 140$ to obtain our best results, which are reported in Appendix \ref{app:boundsC}. 

The bounds produced by the same algorithm for $C_i^2$, with $i>3$, are less precise. In particular, for these OPE coefficients the current setup only produces a nontrivial upper bound. This is compatible in magnitude with the strong coupling results of \cite{Ferrero:2021bsb}. In particular, for the four states with $\Delta_{\text{strong}}^{(0)} = 6$, they found the strong coupling limit $\langle a_{\text{strong}}^{(0)} \rangle_3  = C_4^2 + C_5^2 + C_6^2 + C_7^2 = 10/429\simeq 0.023$, and we found the upper bounds e.g. $C_4^2 < 0.0079$ and $C_5^2 < 0.0123$ at $g = 4$. We reserve further study of these excited states for the future.  We expect that the precision can be improved by including more states in the algorithm and especially by considering more general bootstrap setups as we discuss in section \ref{sec:discussion}.

\section{Analytic Bootstrability}\label{sec:analytical}
In this section we develop a functional analytic bootstrap approach at weak coupling, using input from the QSC solution of the spectrum at weak coupling (collected in Appendix \ref{app:spectrum}). Additionally, we use the two new integral constraints, one of which -- relation (\ref{eq:constr1}) -- is particularly powerful to extract weak coupling data for the first OPE coefficient. 

We start by discussing the subtleties in the weak coupling treatment of (\ref{eq:constr1}), and then  proceed to discuss the weak coupling Bootstrability method. Our main results are summarised at the end of this section. 

\subsection{Weak coupling expansion of the first integral relation }
 We now explain how to interpret the constraint (\ref{eq:constr1}) at weak coupling. 
 In particular, as anticipated in section \ref{sec:integrated}, we need to understand how to 
 regularise correctly the integrals 
 $\int_0^1 dx G^{(\ell)}(x) \frac{\log{x} +1}{x^2}$, $\ell>1$, arising from the perturbative expansion of the integrand. 
These integral has $\log$-divergences, so we define an extraction scheme where the integral is defined by its finite part after dropping $\log^n\epsilon$ terms.
 
 We will now show that
 \beq\begin{split}\label{int1weak}
 \left.\int_0^1 \!\delta G(x)\, \frac{1 \!+\! \log(x)}{x^2}\;dx   \right|_{\text{small }g } \!\sim& 1 \!+\!  \underbrace{\int_{\epsilon}^{\frac{1}{2}}dx \left(\sum_{\ell=1}^{M} g^{2\ell} G_{\text{weak}}^{(\ell)}(x)\right)\frac{\log(x (1-x))}{x^2} }_{\texttt{regularised, }\log(\epsilon)\rightarrow 0}
 \\
&- \underbrace{\left[\frac{C_1^2}{(\Delta _1-1)^2}\right] }_{\texttt{``anomaly''}} + O(g^{2 M + 2})\;,
\end{split}\eeq
for any order $M=1,2,\dots$.
The meaning of this equation is the following. To reproduce the weak coupling expansion of the integral on the l.h.s., we should: i) expand the integrand up to the desired order and integrate term by term; ii) since this produces log-divergences, we introduce a cutoff $\epsilon$ in the integration range and  regularise the result by the prescription $\log^n\epsilon\rightarrow 0$; iii) finally, we should add the  ``anomalous term'' in the second line of  (\ref{int1weak}), which contains the OPE coefficient and scaling dimension of the ground state, expanded to the relevant order. 

In the following subsection we make a digression to prove \ref{int1weak}. Next, we will show that (\ref{eq:constr1}) can be used to deduce several orders of the weak coupling expansion of $C_1^2$ analytically.

\subsubsection{The weak coupling ``anomaly'' }
To understand the regularisation (\ref{int1weak}), it is convenient to break the l.h.s. in the two objects $\int_0^1\frac{\delta G(x)}{x^2}dx + \int_0^1\frac{\delta G(x)}{x^2}\,\log(x)\,dx $. 

\paragraph{The first piece. }
We start by analysing the first of these terms,  which can in fact be evaluated exactly for any $g>0$:
\begin{framed}
\beq\label{eq:specialide}
\int_0^1\frac{\delta G(x)}{x^2}dx = 1 .
\eeq
\end{framed}
\noindent
To prove this identity, notice that the integrand is in fact a total derivative (see  (\ref{eq:totalder})), 
\beq
\frac{\delta G(x)}{x^2}=\d_x\[\left(-\frac{1}{x^2}+\frac{1}{x}-1\right) \delta f(x)\]+1+C^2_{\rm BPS}(g) -2 .
\eeq
This combination is regular both at $x=0$ and $x=1$, so we can apply the integration formula
\beq
\int_0^1\frac{\delta G(x)}{x^2}dx=
C_{BPS}^2(g)-1-
2\[\left(-\frac{1}{x^2}+\frac{1}{x}-1\right) \delta f(x)\]_{x\to 0} \label{eq:firstli}\;\; ,
\eeq
where we used the fact that, due to crossing (\ref{eq:fcrossing}),
\beq
\[\left(-\frac{1}{x^2}+\frac{1}{x}-1\right) \delta f(x)\]_{x\to 1} = -\[\left(-\frac{1}{x^2}+\frac{1}{x}-1\right)\delta f(x)\]_{x\to 0} .
\eeq
 Using the OPE decomposition (\ref{eq:OPEf}), which converges around $x\sim 0$, one can easily evaluate the r.h.s. of  (\ref{eq:firstli}) obtaining $C^2_{BPS}-1 - (C^2_{BPS}-2) = 1$, which proves (\ref{eq:specialide}). 
 
Note, however, that at $g=0$ the integrand of the l.h.s. of (\ref{eq:specialide}) is strictly zero, as we defined $\delta G(x)$ as a difference of $G(x)$ with its tree level value. The discrepancy with the value $1$ taken by the integral is a clear manifestation of the weak coupling anomaly. We will now explain how to  cure this mismatch.

To understand the origin of the anomaly, let us look again at original integral at finite coupling,  which can be rewritten, using crossing, as $\int_0^{1}
    \frac{\delta G(x)}{x^2}dx =2\int_0^{\frac{1}{2}}
    \frac{\delta G(x)}{x^2}dx $. This allows us to use the OPE decomposition, which is fastly convergent on $[0,\frac{1}{2}]$. The contribution of a single conformal block $\mathcal{L}_{0,[0,0]}^{\Delta_n}$ to the integrand is 
 \beq
 \frac{ G(x) }{x^2} = 
 ... - C_n^2(g) \; \partial_x \left[ \left(1-\frac{1}{x} + \frac{1}{x^2} \right)f_{\Delta_n}(x) \right] + ... \,, 
 \eeq
 which behaves, close to the limit of integration $x \sim 0$, as
 \beq\label{eq:relevantsingular}
 \frac{ G(x) }{x^2} = ... + C_n^2(g) x^{\Delta_n(g)-2} \left( 1 + O(x) \right) + ...\;.
 \eeq
 Under the weak coupling expansion  $\Delta_n = \Delta_{n}^{(0)} + \gamma_n(g)$, with $\gamma_n(g) = O(g^2)$, we find
\beq\label{eq:smallx}
 \left.  \frac{ G(x) }{x^2} \right|_{\text{small $g$, $x$}}\!= ...+ C_n^2(g) \! \left[ x^{\Delta_n^{(0)} -2 } ( 1 + g^2 \gamma_n^{(0)} \log(x) + O(g^4) ) \!+\! O(x^{\Delta_n^{(0)}-1} ) \right] \!+...\;.
 \eeq
As long as $\Delta_n^{(0)} > 1 $, the weak coupling expansion thus produces terms that can be safely integrated. The only exception is the block corresponding to the ground state $\Delta_1$: in this case, since $\Delta_1^{(0)} = 1$, we encounter a log-divergence at $x \sim 0$. We can therefore zoom on this term to understand the correct perturbative regularisation of the integral. 

Focusing on the relevant singular behaviour (\ref{eq:relevantsingular}) for $n=1$, we have the integral
\beq
I=2 \int_\epsilon^{1/2} x^{\Delta_1-2}dx ,
\eeq
where $\Delta_1\to 1$ at $g\to 0$. Let us compare its non-perturbative treatment to a na\"ive finite-part regularisation of the weak coupling expansion. At finite coupling, $\Delta_1>1$ and the integral is convergent  giving
\beq\label{eq:Ifinite}
I_{\rm finite}=\frac{2^{\Delta_1 }}{\Delta_1 -1}\;.
\eeq
At weak coupling, $\Delta_1 =1 + \gamma_1$, we expand first at small $\gamma_1$  and then integrate. Choosing the  prescription $\log^n\epsilon\to 0$, and then resumming the result order by order in $\gamma_1$  yields 
\beq
I_{\rm weak}=\frac{2^{\Delta_1 }-2}{\Delta_1 -1} .
\eeq
The discrepancy is
\beq
\delta I = 
I_{\rm finite}-I_{\rm weak} = \frac{2}{\Delta_1-1}\;.
\eeq
Taking into account that the integral (\ref{eq:Ifinite}) comes from the OPE expansion and is multiplied by $C_1^2$, we have established  the weak-coupling identity
\beq\label{fint1}
 \left. \int_0^1 \!\frac{\delta G(x) }{x^2}\;dx   \right|_{ \text{small }g } \!\sim    \underbrace{2 \int_{\epsilon}^{\frac{1}{2}}dx \frac{ \sum_{\ell=1}^{M} g^{2 \ell} G_{\text{weak}}^{(\ell)}(x)}{x^2} }_{\texttt{regularised, }\log(\epsilon)\rightarrow 0} + \underbrace{\left[\frac{2 C_1^2}{\Delta _1-1}\right] }_{\texttt{``anomaly''}} + O(g^{2 M + 2}) ,
\eeq
which is to be interpreted as explained above.
Let us check if this identity is now compatible with the exact result (\ref{eq:specialide}). At leading order $O(1)$, the integral term drops out. This means that the anomaly term, on its own, should match the value $1$ of the integral on the l.h.s. We know that at leading order the anomalous dimension is $\Delta_1 -1 = 4 g^2 + O(g^4)$, which implies that we must have
\beq\label{eq:firstprediction}
C_1^2(g) \simeq 2 g^2 + O(g^4),
\eeq
which is indeed confirmed by the numerical bootstrap data~\cite{Cavaglia:2021bnz} and the analytic bootstrap computation of section \ref{sec:numerology}. At the next-to-leading order, we can plug in (\ref{eq:G1weak}) and fix  one more term in the OPE coefficient. Before doing this, however, we conclude the proof of (\ref{int1weak}), which will allow us to use the constraint (\ref{eq:constr1}) and will prove more powerful. 

\paragraph{The second piece and its anomaly. }
The other integral term on the l.h.s. of  (\ref{eq:constr1}) can be analyzed by repeating the argument above. Now, using crossing, we rewrite $\int_0^1 \frac{\delta G(x)}{x^2} \log(x)\;dx = \int_0^{\frac{1}{2}} \frac{\delta G(x)}{x^2} \log\left(x (1-x)\right)\;dx .$ The troublesome term, again coming from the singular contribution of the block $f_{\Delta_1}$, is the integral
\beq
I'=\int_\epsilon^{1/2} x^{\Delta_1-2} \log\left(x (1-x)\right)\, dx ,
\eeq
which evaluates at finite coupling to 
\beq
I_{\rm finite}' = -\frac{2^{1-\Delta_1 } \left(\Delta_1 +(\Delta_1 -1) (\Delta_1  \log (4)-\, _2F_1(1,1;\Delta_1 +1;-1))\right)}{(\Delta_1 -1)^2 \Delta_1 }.
\eeq
Comparing this with the weak-coupling regularisation,
\beq
I_{\rm weak}' = \sum_{n=0}^{\infty} \underbrace{ \int_{\epsilon}^{\frac{1}{2} } \frac{ (\log x )^{n} (\Delta_1-1)^n}{(n!)\, x} \log\left(x (1-x)\right) dx }_{\texttt{regularised},\;\log\epsilon\rightarrow 0} ,
\eeq
we find, order by order,
\beq
I_{\rm finite}' -I_{\rm weak}' = -\frac{1}{(\Delta_1 - 1)^2} ,
\eeq
which proves the relation (\ref{int1weak}).

\subsubsection{Analytic results for $C_1^2$}\label{sec:predictC1}
We are now ready to study the first integral constraint (\ref{eq:constr1}) at weak coupling. 
Using (\ref{int1weak}), it becomes
\begin{framed}
\beqa\label{int1weak2}
 \left. \frac{3\mathbb{C}-\mathbb{B}}{8 \;\mathbb{B}^2} \right|_{\text{small }g}  &=& 1 \!+\!  \underbrace{\int_{\epsilon}^{\frac{1}{2}}dx \left(\sum_{\ell=1}^{M} g^{2\ell} G_{\text{weak}}^{(\ell)}(x)\right)\frac{\log(x (1-x))}{x^2} }_{\texttt{regularised, }\log(\epsilon)\rightarrow 0} 
- \left. \left[\frac{C_1^2}{(\Delta _1-1)^2}\right] \right|_{\text{small } g}\nn \\ &&+ O(g^{2 M + 2})\;.
\eeqa
\end{framed}
Due to the presence of the anomalous term in this equation,  the knowledge of $G(x)$ at a fixed order at weak coupling allows to produce  nontrivial  predictions for the  leading OPE coefficient.
Using (\ref{eq:B0}), (\ref{eq:weakC0}), the term on the l.h.s. of (\ref{int1weak2}) has the expansion
\beq\begin{split}\label{eq:lhsconstr1}
\frac{3\mathbb{C}-\mathbb{B}}{8 \;\mathbb{B}^2} 
=
&-\frac{1}{8 g^2}+\left(\frac{3}{2}-\frac{\pi ^2}{12}\right)+\left(\frac{\pi ^4}{36}-9 \zeta_3\right) g^2+\left(-\frac{2 \pi ^6}{135}-4 \pi ^2 \zeta_3+135 \zeta_5\right) g^4\\
&+\left(\frac{7 \pi ^8}{810}+\frac{34 \pi ^4 \zeta_3}{15}+78 \pi ^2 \zeta_5-1806 \zeta_7\right) g^6+O\left(g^8\right).
\end{split}\eeq
On the other hand, for general OPE coefficients and scaling dimensions, we expect the weak coupling expansion
\beq
C_i^2(g) = \sum_{\ell=0}^{\infty}  a_i^{(\ell)}  \, g^{2 \ell} ,\qquad
\Delta_i(g) =  \sum_{\ell=0}^{\infty} \Delta_i^{(\ell)} \, g^{2 \ell} ,
\eeq
 which implies that  ``anomaly term'' on the r.h.s. of  (\ref{int1weak2})  starts as $-a_1^{(0)}/(g^2\Delta_1^{(1)})^2$. Since this is not matched on the l.h.s., we must have $a_1^{(0)} = 0$, and the anomaly term expands as
\beq\label{eq:anomalyweak}
-  \left[\frac{C_1^2}{(\Delta _1-1)^2}\right] = -\frac{a_1^{(1)}}{ (\Delta_1^{(1)})^2g^2} + \left(\frac{2 a_1^{(1)} \Delta_1^{(2)} }{(\Delta_1^{(1)})^3} -\frac{a_1^{(2)} }{(\Delta_1^{(1)})^2} \right) + \dots
\eeq
Up to order $O(1)$, the integral on the r.h.s. drops out from (\ref{int1weak2}), and we just need to match (\ref{eq:lhsconstr1}) and (\ref{eq:anomalyweak}). From the QSC, we know (see section  \ref{apd:anylSpec}) $\Delta_1^{(1)} = 4$, $\Delta_1^{(2)} = -16$. The term $O(g^{-2})$ then fixes $a_1^{(1)} = 2$, consistent with  (\ref{eq:firstprediction}),  while at order $O(1)$ we fix
\beq\label{C1w2}
a_1^{(2)} = \frac{4 \pi ^2}{3}-24\,.
\eeq
We can also study the next order $O(g^2)$, which involves the integral over $G_{\text{weak}}^{(1)}$ given in (\ref{eq:G1weak}). Computing the integral and using the next terms in the expansion of (\ref{eq:anomalyweak}), we can now extract 
\beq\label{C1w3}
a_1^{(3)} = 48 \zeta_3+320-16 \pi
   ^2-\frac{76 \pi ^4}{45}\,.
\eeq
In the next section, we will see how, using a functional analytic bootstrap approach, one can push this analysis to one more loop, all in all determining a 4-loop prediction for the OPE coefficient. Our full result is reported below in equation (\ref{eq:C1finalresult}). 

\subsection{Functional bootstrap at weak coupling}\label{sec:numerology}

In this section we use an analytic functional bootstrap  approach, combined with the integrability data, to obtain structure constants in perturbation theory at weak  coupling. 

\subsubsection{General strategy}
\paragraph{Warm up at the first two orders. }
The first two orders of the weak coupling expansion of the reduced correlator $f(x)$ defined in \eqref{fandG} are known. The tree level \eqref{fandGtree} is obtained by free field theory, while the one loop \eqref{f1weak0} was computed in \cite{Kiryu:2018phb}. 

An alternative way to present the perturbative expansion of the reduced correlator is in terms of Harmonic Polylogarithms~\cite{Remiddi:1999ew} (HPL), which are natural functions appearing in the evaluation of Feynman integrals and are implemented in the Mathematica package \texttt{HPL} \cite{Maitre:2005uu,Maitre:2007kp}. 
To do this, it is useful to introduce the new object
\beq\label{defh}
h(x)=\frac{1 - x}{x} f(x) ,
\eeq
for which we have the following HPL representation for the first two orders at weak coupling
\beqa
h_{\text{weak}}^{(0)}(x)&=&1-2x , \label{eq:h0}\\
h_{\text{weak}}^{(1)}(x)&=&-2 {H}_{1,0}+2 {H}_2-\frac{2 \pi ^2}{3}x\label{h1},\label{eq:h1}
\eeqa
with
\beq
{H}_{1,0}=-\text{Li}_2(x)-\log(1-x)\log(x)\,,\qquad
{H}_2={H}_{0,1}=\text{Li}_2(x) ,
\eeq
where we used the HPL property
\beq
{H}_{n_1,...,\underbrace{0,...,0}_{k \text{-times}},n_i,...}=
{H}_{n_1,...,n_i+k,...}\;.
\eeq
Before discussing how to infer an ansatz for the general perturbative order, let us discuss the implications of these results for the conformal data. 

First, one can compare the leading order $h_{\text{weak}}^{(0)}(x)$ with the OPE expansion \eqref{eq:OPEf}, using the leading order for the scaling dimensions, which become degenerate at tree level to the values $\Delta^{(0)} \equiv J = 1,2,3,\dots.$ Comparing order by order the small-$x$ expansion of \eqref{eq:OPEf} and  $h_{\text{weak}}^{(0)}(x) $, one can fix the following constraints on the structure constants
\beq\label{CLOavg}
\langle\; a^{(0)}\;\rangle_J=\frac{4^{-J-1} \sqrt{\pi } (J-1) \Gamma (J+3)}{\Gamma \left(J+\frac{3}{2}\right)} ,
\eeq
where $\langle...\rangle_J$ represents the sum over the state multiplicity at weak coupling for a given $\Delta^{(0)} = J$. 
Such averages are very common in analytic bootstrap approaches, which typically expand around points of degeneracy of the spectrum. For us, however, a big advantage will be the knowledge of the spectrum, which will give us more conditions and allow us to resolve some of these averages much more easily. In particular, at this stage we know the degeneracies, so that for instance $J=1$ counts only one state, and then (\ref{CLOavg}) means $\langle\; a^{(0)}\;\rangle_1=a_1^{(0)}=0$, which is the same result we found in the previous section. 
At the next level $J=2$ we have two states implying $\langle\; a^{(0)}\;\rangle_2=a_2^{(0)}+a_3^{(0)}=1/5$, and so on. 

The case of the ground state $\Delta_1$, with $\Delta_1^{(0)} = 1$, is special. In fact, due to the factor $1/(\Delta - 1)$ in the definition of the superconformal blocks (\ref{superblocks}), the scaling dimension and OPE coefficient  \emph{at one loop}  enter the OPE expansion of the correlator at tree level. In particular, the leading behaviour at small $x$ is determined by the block $C_1^2 f_{\Delta_1}(x) \sim x^{\Delta_1 + 1}/(\Delta_1 - 1)\sim  a_1^{(1)} x^{2} /\Delta_1^{(1)} + O(g^2)$. Matching this with the small-$x$ behaviour of $h_{\text{weak}}^{(0)}(x)$, and using  $\Delta_1^{(1)} = 4$ coming from the QSC, we  can read off $a^{(1)}_1=2$, in agreement with the independent derivation in \eqref{eq:firstprediction}. This obviously extends to higher loops: the small-$x$ behaviour of $h_{\text{weak}}^{(\ell)}(x)$ is determined by conformal data of the ground state up to $\ell + 1$ loops. 

Repeating the same procedure for $h_{\text{weak}}^{(1)}(x)$, and including the results of the previous order, we can read $a_1^{(2)} = 4 \pi ^2/3-24$, matching  the result obtained by the integral relation in \eqref{C1w2}. Expanding the conformal blocks at higher order in $x$, it is possible to disentangle the average appearing in \eqref{CLOavg} obtaining 
\beq\begin{split}\label{cleading}
a_2^{(0)}&=a_3^{(0)}=\frac{1}{10},\qquad
a_4^{(0)}=a_5^{(0)}=a_7^{(0)}=a_9^{(0)}=0,\\
a_6^{(0)}&=\frac{1}{14}+\frac{2}{7\sqrt{37}},\qquad\;\;
a_8^{(0)}=\frac{1}{14}-\frac{2}{7\sqrt{37}} ,
\end{split}\eeq
together with the following constraints for the sub-leading orders
\beq\label{csubleading}
\langle\,a^{(1)}\,\rangle_2=\frac{2\pi^2}{15}-1,\qquad
\langle\,a^{(1)}\,\rangle_3=\frac{2\pi^2}{21}-\frac{1159}{882},\qquad
\langle\,a^{(1)}\,\rangle_4=-\frac{29}{36}+\frac{\pi ^2}{21}.
\eeq
\paragraph{General ansatz and strategy. }
The reduced correlator at order $g^4$ is unknown. To proceed, we will formulate an ansatz for the generic term based on the  form of the first two orders (and also inspired by the functional forms observed at strong coupling~\cite{Ferrero:2021bsb}). Consistent with (\ref{eq:h0}),(\ref{eq:h1}), we assume that, at $\ell$ loops, $h$ is given in terms of a complete basis of HPL's with transcendentality\footnote{For an HPL function $H_{n_1,n_2,\dots, n_{\tau}}(x)$, with $n_i\in\left\{0,1\right\}$, the transcendentality is the number of indices.} up to $\tau = 2 \ell$. Besides,  we assume the same transcendentality for all terms, i.e.,  
\beq\label{basis}
h_{\text{weak}}^{(\ell)}(x)=\beta_0^{(2\ell)}+\beta_1^{(2\ell)} x+\sum_{\tau=1}^{2\ell}\beta_{n_1,...,n_\tau}^{(2\ell-\tau)}\,H_{n_1,...,n_\tau},
\eeq
where the coefficients $\beta^{(m)}$ are transcendental numbers of weight $m$ and the indices $n_i=0,1$. For example, the basis used for $\ell=1$ according to this ansatz contains eight terms:
\beq\label{eq:hansatz}
h_{\text{weak}}^{(1)}(x)=\beta_0^{(2)}+\beta_1^{(2)} x+\beta_{0}^{(1)}H_{0}+\beta_{1}^{(1)}H_{1}
+\beta_{0,0}^{(0)}H_{0,0}
+\beta_{1,0}^{(0)}H_{1,0}
+\beta_{0,1}^{(0)}H_{0,1}
+\beta_{1,1}^{(0)}H_{1,1} .
\eeq
Comparing it with \eqref{f1weak0}, one can conclude that the only non-vanishing coefficients are the following
\beq\label{eq:1loopfixing}
\beta_1^{(2)}=-\frac{2}{3}\pi^2\,\qquad
\beta_{1,0}^{(0)}=-2\,\qquad
\beta_{0,1}^{(0)}=2\,,
\eeq
leading to \eqref{h1}. 

In order to fix the coefficients of the ansatz \eqref{basis} for $\ell>1$, we apply the following strategy
\begin{itemize}
    \item \texttt{Crossing equation.} In terms of $h(x)$, crossing symmetry translates to
\beq\label{hcrossing}
h(1-x)+h(x)=0 .
\eeq    
To impose this equation, in practice is enough to study some terms of its expansion around $x \sim 0$. Once these terms are set to zero by fixing some of the $\beta$ coefficients, the equation is satisfied for any $x$. 
    \item \texttt{Cancelling logarithms.} In order to come from an OPE expansion (\ref{eq:OPEf}) at weak coupling, the function $h(x)$ has to satisfy certain constraints on its $x\sim 0$ behaviour. At $\ell$ loops, $h^{\ell}(x)$ can only contain terms $\log^{m}(x)\, x^n$, with $n\in \mathbb{N}$ and $m = 1,\dots, \ell$. We compare this with the expansion of the ansatz, and  impose the cancellation of the logarithms with higher powers. 
    \item \texttt{Conformal data matching.} The OPE expansion gives  infinitely many relations between terms in a small $x$ expansion of $h^{\ell}(x)$ and the conformal data. Equating these predictions with the expansion of the ansatz (\ref{eq:hansatz}), we find relations between the $\beta^{(\ell)}$ coefficients and the conformal data $a_i^{(m)}$, $\Delta_i^{(m)}$, with $m\leq \ell$ for $i>1$ and $m\leq \ell+1$ for $i = 1$. Using the \emph{knowledge of the spectrum} from integrability, we can use these relations to fix some $\beta$'s as well as some OPE coefficients. 
 \item \texttt{Integral constraints. }
 We impose the two integral relations described in section \ref{sec:integrated}.  At a given order, the constraint (\ref{eq:constr2}) fixes further information on the coefficients of the ansatz. The constraint (\ref{eq:constr1}), due to the ``weak coupling anomaly'' effect described in section \ref{sec:weakcouplingintegrals}, can be used to extract the structure constant $a_1^{(\ell+2)}$ in terms of the coefficients $\beta^{(\ell)}$ of the ansatz.  
 \item \texttt{Transcendentality. } Our assumption is that the coefficients $\beta^{(2 \ell - \tau)}_{n_1,\dots, n_{\tau}}$ are combinations with rational coefficients\footnote{For many of them it turns out the coefficients are integer.}
  of numbers of uniform transcendentality   $2 \ell -\tau$. 
  In particular, up to the perturbative order we considered, we did not encounter Multiple Zeta numbers, but only numbers which are products of elements of the basis
  \beq\label{eq:basis}
  \left\{
\pi^2 , \zeta_3, \zeta_5, \zeta_7, \dots, \zeta_{2n+1} , \dots  
  \right\}.
 \eeq
 The element of this basis have transcendental weight $2$,$3$,$5$, $7,\dots,2n+1,\dots$, and their products have weight equal to the sum of the weights of the factors. 
 There are no linear relations with rational coefficients between the numbers in the basis (\ref{eq:basis}). 
  This implies that some linear relations generated by the other ``axioms'' listed above split into more constraints, as terms of different transcendentality should vanish individually.  
\end{itemize}
\subsubsection{Higher loops}
\paragraph{Fixing the correlator at 2 loops. }
For $\ell=2$, the basis \eqref{basis} with maximal transcendentality 4 counts $32$ elements. 
Imposing the crossing equation \eqref{hcrossing}
one can fix $16$ of the coefficients $\beta^{(2)}$. Furthermore, requiring that in the small $x$ expansion there are no $\log^3x$ and $\log^4x$ terms, we obtain $3$ coefficients more.
The remaining constraints can be found by injecting in the analytical bootstrap method the integrability data, namely the spectrum obtained with the QSC, the structure constants fixed at previous orders using $h_{\text{weak}}^{(0)}$ and $h_{\text{weak}}^{(1)}$, as well as using the integrated correlator \eqref{eq:constr1} at weak coupling,~{\it cf.}~\eqref{int1weak}. Let us describe these steps in more detail.

First, we expand at small $x$ and compare the terms $x^n\log^m x$ for $n,m=0,1,2$ with the same expansion of \eqref{eq:OPEf}. At this order, only the first $3$ lowest non-trivial states (i.e., the ground state with $J=1$
and the two states with $J=2$) contribute to these coefficients. Then, using the spectral data\footnote{Because of the pole in the conformal block at $\Delta=1$, we actually need to use all terms up to $O(g^6)$ for $\Delta_1$.} for these states reported in \eqref{D1w}-\eqref{D3w},  
together with their structure constants at leading~\eqref{cleading} and subleading order~\eqref{csubleading}, we fix $9$ more parameters.\footnote{Of these, 3  parameters are reduced by using the assumption of fixed transcendentality of the $\beta$ coefficients.}  At this stage, we are left with 4 unfixed coefficients $\beta^{(2)}$.  

As a by-product of the procedure explained above, and also including data for states with $J>2$, we generate new constraints for the structure constants. One of these constraints is particularly useful, since it expresses the $O(g^6)$ term in the expansion of $C_1^2$ in terms of the 4 remaining parameters $\beta$ as follows
\beq\begin{split}\label{a13coeff}
a_{1}^{(3)}&=-\frac{28 \pi ^4}{15}-16 \pi ^2+320+64 \zeta_3+ \left(\frac{2 \pi ^2}{3}+8 \zeta_3-\frac{7 \pi ^4}{45}\right)\beta^{(0)}_{0,1,0,1}\\
&-\!\left(\!4\!+\!4 \zeta_3-\frac{4 \pi ^4}{45}\right) \!\beta_{0,0,0,1}^{(0)}
- \left(\frac{2 \pi ^2}{3}+4 \zeta_3\!-\!\frac{\pi ^4}{9}\right)\!\beta_{0,1,1,0}^{(0)}\!-\!\left(\!4\!+\!\frac{2 \pi ^2}{3}-8 \zeta_3\right)\!\beta_{0,0,1}^{(1)} . 
\end{split}\eeq
As shown in section \ref{sec:predictC1}, using the integral relation \eqref{eq:constr1} in its weak coupling fashion one can generate  terms for $C_1^2$ at higher orders exploiting the anomaly. We have already used this method to give a prediction for $a_1^{(3)}$ from the knowledge of the correlator at 1 loop. Comparing this result~\eqref{C1w3} with \eqref{a13coeff}, and using that the $\beta$ coefficients are rational, we conclude that 
\beq
\beta_{0,0,1}^{(1)}=\beta_{0,0,0,1}^{(0)}=0\,,\qquad\text{and}\qquad
\beta_{0,1,1,0}^{(0)}=\beta_{0,1,0,1}^{(0)}=-4 ,
\eeq
leading to the following expression for $h(x)$ at order $g^4$:
\beq\begin{split}\label{h2}
h^{(2)}_{\text{weak}}(x)&=4 H_{1,3}-4 H_{2,2}+8 H_{3,0}+8 H_{3,1}-8 H_{1,1,2}+4 H_{1,2,0}+8 H_{1,2,1}-8 H_{2,0,0}\\
&-4 H_{2,1,0}-8 H_{1,1,0,0}+\frac{4}{3} \pi ^2 H_2-\frac{4}{3} \pi ^2 H_{1,0}+\frac{4}{3} \pi ^2 H_{1,1}-12 \zeta_3 H_{1}+\frac{8 \pi ^4}{15}x .
\end{split}\eeq
As a cross check, this formula can be tested using the integral relation \eqref{eq:constr2}. Indeed, plugging \eqref{h2} into \eqref{defh} and then performing the integral, we obtain
\beq
 g^4\int_{0}^1 dx \frac{h_{\text{weak}}^{(2)}(x)}{1-x} = 
 g^4\left(-\frac{8 \pi ^4}{15}-\frac{8 \pi ^2 \zeta_3}{3}+90 \zeta_5\right)
 =        \left.\frac{\mathbb{C}}{4\;\mathbb{B}^2} + \mathbb{F}-3\right|_{\text{order }g^4} ,
\eeq
in agreement with the expansion of the r.h.s. of \eqref{eq:constr2}. 

\paragraph{Structure constants from $h_{\text{weak}}^{(2)}$. }

Having fixed the reduced correlator at 2 loops, we can mine new data for the structure constants. Indeed, comparing this answer to the OPE, and using the knowledge for the spectrum at one loop, one can disentagle the first relation of \eqref{csubleading}, obtaining the two separate subleading terms as follows
\beq
a_{2}^{(1)}=\frac{1}{150} \left(10 \pi ^2-75-9 \sqrt{5}\right)\,,\qquad
a_{3}^{(1)}=\frac{1}{150} \left(10 \pi ^2-75+9 \sqrt{5}\right).
\eeq
Similarly, the 3-point functions associated to the leading twist states\footnote{Leading twist refers to states with oscillator content $[2,2|3,3,2,2|2,2]$, in the classification introduced in \cite{Cavaglia:2021bnz}. Notice that in this case the indices $11$, $12$ and $13$ do not represent the order in which they appear at weak coupling, and they differ from the notation used in \cite{Cavaglia:2021bnz}. The scaling dimensions of those states are computed with the QSC and read
\beq
\Delta_{11}=4+w_1 g^2+z_3 g^4+O(g^6),\qquad
\Delta_{12}=4+w_2 g^2+z_2 g^4+O(g^6),\qquad
\Delta_{13}=4+w_3 g^2+z_1 g^4+O(g^6),\nn
\eeq
with $w_{1,2,3}$ and $z_{1,2,3}$ the solutions of the following equations
\beqa
&& 3 w_i^3-73 w_i^2+553 w_i-1274=0\qquad\text{with}\qquad w_1<w_2<w_3,\nn\\
&& 32399568 z_j^3+3007792328 z_j^2+66692590965 z_j+57114350496=0
\qquad\text{with}\qquad z_1<z_2<z_3.\nn
\eeqa} at $J=4$ reads
\beq
a_{11}^{(0)}=s_1,\qquad
a_{12}^{(0)}=s_2,\qquad
a_{13}^{(0)}=s_3,
\eeq
where $s_{1,2,3}$ are the three solutions of the following polynomial equation
\beq\label{eq:defpi}
794584 \,s_i^3 - 56756 \,s_i^2 + 1066 \,s_i - 3=0\qquad \text{with}\qquad p_1<p_2<p_3 ,
\eeq
which are numbers deriving from the form of the one loop anomalous dimensions. 
Notice that $a_{11}^{(0)}+a_{12}^{(0)}+a_{13}^{(0)}=1/14$ as expected by \eqref{CLOavg}, which implies that all the other states at $J=4$ have vanishing structure constants at leading order. 
Together with the previous results, we obtain also new constraints on the averages for $J=2,3,4$, given by
\beq\begin{split}
\langle \,a^{(2)}\,\rangle_2&=9-\frac{4 \pi ^2}{3}+6 \zeta_3-\frac{8 \pi ^4}{75},\\
\langle \,a^{(2)}\,\rangle_3&=\frac{166907}{9261}-\frac{2041 \pi ^2}{1323}+\frac{38 \zeta_3}{7}-\frac{8 \pi ^4}{105},\\
\langle \,a^{(2)}\,\rangle_4&=\frac{4919}{432}-\frac{28 \pi ^2}{27}+3 \zeta_3-\frac{4 \pi ^4}{105}.
\end{split}\eeq

Finally, following the same logic of section \ref{sec:predictC1}, we can use the integral relation \eqref{eq:constr1} to generate more data for $C_1^2$. First, from our result \eqref{h2} and \eqref{defh}, \eqref{pt4} 
we assemble the 4-point function at order $g^4$ obtaining 
\beqa\label{eqn:Gweakg4}
&&G_{\text{weak}}^{(2)}(x)\!=\!\frac{4(1\!-\!2x)}{(1\!-\!x)^2}\biggl[
 \frac{\pi^2}{3}(  H_{2}\!-\! H_{1, 0} \!+\! H_{1, 1})\! -\! 
    3 \zeta_3 H_{1}\!+\! 2 (H_{1, 2, 1}\! - \!
  H_{1, 1, 2}\!-\! 
 H_{2, 0, 0}\!-\! 
  H_{1, 1, 0, 0})\nn\\
 &&
 + \!\frac {(x \!-\! 1)x \!+\! 1} {2 x \!-\! 1} \!\left(\!2H_{2, 1}\!-\! 2H_{1, 0, 0}\!+\! (x \!-\! 1)H_{1, 1, 0}\!-\! (x\! -\! 2)H_{2, 0}\!+\!  xH_{3}\!+\!\frac{\pi^2}{3}(H_{1}\!-\!    xH_{0})\!\right)\label{eq:G2loop}\\
 && +\!\frac{ (x^3 \!+\! 1)}{1\! -\! 2 x} H_{1, 2}  \!+\! 
  H_{1, 3}\! -\! 
 H_{2, 2} \!+\! 2 H_{3, 0} \!+\! 
 2 H_{3, 1}\! +\! H_{1, 2, 0}   \!-\!  H_{2, 1, 0}  \!+\! x\frac{
  2 \pi^4 x\! -\! 45 ((x \!-\! 1) x\! +\! 1) \zeta_3}{15(2 x \!- \!1)}\biggr].\nn
\eeqa
Then, plugging $G_{\text{weak}}^{(2)}$ into the integral relation \eqref{int1weak2}, performing the integrals and solving for $a_1^{(4)}$ we get the 4 loop prediction
\beq\label{a14}
a_1^{(4)}=-4480+\frac{832 \pi ^2}{3}-256 \zeta_3+\frac{224 \pi ^4}{15}-880 \zeta_5+\frac{64 \pi ^6}{45} .
\eeq

\paragraph{The correlator at 3 loops. }
Fixing the correlator at 3 loops is way more involved than the previous case, and it is the first order at which the algorithm described above fails to fix the answer completely. 
For $\ell=3$, the basis \eqref{basis} with maximal transcendentality 6 counts 128 elements. 
We will assume, as suggested by the previous orders, that the  $\beta^{(3)}$ constants are linear combinations of numbers built as products of the basis elements (\ref{eq:basis}), with rational coefficients. This assumption reduces the ansatz to 100  unknown rational parameters. 

Requiring that the small $x$ expansion of $h_{\text{weak}}^{(3)}$ does not contains terms proportional to $\log^nx$ with $n>3$, we get 5 additional $\beta$'s. Then, using the crossing equation \eqref{hcrossing}, we further constrain the system obtaining 50 additional coefficients.

The next step is to include integrability data in the derivation.
Expanding $h_{\text{weak}}^{(3)}$ at small $x$, we compare terms proportional to $x^n\log^m x$ with the same terms appearing in the expansion of \eqref{eq:OPEf} where we have injected spectral data from the QSC (see appendix \ref{apd:anylSpec}) and structure constants from the previous orders. Inspecting the contribution of the $J=1$ state fixes 15 coefficients, while the contributions of higher states are less constraining since they are more and more degenerate. Indeed, $J=2$ states gives 3 constraints and $J=3$ and $J=4$ only 1 each. As before, this procedure generates several constraints on structure constants functions and  in particular we obtain $a_1^{(4)}$ in terms of $\beta^{(3)}$ coefficients. Since this quantity was previously computed in \eqref{a14} exploiting the integral relation \eqref{eq:constr1}, one can use it to fix 8 additional coefficients. Finally, plugging  $h_{\text{weak}}^{(3)}$ in the second integral relation \eqref{eq:constr2}, performing the integrals and comparing with the $O(g^6)$ term of the r.h.s., we get 3 more constraints\footnote{A partial result for the $4$-point function at $3$ loops is available upon request.}.

The remaining 14 coefficients are unconstrained. 
It is possible that more sophisticated analytical bootstrap techniques, such as the inversion formula developed in 1D in \cite{Mazac:2018qmi}, might be helpful in this context. In any case, our finding  seems to suggest that using analytical bootstrap equations for a single correlators may not be enough to pinpoint the solution completely, even with a full knowledge of the spectrum. One expects that studying multiple correlators would have the biggest impact on fixing the solution at higher orders, as also observed at strong coupling~\cite{Ferrero:2021bsb}.

\subsection{Results} To summarise, we collect here the analytic results for structure constants obtained with the above approach. For the ground state:
\beq\begin{split}\label{eq:C1finalresult}
C_1^2(g) = &2 g^2 - \left(24-\frac{4 \pi ^2}{3}\right)g^4+
\left(320-16 \pi
   ^2+48 \zeta_3-\frac{76 \pi ^4}{45}\right)g^6\\
   &-\left(4480-\frac{832 \pi ^2}{3}+256 \zeta_3-\frac{224 \pi ^4}{15}+880 \zeta_5-\frac{64 \pi ^6}{45}\right)g^8+
O(g^{10}),
\end{split}\eeq
and for excited states at $J=2$
\beqa
C_2^2&=&\frac{1}{10}+\frac{1}{150} \left(10 \pi ^2-75-9 \sqrt{5}\right)g^2+
O(g^{4}),\\
C_3^2&=&\frac{1}{10}+\frac{1}{150} \left(10 \pi ^2-75+9 \sqrt{5}\right)g^2+
O(g^{4}),
\eeqa
and the non-vanishing excited states at leading order at $J=3,4$
\beqa
C_6^2=\frac{1}{14}+\frac{2}{7\sqrt{37}}+
O(g^{2})\qquad
C_8^2=\frac{1}{14}-\frac{2}{7\sqrt{37}}+
O(g^{2}),
\eeqa
\beqa
C_{11}^2=s_1+
O(g^{2})\qquad
C_{12}^2=s_2+
O(g^{2})\qquad
C_{13}^2=s_3+
O(g^{2}),
\eeqa
where the constants $s_i$ are the roots of (\ref{eq:defpi}). Our 2-loop result for the 4-point function is given in (\ref{eq:G2loop}).  Weak coupling data for scaling dimensions, coming from the QSC, are collected  in Appendix \ref{app:spectrum}.

\section{Discussion}\label{sec:discussion}
In this paper we have continued experimenting with a combination of integrability and conformal bootstrap methods to study observables in $\mathcal{N}$=4 SYM. This has led to the most accurate results to date for a non-supersymmetric  OPE coefficient of short operators at finite coupling. For instance,  as highlighted in fig. \ref{fig:zoom}, with the methods of this paper we determine one such structure constant with error $10^{-8}$ for 't Hooft coupling $\lambda \sim 24 \pi^2 $. Presently, this would  not be  achievable either with integrability or conformal bootstrap methods on their own. 

 In this work, we have introduced new constraints on integrated correlators in the 1D defect CFT,  connecting them to another quantity available from integrability, the cusp anomalous dimension. 
The addition of these constraints was shown to greatly enhance the precision of the numerical bootstrap algorithm, allowing us to reach at least 7 digits of precision for $C_1^2$ over a wide range of the coupling for $g>1$ (becoming 9 digits for $g\gtrsim 3$), and at least 2 digits precision for the next two OPE coefficients for $g>1$ (see fig.~\ref{fig:plotintboth2}). 

These new constraints were also very powerful in an analytic functional bootstrap approach, which allowed us to fix the form of the 4-point function at 2 loops, fix 4 loops for $C_1^2$, and 2 loops for the next two excited states OPE coefficients.

This development resonates nicely with the recent discovery of integrated correlators constraints for the bulk 4D theory, in that case arising from localisation~\cite{Binder:2019jwn} (see also \cite{Dorigoni:2021bvj,Dorigoni:2021guq}), which were also shown to have a great impact on the bootstrap~\cite{Chester:2021aun}. 

It would be interesting to see if more general deformations of the MWL can lead to even more constraints of the type studied here. At the very least,  
 there should be generalisations of the present identities involving integrated $n$-point functions with $n>4$. Those would be related to higher-orders in the  near-BPS expansion of  the cusp anomalous dimension, which are also in principle accessible with integrability. Such generalised constraints might be useful in the bootstrap. 

An important question is whether using Bootstrability it is  possible to compute the OPE coefficients with (ideally) arbitrary precision, or if there is a fundamental limit. One way to improve the precision is certainly to include input from more states in the spectrum in the Numerical Boostrap algorithms (here we used only 10 states). Another possibility is to use analytical bootstrap techniques such as the ones developed for 1D CFTs in \cite{Mazac:2016qev,Mazac:2018mdx,Mazac:2018qmi,Mazac:2018ycv,Paulos:2019fkw,Ferrero:2019luz,Bianchi:2021piu}, which might  prove to be advantageous. 

However, we believe it is very unlikely that the study of a single correlator will be enough to fix all OPE coefficients, even from a completely known spectrum. This is in fact what we observe analytically at weak coupling, where the functional bootstrap approach did not fix completely the 4-point function at 3 loops. One can argue that this is in fact to be expected, as the CFT is defined not by one, but by all its correlation functions. Also in the Numerical Conformal Bootstrap, it is only the study of multiple correlators that allows to find small islands for allowed conformal data~\cite{Kos:2016ysd}. Thus, we believe that multi-correlator Bootstrability is a very promising direction for the future. 

Our present setup is the 1D defect CFT, which presents some simplifications. In particular, it is a consistent CFT at the planar level. 
This is not the case for the bulk 4D CFT, where there is intermingling of single and double traces contributing to planar 4-point functions. This presents a challenge, since the QSC does not know about the anomalous dimensions of double traces. However, we expect that analytic conformal bootstrap techniques such as the ones in \cite{Caron-Huot:2017vep,Alday:2017vkk,Caron-Huot:2020adz,Bissi:2022mrs,Bissi:2021spj} might help to resolve this problem. In particular, it is inspiring that the double discontinuity of a 4-point function, which parametrises  the full correlator thanks to the Lorentzian inversion formula~\cite{Caron-Huot:2017vep}, is determined only by single trace operators in a large $N$ theory~\cite{Alday:2017vkk}. 
For instance, at strong coupling $\lambda \sim \infty$, the correlator was reconstructed, at several orders in $1/N$, starting from information on single-trace protected operators, which are the only ones contributing in this regime~\cite{Aharony:2016dwx,Alday:2017xua}. 
Our hope is that Bootstrability could be the way to extend these beautiful results to the full finite $\lambda$ region. 

It is also an interesting direction for the future to extend these techniques to other integrable gauge theories. A natural setup would be the one of the Wilson line defect CFTs living in ABJM theory defined in \cite{Bianchi:2017ozk,Bianchi:2018scb}, recently studied from the Bootstrap approach in \cite{Bianchi:2020hsz}. For this setup, the cusp is also intensively studied in \cite{Griguolo:2012iq,Bonini:2016fnc} and the Bremsstrahlung function is known exactly \cite{Bianchi:2014laa,Correa:2014aga,Bianchi:2017svd,Bianchi:2018scb} (see also \cite{Drukker:2019bev}), although an integrability formulation is still lacking. The QSC for the spectrum of local operators was found for this theory in \cite{Cavaglia:2014exa,Bombardelli:2017vhk} (for its numerical solution see \cite{Bombardelli:2018bqz}). Finding its deformation capturing the  defect CFT would be very interesting and would open the way to applying the methods presented here. 
The AdS$_3$/CFT$_2$ duality for which a QSC was recently proposed in \cite{Cavaglia:2021eqr,Ekhammar:2021pys} is also a fascinating laboratory to develop Bootstrability in the context of 2D CFTs. 

 Finally, the fishnet limit of $\mathcal{N}$=4 SYM ~\cite{Gurdogan:2015csr} and ABJM \cite{Caetano:2016ydc} theories are also very interesting playgrounds for combining integrability and bootstrap techniques, which present additional challenges due to the non-unitarity. We should mention that the Fishnet theories are a very  promising setting to understand correlation functions analytically as shown in many recent works 
 \cite{Grabner:2017pgm,Basso:2017jwq,Gromov:2018hut,Kazakov:2018gcy,Pittelli:2019ceq,Derkachov:2019tzo,Derkachov:2020zvv,Shahpo:2021xax}. The QSC is also under control in this limit  \cite{Gromov:2017cja,Gromov:2019jfh,Cavaglia:2020hdb,Levkovich-Maslyuk:2020rlp} (for the open spin chain case, relevant for the fishnet limit of the present setup, see \cite{Gromov:2021ahm}), and for their  simplicity the Fishnet theories appear to be the  
 perfect laboratory to understand the expected connection between QSC and  correlators. 
 There has been progress in this direction~\cite{Cavaglia:2018lxi,Cavaglia:2021mft} (for other limits see e.g. \cite{Jiang:2015lda,Giombi:2018qox,Giombi:2022anm}) but a full solution is still missing. Having  non-perturbative data obtained with the help of bootstrap methods could be important to inform these efforts. Moreover, perhaps the study of these limits will reveal new ways in which integrability and bootstrap should be fused at a more fundamental level to study AdS/CFT-related theories. 

\acknowledgments
We thank Simon Caron-Huot, Nadav Drukker, Pietro Ferrero, Alessandro Georgoudis, Gregory Korchemsky, Petr Kravchuk, Andrea Manenti, Carlo Meneghelli, Amit Sever, Evgeny Sobko, Nika Sokolova, Andreas Stergiou, Roberto Tateo, Emilio Trevisani and Pedro Vieira for inspiring discussions. 
The work of AC, NG and MP is
supported by European Research Council (ERC) under
the European Union’s Horizon 2020 research and innovation programme (grant agreement No. 865075) EXACTC. NG is also partially supported by the STFC grant (ST/P000258/1).

\appendix

\section{Details on the curvature function from the QSC}\label{app:curvature}
We collect some details in the determination of the Curvature function from integrability~\cite{Gromov:2015dfa}. 
The kernels appearing in \eqref{Gamma2res} are given by
\beqa
\label{Gamma0}
K_0(u)&=&\d_u\log\frac{\Gamma(i u+1)}{\Gamma(-i u+1)},\\
\label{Gammam}
K_{-\phi}(u)&=&
e^{-2\phi u}
\[
-i e^{-2i\phi}\Phi(e^{-2 i\phi},1,1-iu)
-i e^{+2i\phi}\Phi(e^{+2 i\phi},1,1+iu)
\],\\
\label{Gammap}
K_{+\phi}(u)&=&
e^{+2\phi u}
\[
-i e^{+2i\phi}\Phi(e^{+2 i\phi},1,1-iu)
-i e^{-2i\phi}\Phi(e^{-2 i\phi},1,1+iu)
\]\ ,
\eeqa
where $\Phi$ is the Hurwitz-Lerch transcendent function\footnote{In Wolfram Mathematica it is the function
\texttt{HurwitzLerchPhi}.} defined through the following infinite sum
\beq
\Phi(z,s,a)=\sum_{k=0}^\infty \frac{z^k}{(a+k)^s}.
\eeq
The remainder of the integrand of \eqref{Gamma2res} is organised in the following functions
\beqa\label{D+-0}
	D_+(x, y)&=&
	\frac{iS_+(y)e^{2g\phi (y-\frac{1}{y})}}{g^3I_1^\phi e^{2g\phi (x-\frac{1}{x})}}
	\left(\frac{I^\phi_2 x S_+(y)}
	{(I^\phi_1)^2(x^2-1)}-\frac{2S_+(y)}{gI^\phi_1}
	-\frac{2S_+(x)e^{4g\phi (x-\frac{1}{x})}}{gI^\phi_1}
	-\frac{2(x\!+\!y)(1\!-\!x y)}{(x^2\!-\!1)(y^2\!-\!1)}\right),\nn\\
	D_0(x,y)&=&
	\frac{2iS_+(y)}{g^3I_1^\phi}\left(\frac{S_+(x)}{gI^\phi_1}
	-\frac{x I^\phi_2 S_+(x)}{(I^\phi_1)^2(x^2-1)}
    -\frac{2 x^2}{(x+1/x)(x^2-1)}\right),\\
	D_-(x,y)&=&\frac{iI^\phi_2}{g^3 (I^\phi_1)^3}
	\frac{x (S_+(x))^2e^{2g\phi (x-\frac{1}{x})}}
	{(x^2-1)e^{2g\phi (y-\frac{1}{y})}}.\nn
\eeqa
These are functions of $x$, $y$, related to $u_x$, $u_y$ by the Zhukovsky map
\beq\label{zhu}
	x+\frac{1}{x}=\frac{u_x}{g},\ \ \ |x|\geq 1\ ,
\eeq
(analogous for $y$, $u_y$), which resolves the cut $[-2g,2g]$ around which the integrals in \eqref{Gamma2res} run. The other objects appearing in \eqref{D+-0} are the deformed Bessel functions
\beq
	I_n^\phi=\frac{1}{2}I_{n}\(4\pi g\sqrt{1-\frac{\phi^2}{\pi^2}}\)\[
	\(\sqrt{\frac{\pi+\phi}{\pi-\phi}}\)^{n}-
	(-1)^n\(\sqrt{\frac{\pi-\phi}{\pi+\phi}}\)^{n}
	\]\;,
\eeq
with $I_n$ being the modified Bessel function, and the following sum 
\beq
\label{Sdef}
S_+(x)\equiv \sum_{n=1}^\infty I^{+\phi}_n x^{-n}.
\eeq

In the $\phi\rightarrow 0$ limit, \eqref{Gamma2res} reduces to the Curvature function defined in \eqref{curvaturedef} where the kernel $K_0$ is given in \eqref{Gamma0} and the function $F$ is
\beqa\label{Fxy}
	F[x,y]&&=-\frac{8 i \sinh \left(2 \pi  u_x\right) u_x  u_y x^2S_0(y) }{I_1(4 g \pi ){}^2}
	\\ \nn
	&&+S_0(y){}^2\left[\frac{8 i x y I_2(4 g \pi ) u_x u_y}{g
   \pi  \left(x^2-1\right) I_1(4 g \pi ){}^3}-\frac{8 i x y I_2(4 g \pi ) u_x u_y}{g \pi
   \left(y^2-1\right) I_1(4 g \pi ){}^3}+\frac{32 i x y u_x u_y}{I_1(4 g \pi ){}^2}\right]
	\\ \nn
	&&+\sinh ^2\left(2 \pi  u_y\right) \left[\frac{4 i x y I_2(4 g \pi ) u_x u_y}{g
   \pi  \left(x^2-1\right) I_1(4 g \pi ){}^3}+\frac{16 i x y u_x u_y}{I_1(4 g \pi
   ){}^2}\right]
	\\ \nn
	&&+\sinh \left(2 \pi  u_y\right) \left[\frac{4 i x u_x u_y
   y^2}{\left(x^2-1\right) I_1(4 g \pi )}
	-\frac{8 i x \sinh \left(2 \pi  u\right) u_x u_y
   y}{I_1(4 g \pi ){}^2}
	-\frac{8 i u_x u_y S_1(x) y}{g I_1(4 g \pi ){}^2}
	\right.
	\\ \nn
	&& \left.
	-\frac{16 i x u_x 
   u_y}{\left(y^2-1\right) I_1(4 g \pi )}+\left(-\frac{8 i x y I_2(4 g \pi ) u_x u_y}{g \pi
   \left(x^2-1\right) I_1(4 g \pi ){}^3}-\frac{32 i x y u_x u_y}{I_1(4 g \pi ){}^2}\right)
   S_0(y)\right]
	\\ \nn
	&&+S_1(y)\left[\frac{8 i x y u_x u_y}{g \left(x^2-1\right) I_1(4 g \pi )}-\frac{8 i x
   y u_x u_y}{g \left(y^2-1\right) I_1(4 g \pi )}\right]
	\\ \nn
	&&
	+S_0(x) \left[S_0(y)\left(\frac{16 i
   u_x u_y}{I_1(4 g \pi ){}^2}-\frac{16 i y^2 u_x u_y}{I_1(4 g \pi ){}^2}\right)
   -\frac{4 i x I_2(4 g \pi ) u_x u_y S_1(y)}{g^2 \pi  \left(x^2-1\right) I_1(4 g \pi
   ){}^3}\right]
	\\ \nn
	&+&S_0(y) \left[\frac{8 i x u_x u_y y^2}{\left(x^2-1\right) I_1(4 g \pi
   )}
	+\frac{8 i x u_x u_y}{I_1(4 g \pi )}
	-\frac{8 i x u_x u_y}{\left(x^2-1\right) I_1(4 g \pi
   )}
	+\frac{32 i x u_x u_y}{\left(y^2-1\right) I_1(4 g \pi )}
	\right. \\ \nn
	&&\left. +S_1(x)\left(-\frac{4 i x I_2(4 g \pi
   ) u_x u_y}{g^2 \pi  \left(x^2-1\right) I_1(4 g \pi ){}^3}-\frac{16 i x u_x u_y}{g I_1(4 g
   \pi ){}^2}\right)
	\right. \\ \nn
	&& \left.
	+S_1(y)\left(\frac{4 i x I_2(4 g \pi ) u_x u_y}{g^2 \pi
   \left(x^2-1\right) I_1(4 g \pi ){}^3}+\frac{16 i x u_x u_y}{g I_1(4 g \pi ){}^2}\right)
   \right],
\eeqa
where
\beq
	S_0(x)=\sum\limits_{n=1}^\infty \frac{I_{2n+1}(4\pi g)}{x^{2n+1}},\ \ \
	S_1(x)=\sum\limits_{n=1}^\infty \frac{2nI_{2n}(4\pi g)}{\pi x^{2n}}\textbf{}.
\eeq
\section{Four-point function with generic polarisations}\label{app:covariant}
The 4-point function was  for any combination of operators from the $\mathcal{B}_1$  multiplet  in \cite{Liendo:2018ukf}, using a superspace formalism. By superconformal symmetry, all such correlators depend on the same reduced function of the cross ratio $f(x)$, entering the bootstrap problem of section \ref{sec:bootstrapsetup}. In the case of scalar primaries with generic  polarisations, we have explicitly
\beq
\langle \langle \Phi_{\perp}^{i_1}(x_1) \Phi_{\perp}^{i_2}(x_2) \Phi_{\perp}^{i_3}(x_3) \Phi_{\perp}^{i_4}(x_4)\rangle\rangle = G_{{i_1}i_2 i_3 i_4}(x)\times \langle \langle \Phi_{\perp}^1(x_1) \Phi_{\perp}^1(x_2) \rangle\rangle \, \langle \langle \Phi_{\perp}^1(x_3) \Phi_{\perp}^1(x_4) \rangle\rangle \;, 
\eeq
with $x = x_{12} x_{34}/(x_{13} x_{24} )$. Depending on polarisations, the amplitude can be split in three contributions,
\beq
G_{{i_1}i_2 i_3 i_4}(x)  =
\delta_{i_1,i_2}\delta_{i_3,i_4
}G_1(x) +\delta_{i_1 i_3} \delta_{i_2 i_4 } G_2(x) 
+
\delta_{i_1 i_4} \delta_{i_2,i_3}G_3(x)\;,
\eeq
which are related to $f(x)$ as
\beqa
G_1(x) &=&  \left(\frac{2}{x}-1\right) f(x)+(x-1) f'(x) \;,\\
G_2(x) &=&  \mathbb{F}\,x^2 -(x-1) f'(x) x-f(x)\;, \\
G_3(x) &=&f(x)-x f'(x) \;.\label{eq:G3}
\eeqa
For homogeneous polarisation one recovers $G(x)$ used in section \ref{sec:bootstrapsetup}, as \beq
G(x) = G_1(x) + G_2(x)+G_3(x).
\eeq
\section{Rewriting the integral constraints}\label{app:rewriteconstr}
Consider the second constraint (\ref{eq:constr2}) first. Using the crossing equation, it can be rewritten as
\begin{align}\label{constr2half}
    \text{Constraint 2:   }\;\;    \int_{0}^{\frac{1}{2}} dx \frac{\delta f(x) \; ( 2 x - 1) }{x^2} =     \frac{\mathbb{C}}{4\;\mathbb{B}^2} + \mathbb{F}-3,
\end{align}
which is convenient since on the interval $x\in (0,\frac{1}{2})$ the OPE decomposition (\ref{eq:OPEf}) converges very fast and can be safely exchanged with the integration. Plugging in $\delta f = \sum_n C^2_n f_{\Delta_n} + f_{\mathcal{I}} + C^2_{\text{BPS}} f_{\mathcal{B}_2 } - f_{\text{tree}}$, we can now rewrite (\ref{constr2half}) as
\beq
\sum_{\Delta_n} C^2_n \; \texttt{Int}_2\left[\, f_{\Delta_n} \,\right] + \texttt{RHS}_2 = 0 ,
\eeq
where $\texttt{Int}_2$ is an integral operator defined as 
\beq
\texttt{Int}_2\left[ F(x)\right]\equiv  \int_{0}^{\frac{1}{2}} dx \frac{ F(x) \; ( 2 x - 1) }{x^2} ,
\eeq
and where  $\texttt{RHS}_2$ is defined by $\texttt{RHS}_2 \equiv  \texttt{Int}_2\left[ f_{\mathcal{I}} + C^2_{\text{BPS}}\, f_{\mathcal{B}_2} - f_{\text{tree}}\right] -\frac{\mathbb{C}}{4\;\mathbb{B}^2} - \mathbb{F}+3 $. Plugging in the values of the blocks, we find  it explicitly:
\beq
\texttt{RHS}_2 = \frac{1-\mathbb{F}}{6}+(2-\mathbb{F}
   ) \log (2)+1 -\frac{\mathbb{C}}{4\;\mathbb{B}^2} .
\eeq
We can recast (\ref{eq:constr1}) in a similar form. Using the identity (\ref{eq:specialide}), the constraint can be written as
\beq
\text{Constraint 1:   }\;\;\; 1 +  \int_0^{\frac{1}{2}}
    \delta G(x)\frac{\log\left( x (1-x)\right)}{x^2} dx  = \frac{3\mathbb{C}-\mathbb{B}}{8 \;\mathbb{B}^2}.
\eeq
Given identity (\ref{eq:totalder}), we see that $\delta G(x)/x^2$ is a total derivative, so that we can integrate by parts. For this, it is useful to notice that, due to crossing,  $f(\frac{1}{2}) = 0$. Plugging in the OPE expansion of $f$, integrating by parts the pieces corresponding to the long blocks and performing the remaining integrals, we can massage the constraint to the form:
\beq
\text{Constraint 1:} \;\;\;\sum_{\Delta_n} C^2_n \; \texttt{Int}_1\left[ f_{\Delta_n} \right] + \texttt{RHS}_1 = 0 ,
\eeq
where we defined the integral operator
\beq
\texttt{Int}_1\left[ F(x) \right] \equiv - \int_0^{\frac{1}{2}} (x-1-x^2)\frac{F(x)}{x^2} \partial_x\log\left( x (1-x)\right)\, dx ,
\eeq
and the explicit term $\texttt{RHS}_1$ is obtained as
\beqa
\texttt{RHS}_1 &\equiv& 1 -\frac{3\mathbb{C}-\mathbb{B}}{8 \;\mathbb{B}^2}-  \left.   (x-1-x^2)\frac{f_{\mathcal{I}}(x) + C^2_{\text{BPS}} f_{\mathcal{B}_2}(x) }{x^2} \, \log\left( x (1-x)\right)\right|_{x = \frac{1}{2} } \nn\\
&&+  \int_0^{\frac{1}{2}} \partial_x \left( (\mathbb{F}-2) x + (x-1-x^2)\frac{f_{\mathcal{I}}(x) + C^2_{\text{BPS}} f_{\mathcal{B}_2}(x) - f_{\text{tree}}(x)}{x^2} \right)\, \log\left( x (1-x)\right)\, dx \nn ,\\
\eeqa
which evaluates explicitly to
\beq
\texttt{RHS}_1=\frac{\mathbb{B}-3
   \mathbb{C}}{8\mathbb{B}^2}+\left(7 \log(2) -\frac{41}{8}\right) (\mathbb{F}-1)+  \log
   (2).
\eeq

\section{Spectral data}\label{app:spectrum}

In this appendix we present some analytical and numerical data for the ten states in our spectrum.
This data is also shared in a~\texttt{Mathematica} notebook attached to this paper. 

\subsection{Perturbative }\label{apd:anylSpec}

Here we present analytical weak coupling data for the first 10 states. States are labelled in the order they appear as $g\to0\;$. 
\begin{multline}\label{D1w}
    \Delta_{1}=1+4 g^2-16 g^4 \\ +\bigg[128-\frac{56 \pi ^4}{45}\bigg] g^6 
    +\bigg[
    \frac{272}{135} \pi^{6}+\frac{128}{3} \pi^{2}-\frac{64}{3} \pi^{2} \zeta_{3}+128 \zeta_{3}-160 \zeta_{5}-1280
   \bigg]g^8\\
   +\bigg[-\frac{7328}{2835} \pi^{8}-\frac{64}{2835} \pi^{6}-\frac{896}{45} \pi^{4}-\frac{2560}{3} \pi^{2}+\frac{64}{3} \pi^{4} \zeta_{3}+\frac{512}{3} \pi^{2} \zeta_{3}+\frac{448}{3} \pi^{2} \zeta_{5} \\
-384\left(\zeta_{3}\right)^{2}-1024 \zeta_{3}-640 \zeta_{5}+2688 \zeta_{7}+14336\bigg] g^{10}+O\left(g^{12}\right)\;,
\end{multline}
from~\cite{Grabner:2020nis}. Solving the same QSC equations, for excited states we find
\begin{multline}\label{D2w}
    \Delta_{2}
    =
    2
    +\bigg[5-\sqrt{5}\bigg] g^2
    -2\bigg[9-4 \sqrt{5}\bigg] g^4
    \\
    +\bigg[185-\frac{437}{\sqrt{5}}+\frac{2}{3} \left(13-5 \sqrt{5}\right) \pi
   ^2+\frac{14}{45} \left(3 \sqrt{5}-5\right) \pi ^4\bigg]
   g^6+O\left(g^8\right)\;,
\end{multline}

\begin{multline}\label{D3w}
    \Delta_{3}
    =
    2
    +\bigg[5+\sqrt{5}\bigg] g^2
    -2 \bigg[9+4 \sqrt{5}\bigg]
    g^4
    \\
    +\frac{1}{45} \bigg[9 \left(925+437 \sqrt{5}\right)+30 \left(13+5 \sqrt{5}\right) \pi ^2-14
   \left(5+3 \sqrt{5}\right) \pi ^4\bigg]
   g^6+O\left(g^8\right)\;,
\end{multline}

\beq\label{D4w}
\Delta_{4}=3 + p_1 \,g^2 + q_3 \,g^4+O\left(g^6\right)\;,
\eeq

\beq\label{D5w}
\Delta_{5}=3 + p_2 \,g^2 + q_2 \,g^4 + O\left(g^6\right)\;,
\eeq

\beq\label{D6w}
\Delta_{6}=3+\frac{1}{3} \bigg[23-\sqrt{37}\bigg] g^2-\frac{2}{10989} \bigg[138824-21943 \sqrt{37}\bigg] g^4+O\left(g^6\right)\;,
\eeq

\beq\label{D7w}
\Delta_{7}=
3+9 g^2-\frac{69}{2}g^4
+ \bigg[\frac{501}{2}-\frac{7 \pi ^2}{2}\bigg] g^6
+O\left(g^8\right)\;,
\eeq

\beq\label{D8w}
\Delta_{8}=
3+\frac{1}{3} \bigg[23+\sqrt{37}\bigg] g^2-\frac{2}{10989} \bigg[138824+21943 \sqrt{37}\bigg] g^4
+O\left(g^6\right)\;,
\eeq

\beq\label{D9w}
\Delta_{9}=3 + p_3 \,g^2 + q_1 \,g^4+O\left(g^6\right)\;,
\eeq

\begin{align}\label{D10w}
    \Delta_{10} = 4 + r_1 \,g^2 + O(g^4)\;,
\end{align}
The coefficients $p_i$ and $q_j$ for the states $\Delta_4$, $\Delta_5$ and $\Delta_6$ are respectively, solutions of the algebraic equations
\beq
p_i^3 - 19 \,p_i^2 + 96\, p_i -128=0\;, \quad\text{with}\quad p_1<p_2<p_3\;.
\eeq
and
\beq
781\, q_j^3 + 59143\, q_j^2 + 1008864\, q_j + 12200=0\;, \quad\text{with}\quad q_1<q_2<q_3\;.
\eeq
Finally, the coefficient $r_1$ in $\Delta_{10}$ is the lowest magnitude solution of
\begin{multline}
    r_k^{10}
    -88 \,r_k^9
    +3388 \,r_k^8
    -74980 \,r_k^7
    +1053428 \,r_k^6
    -9783816 \,r_k^5
    +60570976 \,r_k^4\\
    -245427424 \,r_k^3
    +618124224 \,r_k^2
    -864113152 \,r_k
    +500028928 = 0\;.
\end{multline}
Indeed, other solutions $r_k$ are one-loop anomalous dimensions of higher states with $\Delta_0 = 4$.  
One-loop anomalous dimensions for $\Delta_{2}$, $\Delta_{3}$, $\Delta_{4}$, $\Delta_{5}$, $\Delta_{7}$, $\Delta_{9}$ and $\Delta_{10}$ can also be extracted from diagonalising the mixing matrix/spin-chain Hamiltonian in~\cite{Correa:2018fgz}.

\subsection{Numerical}\label{apd:numSpec}
In Tables \ref{tab:specDel1}-\ref{tab:specDel10} we present numerical values, with at least 12  significant digits, for $\Delta_i$, $1\leq i \leq 10$ obtained with the QSC.
States are labelled in the order they appear as $g\to0\;$. Note that at different values of $g$, the ordering of the states could change and one should keep this in mind while using these data for NCB applications like in section~\ref{sec:numerical}.
The complete set of data is provided in the notebook supplementing this paper. 

\begin{table}[H]
    \centering
    \begin{tabular}{||cc|cc|cc|cc||}
    \hline
    $g$ & $\Delta$ & $g$ & $\Delta$ & $g$ & $\Delta$ & $g$ & $\Delta$ \\
    \hline\hline
 0.2 & 1.136002453 & 1.2 & 1.716822613 & 2.2 & 1.834303271 & 3.2 & \
1.882963221 \\
 0.4 & 1.358591986 & 1.4 & 1.751935390 & 2.4 & 1.847018106 & 3.4 & \
1.889459116 \\
 0.6 & 1.509952145 & 1.6 & 1.779330126 & 2.6 & 1.857923582 & 3.6 & \
1.895272465 \\
 0.8 & 1.605510910 & 1.8 & 1.801292632 & 2.8 & 1.867379766 & 3.8 & \
1.900505392 \\
 1.0 & 1.670227842 & 2.0 & 1.819289288 & 3.0 & 1.875657233 & 4.0 & \
1.905240630 \\
    \hline
    \end{tabular}
    \caption{$\Delta_1$ taken from~\cite{Grabner:2020nis}.}
    \label{tab:specDel1}
\end{table}

\begin{table}[H]
    \centering
    \begin{tabular}{||cc|cc|cc|cc||}
    \hline
    $g$ & $\Delta$ & $g$ & $\Delta$ & $g$ & $\Delta$ & $g$ & $\Delta$ \\
    \hline\hline
 0.2 & 2.111484971 & 1.2 & 3.244972034 & 2.2 & 3.548950040 & 3.2 & \
3.678681415 \\
 0.4 & 2.419585353 & 1.4 & 3.334493736 & 2.4 & 3.582628334 & 3.4 & \
3.696173111 \\
 0.6 & 2.740095285 & 1.6 & 3.405114865 & 2.6 & 3.611638194 & 3.6 & \
3.711861467 \\
 0.8 & 2.968232142 & 1.8 & 3.462235845 & 2.8 & 3.636885730 & 3.8 & \
3.726011360 \\
 1.0 & 3.127846278 & 2.0 & 3.509381656 & 3.0 & 3.659057111 & 4.0 & \
3.738838260 \\
    \hline
    \end{tabular}
    \caption{$\Delta_2$, only this state and $\Delta_1$ were used to obtain the results of~\cite{Cavaglia:2021bnz}.}
    \label{tab:specDel2}
\end{table}

\begin{table}[H]
    \centering
    \begin{tabular}{||cc|cc|cc|cc||}
    \hline
    $g$ & $\Delta$ & $g$ & $\Delta$ & $g$ & $\Delta$ & $g$ & $\Delta$ \\
    \hline\hline
 0.2 & 2.241489897 & 1.2 & 3.318405927 & 2.2 & 3.577014843 & 3.2 & \
3.693284686 \\
 0.4 & 2.637745222 & 1.4 & 3.392795505 & 2.4 & 3.606820906 & 3.4 & \
3.709271389 \\
 0.6 & 2.914439582 & 1.6 & 3.452459836 & 2.6 & 3.632703586 & 3.6 & \
3.723675243 \\
 0.8 & 3.095254939 & 1.8 & 3.501413234 & 2.8 & 3.655390386 & 3.8 & \
3.736720187 \\
 1.0 & 3.222893829 & 2.0 & 3.542317830 & 3.0 & 3.675439099 & 4.0 & \
3.748589811 \\
    \hline
    \end{tabular}
    \caption{$\Delta_3$}
    \label{tab:specDel3}
\end{table}

\begin{table}[H]
    \centering
    \begin{tabular}{||cc|cc|cc|cc||}
    \hline
    $g$ & $\Delta$ & $g$ & $\Delta$ & $g$ & $\Delta$ & $g$ & $\Delta$ \\
    \hline\hline
 0.2 & 3.085668899 & 1.2 & 4.611262730 & 2.2 & 5.154254766 & 3.2 & \
5.392449421 \\
 0.4 & 3.354401576 & 1.4 & 4.769123097 & 2.4 & 5.215698984 & 3.4 & \
5.424879036 \\
 0.6 & 3.762820692 & 1.6 & 4.894817892 & 2.6 & 5.268844551 & 3.6 & \
5.454029130 \\
 0.8 & 4.133185001 & 1.8 & 4.997280771 & 2.8 & 5.315263664 & 3.8 & \
5.480372469 \\
 1.0 & 4.407035545 & 2.0 & 5.082408648 & 3.0 & 5.356155220 & 4.0 & \
5.504295213 \\
    \hline
    \end{tabular}
    \caption{$\Delta_4$}
    \label{tab:specDel4}
\end{table}

\begin{table}[H]
    \centering
    \begin{tabular}{||cc|cc|cc|cc||}
    \hline
    $g$ & $\Delta$ & $g$ & $\Delta$ & $g$ & $\Delta$ & $g$ & $\Delta$ \\
    \hline\hline
0.2 & 3.179538060 & 1.2 & 4.730691980 & 2.2 & 5.202154921 & 3.2 & \
5.417927790 \\
 0.4 & 3.591435315 & 1.4 & 4.865213974 & 2.4 & 5.257226150 & 3.4 & \
5.447799149 \\
 0.6 & 4.017590943 & 1.6 & 4.973723374 & 2.6 & 5.305182235 & 3.6 & \
5.474756134 \\
 0.8 & 4.332109863 & 1.8 & 5.063185037 & 2.8 & 5.347321142 & 3.8 & \
5.499205454 \\
 1.0 & 4.559164233 & 2.0 & 5.138250140 & 3.0 & 5.384642229 & 4.0 & \
5.521481452 \\
    \hline
    \end{tabular}
    \caption{$\Delta_5$}
    \label{tab:specDel5}
\end{table}

\begin{table}[H]
    \centering
    \begin{tabular}{||cc|cc|cc|cc||}
    \hline
    $g$ & $\Delta$ & $g$ & $\Delta$ & $g$ & $\Delta$ & $g$ & $\Delta$ \\
    \hline\hline
 0.2 & 3.210615426 & 1.2 & 4.710076359 & 2.2 & 5.190012398 & 3.2 & \
5.410803027 \\
 0.4 & 3.640914338 & 1.4 & 4.846040453 & 2.4 & 5.246398949 & 3.4 & \
5.441316835 \\
 0.6 & 4.025590582 & 1.6 & 4.956452208 & 2.6 & 5.295490975 & 3.6 & \
5.468836295 \\
 0.8 & 4.317896637 & 1.8 & 5.047804538 & 2.8 & 5.338610706 & 3.8 & \
5.493780214 \\
 1.0 & 4.538908388 & 2.0 & 5.124591891 & 3.0 & 5.376780526 & 4.0 & \
5.516492991 \\
    \hline
    \end{tabular}
    \caption{$\Delta_6$}
    \label{tab:specDel6}
\end{table}

\begin{table}[H]
    \centering
    \begin{tabular}{||cc|cc|cc|cc||}
    \hline
    $g$ & $\Delta$ & $g$ & $\Delta$ & $g$ & $\Delta$ & $g$ & $\Delta$ \\
    \hline\hline
 0.2 & 3.315020835 & 1.2 & 5.376484671 & 2.2 & 5.972956517 & 3.2 & \
6.248917413 \\
 0.4 & 3.941060112 & 1.4 & 5.545686923 & 2.4 & 6.043229749 & 3.4 & \
6.287246836 \\
 0.6 & 4.490146402 & 1.6 & 5.682758918 & 2.6 & 6.104515476 & 3.6 & \
6.321860864 \\
 0.8 & 4.879729434 & 1.8 & 5.796153401 & 2.8 & 6.158435062 & 3.8 & \
6.353274253 \\
 1.0 & 5.161911422 & 2.0 & 5.891557969 & 3.0 & 6.206241201 & 4.0 & \
6.381910882 \\
    \hline
    \end{tabular}
    \caption{$\Delta_7$}
    \label{tab:specDel7}
\end{table}

\begin{table}[H]
    \centering
    \begin{tabular}{||cc|cc|cc|cc||}
    \hline
    $g$ & $\Delta$ & $g$ & $\Delta$ & $g$ & $\Delta$ & $g$ & $\Delta$ \\
    \hline\hline
 0.2 & 3.323455425 & 1.2 & 4.848038537 & 2.2 & 5.251786153 & 3.2 & \
5.444974771 \\
 0.4 & 3.867435619 & 1.4 & 4.961086152 & 2.4 & 5.300520446 & 3.4 & \
5.472212501 \\
 0.6 & 4.259504217 & 1.6 & 5.053416701 & 2.6 & 5.343270686 & 3.6 & \
5.496900789 \\
 0.8 & 4.519240721 & 1.8 & 5.130417567 & 2.8 & 5.381082718 & 3.8 & \
5.519382197 \\
 1.0 & 4.705712531 & 2.0 & 5.195697383 & 3.0 & 5.414769846 & 4.0 & \
5.539940361 \\
    \hline
    \end{tabular}
    \caption{$\Delta_{8}$}
    \label{tab:specDel8}
\end{table}

\begin{table}[H]
    \centering
    \begin{tabular}{||cc|cc|cc|cc||}
    \hline
    $g$ & $\Delta$ & $g$ & $\Delta$ & $g$ & $\Delta$ & $g$ & $\Delta$ \\
    \hline\hline
 0.2 & 3.406279873 & 1.2 & 5.951763518 & 2.2 & 6.704379301 & 3.2 & \
7.053178455 \\
 0.4 & 4.185347283 & 1.4 & 6.164782791 & 2.4 & 6.793212920 & 3.4 & \
7.101604295 \\
 0.6 & 4.852983460 & 1.6 & 6.337711381 & 2.6 & 6.870682871 & 3.6 & \
7.145329080 \\
 0.8 & 5.331457258 & 1.8 & 6.480927012 & 2.8 & 6.938835086 & 3.8 & \
7.185004502 \\
 1.0 & 5.682681998 & 2.0 & 6.601489711 & 3.0 & 6.999252357 & 4.0 & \
7.221167257 \\
    \hline
    \end{tabular}
    \caption{$\Delta_9$}
    \label{tab:specDel9}
\end{table}

\begin{table}[H]
    \centering
    \begin{tabular}{||cc|cc|cc|cc||}
    \hline
    $g$ & $\Delta$ & $g$ & $\Delta$ & $g$ & $\Delta$ & $g$ & $\Delta$ \\
    \hline\hline
   0.2 & 4.069718669 & 1.2 & 5.836379947 & 2.2 & 6.659463686 & 3.2 & \
7.029223098 \\
 0.4 & 4.285598558 & 1.4 & 6.073247869 & 2.4 & 6.754278507 & 3.4 & \
7.080032639 \\
 0.6 & 4.668075698 & 1.6 & 6.263152913 & 2.6 & 6.836601572 & 3.6 & \
7.125801571 \\
 0.8 & 5.135128838 & 1.8 & 6.418930018 & 2.8 & 6.908748136 & 3.8 & \
7.167243334 \\
 1.0 & 5.532468509 & 2.0 & 6.549077966 & 3.0 & 6.972493691 & 4.0 & \
7.204942895 \\
    \hline
    \end{tabular}
    \caption{$\Delta_{10}$}
    \label{tab:specDel10}
\end{table}

\section{Numerical bounds for OPE coefficients}\label{app:boundsC}
The data obtained with the numerical bootstrap are listed in Tables \ref{tab:C1bounds}-\ref{tab:C3bounds}. The format is $\frac{1}{2}\left(C_{i\text{ lower}} + C_{i\text{ upper}}^2\right) \pm \frac{1}{2}\left(C_{i\text{ upper}} - C_{i\text{ lower}}^2\right)$. The results are obtained as explained in section \ref{sec:numerical}, with the input from the spectrum of the first 10 states, exploiting the two integrated correlator constraints and with $N_{\text{der}} = 140$. These results can be extracted from the \texttt{Mathematica} notebook attached to this paper. 
\begin{table}[H]
   \centering
    \begin{tabular}{||cc|cc||}
    \hline
    $g$ & $C_1^2$ & $g$ & $C_1^2$  \\
    \hline\hline
 0.2 & 0.065679029
\,$\pm$\,6.95
 \,$10^{-7}$ & 2.2 & 0.34963125312
 \,$\pm$\,   1.45
 \, $10^{-9}$ \\
 0.4 & 0.16838882
 \,$\pm$\,1.29 
 $10^{-6}$ & 2.4 & 0.353696925390
 \,$\pm$\,  9.93
 \, $ 10^{-10}$ \\
 0.6 & 0.233041731
 $\pm  $ 4.49
 \,$10^{-7}$ & 2.6 & 0.357157434333
 \,$\pm$\, 7.00
 \,$10^{-10}$ \\
 0.8 & 0.270286735
 \,$\pm$\, 1.32
 \,$10^{-7}$ & 2.8 & 0.360138240651
 \,$\pm$\, 5.08
 \,$10^{-10}$ \\
 1.0 & 0.2940148737
 \,$\pm$\, 4.88
 \,$10^{-8}$ & 3.0 & 0.362732415360
 \,$\pm$\, 3.78
 \,$10^{-10}$\\
 1.2 & 0.3104333079
 \,$\pm$\, 2.16
 \,$10^{-8}$ & 3.2 & 0.365010449531
 \,$\pm$\, 2.87
 \,$10^{-10}$ \\
 1.4 & 0.3224668639
 \,$\pm$\, 1.08
 \,$10^{-8}$ & 3.4 & 0.367026704055
 \,$\pm$\, 2.23
 \,$10^{-10}$ \\
 1.6 & 0.33166329164
 \,$\pm$\, 5.97
 \,$10^{-9}$ & 3.6 & 0.368823769320
 \,$\pm$\, 1.75
 \,$10^{-10}$ \\
 1.8 & 0.33891847883
 \,$\pm$\, 3.53
 \,$10^{-9}$ & 3.8 & 0.370435484280
 \,$\pm$\, 1.40
 \,$10^{-10}$ \\
 2.0 & 0.34478716132
 \,$\pm$\, 2.21
 \,$10^{-9}$ & 4.0 & 0.371889072242
\,$\pm$\, 1.14
\,$10^{-10}$ \\
    \hline
\end{tabular}
\caption{Bounds for the OPE coefficient $C_1^2$}
    \label{tab:C1bounds}
\end{table}

\begin{table}[H]
   \centering
    \begin{tabular}{||cc|cc||}
    \hline
    $g$ & $C_2^2$ & $g$ & $C_2^2$  \\
    \hline\hline
 0.2 & 0.09452
 \,$\pm$\, 7.25
 \, $10^{-3}$  & 2.2 & 0.0311296
 \,$\pm$\,   6.02
 \, $10^{-5}$ \\
 0.4 &  0.06925
 \,$\pm$\, 2.80
 \, $10^{-3}$   & 2.4 & 0.0305818
 \,$\pm$\, 4.90
 \, $10^{-5}$    \\
 0.6 & 0.05246
 \,$\pm$\, 1.47
 \, $10^{-3}$ & 2.6 &  0.0301230
 \,$ \pm  $\,4.06
 \, $10^{-5}$ \\
 0.8 &  0.044285
 \,$\pm$\, 7.18
 \, $10^{-4}$  & 2.8 & 0.0297329
 \,$\pm$\, 3.42
 \, $10^{-5}$  \\
 1.0 &  0.039788
 \,$\pm$\,  4.10
 \, $10^{-4}$ & 3.0 &  0.0293973
 \,$\pm$\, 2.92
 \, $10^{-5}$ \\
 1.2 &  0.036979
 \,$\pm$\,  2.62
 \, $10^{-4}$  & 3.2 &  0.0291054
 \,$\pm$\, 2.52
 \, $10^{-5}$  \\
 1.4 & 0.035063
 \,$\pm$\, 1.79
 \, $10^{-4}$  & 3.4 & 0.0288492
 \,$\pm$\, 2.20
 \, $10^{-5}$ \\
 1.6 &  0.033675
 \,$\pm$\, 1.30
 \, $10^{-4}$  & 3.6 &  0.0286224
 \,$\pm$\, 1.94
 \, $10^{-5}$ \\
 1.8 & 0.0326214
 \,$\pm$\, 9.75
 \, $10^{-5}$ & 3.8 &  0.0284203
 \,$\pm$\, 1.73
 \, $10^{-5}$ \\
 2.0 & 0.0317952
 \,$\pm$\, 7.56
 \, $10^{-5}$ & 4.0 & 0.0282390
 \,$\pm$\, 1.55
 \, $10^{-5}$ \\
    \hline
\end{tabular}
\caption{Bounds for the OPE coefficient $C_2^2$}
    \label{tab:C2bounds}
\end{table}

\begin{table}[H]
   \centering
    \begin{tabular}{||cc|cc||}
    \hline
    $g$ & $C_3^2$ & $g$ & $C_3^2$  \\
    \hline\hline
 0.2 & 0.1101
 \,$\pm$\, 1.27
 \, $10^{-2}$   & 2.2 &  0.1361104
 \,$\pm$\, 6.60
 \, $10^{-5}$\\
 0.4 & 0.13196
 \,$\pm$\,  7.16
  \, $10^{-3}$& 2.4 & 0.1349397
  \,$\pm$\,   5.30
  \, $10^{-5}$\\
 0.6 & 0.14546
 \,$\pm$\,2.99
  \, $10^{-3}$ & 2.6 & 0.1339028
  \,$\pm$\, 4.34
   \, $10^{-5}$\\
 0.8 & 0.14798
 \,$\pm$\,  1.17
  \, $10^{-3}$& 2.8 &  0.1329800
  \,$\pm$\,  3.63
   \, $10^{-5}$\\
 1.0 &  0.146757
 \,$\pm$\, 5.82
 \, $10^{-4}$ & 3.0 & 0.1321546
 \,$\pm$\, 3.07
  \, $10^{-5}$\\
 1.2 &  0.144696
 \,$\pm$\, 3.40
  \, $10^{-4}$& 3.2 & 0.1314126
  \,$\pm$\, 2.64
   \, $10^{-5}$\\
 1.4 &  0.142594
 \,$\pm$\, 2.20
  \, $10^{-4}$ & 3.4 &  0.1307425
  \,$\pm$\, 2.30 
   \, $10^{-5}$\\
 1.6 & 0.140664
 \,$\pm$\, 1.52
  \, $10^{-4}$ & 3.6 & 0.1301347
  \,$\pm$\, 2.02
   \, $10^{-5}$\\
 1.8 & 0.138948
 \,$\pm$\, 1.11
  \, $10^{-4}$ & 3.8 &
  0.1295811
  \,$\pm$\, 1.79
   \, $10^{-5}$\\
 2.0 & 0.1374382
 \,$\pm$\, 8.44
\, $10^{-5}$ & 4.0 & 0.1290748
\,$\pm$\,  1.60
   \, $10^{-5}$ \\
    \hline
\end{tabular}
\caption{Bounds for the OPE coefficient $C_3^2$}
    \label{tab:C3bounds}
\end{table}

\section{Fixing $C^2_\text{BPS}$ from integrability}

\label{app:CBPS}\label{sec:CBPS}

In this appendix we show that $C_\text{BPS}^2$, which was originally computed in~\cite{Liendo:2018ukf} by comparison with a localisation result~\cite{Erickson:2000af,Drukker:2000rr,Pestun:2009nn,Giombi:2018qox}, can also be obtained solely by matching with observables accessible with integrability. The argument we present here is an anticipation of the techniques we use in \cite{upcomingAJMNderivation} to prove the integrated correlator constraints.

We consider the following setup of an infinite-straight MWL with two insertions of $\Phi_\perp^2$, at points $t_1$ and $t_2$. For $t > t_2$ and $t<t_1$, the scalar coupled to the line is $\Phi_{||}$. Between these points, {\it i.e.}~$t_1<t<t_2$, we couple to the rotated scalar $\Phi_{||}\cos\theta + \Phi_\perp^1\sin\theta$. Using the notation~\eqref{eqn:thetaphinotation}, we consider the operator
\begin{align}\label{eqn:insert2phiperp2}
    {\cal W}_{\theta} (t_1,t_2) \equiv \operatorname{Tr}\left[ W_{-\infty}^{t_1}(0,0)\Phi_{\perp}^2(t_1)
    W_{t_1}^{t_2}(0,\theta)\Phi_{\perp}^2(t_2) W_{t_2}^{+\infty}(0,0)\right]
    \;,
\end{align}
where we remind that
\begin{align*}
    {W}_{t_1}^{t_2}(0,\theta) &= \operatorname{P}\exp\int_{t_1}^{t_2}dt\bigg[i\, A_\mu \dot{x}^{\mu}(t) + (\Phi_{||}\cos\theta + \Phi^1_\perp\sin\theta)\,|\dot{x}(t)|\bigg]\;,\\
    x(t) &= \big(t,0,0,0\big)\; .
\end{align*}
One can view the expectation value of ${\cal W}_\theta$ as a 2-point function in the defect CFT. 
At $\theta = 0$, we have
\begin{align}
    {\cal W}_{\theta = 0}(t_1,t_2) =  \langle \langle \Phi_{\perp}^2(t_1) \Phi_{\perp}^2(t_2) \rangle\rangle = \frac{2\mathbb{B}}{t_{12}^2}\;,
\end{align}
where $t_{12} \equiv (t_1 - t_2)$.
The normalisation of this correlator in terms of the Bremsstrahlung function is a general result for the 2-point functions of the line deformation operators in the displacement multiplet~\cite{Correa:2012at}. 
At general $\theta$, on the other hand, we can view this as the correlator of two ``defect-changing'' operators (which live outside the 1D CFT):
\begin{align}
    \langle {\cal W}_\theta(t_1,t_2) \rangle = \frac{C(g,\theta)}{(t_1 - t_2)^{2\,\Gamma_1(g,\theta)}}\;,
\end{align}
where $\Gamma_1$ is a generalisation of the cusp anomalous dimension computed by integrability techniques in~\cite{Correa:2012hh,Drukker:2012de,Gromov:2015dfa}. For small $\theta$, it admits the expansion
\begin{align}
    \Gamma_1(g, \theta) = 1 + \sin^2\theta\,\mathbb{B}_1(g) + {\rm O}(\sin^4\theta)\;.
\end{align}
where $\mathbb{B}_1$ is known exactly from integrability~\cite{Gromov:2012eu}
\begin{align}
   \mathbb{B}_1 = \frac{g^2 \left(-2 I_3(4 g \pi ){}^2+I_1(4 g \pi ) I_3(4 g \pi )+I_1(4 g \pi ) I_5(4 g
   \pi )\right)}{I_1(4 g \pi ){}^2-I_1(4 g \pi ) I_3(4 g \pi )}\;.
\end{align}
Normalising with respect to the $\theta = 0$ case, taking log, differentiating with respect to $t_1$ and $t_2$, and expanding in small $\theta$, we get
\begin{align}\label{eqn:2ptB1}
   \partial_{t_1}\partial_{t_2}\log \left[\frac{\langle{\cal W}_\theta(t_1,t_2)\rangle}{\langle{\cal W}_{\theta = 0}(t_1,t_2)\rangle}\right]  
   = 
   - \frac{2 \mathbb{B}_1}{t_{12}^2}\sin^2\theta + {\rm O}(\sin^4\theta)
   \;.
\end{align}
We will use this equation to compute $C^2_{BPS}$. The strategy is to relate the
 LHS of~\eqref{eqn:2ptB1}, at leading order at small $\sin^2\theta$, to an integrated correlation function in the 1D CFT. As we now show, for the configuration under study the relevant contribution from this integral is expressed in terms of $C^2_{BPS}$.

First, we notice that, by differentiating the explicit $\theta$ dependence in the path-ordered exponentials in~\eqref{eqn:insert2phiperp2}, one can write 
\begin{multline}
    \log \left[\frac{\langle{\cal W}_\theta(t_1,t_2)\rangle}{\langle{\cal W}_{\theta = 0}(t_1, t_2)\rangle}\right]
    \\= \sin^2 \theta \bigg[ \int_{t_1<s_1<s_2<t_2} ds_1\,ds_2  \frac{\langle\langle\Phi_{\perp}^2(t_1)\Phi_{\perp}^1(s_1)\Phi_{\perp}^1(s_2)\Phi_{\perp}^2(t_2)\rangle\rangle}{\langle{\cal W}_{\theta = 0}(t_1, t_2)\rangle}
    + \texttt{counterterms} \bigg]
    + {\rm O}(\sin^4\theta)
    \;,
\end{multline}
Here the counterterms are contributions from integrated 3-point functions\footnote{Full details on the form of such counterterms in similar calculations will be provided in~\cite{upcomingAJMNderivation}. Here we do not elaborate on them since they drop out from the final result. }. 
Next, we differentiate with respect to $t_1$ and $t_2$. This operation kills all the $t_1$- and $t_2$-independent terms. In particular, this means that the {counterterm} contributions drop out. We are left only with the normalised integrated 4-point function. Plugging this into the LHS of~\eqref{eqn:2ptB1}, we get
\begin{align}\label{eqn:int4pt2phperp2}
   \partial_{t_1}\partial_{t_2} \bigg[ \int_{t_1<s_1<s_2<t_2} ds_1\,ds_2  \frac{\langle\langle\Phi_{\perp}^2(t_1)\Phi_{\perp}^1(s_1)\Phi_{\perp}^1(s_2)\Phi_{\perp}^2(t_2)\rangle\rangle}{\langle{\cal W}_{\theta = 0}(t_1, t_2)\rangle} \bigg]
   =
   - \frac{2 \mathbb{B}_1}{t_{12}^2}
   \;.
\end{align}
The 4-point function takes the form (cf. appendix~\ref{app:covariant})
\beq\label{eqn:4ptG3}
\langle\langle\Phi_{\perp}^2(t_1)\Phi_{\perp}^1(s_1)\Phi_{\perp}^1(s_2)\Phi_{\perp}^2(t_2)\rangle\rangle = \frac{4 \mathbb{B}^2}{(t_1 - s_1)^2 (s_2 - t_2)^2 } \; G_3(x) , \;\;\; x = \frac{(t_1 - s_1) (s_2 - t_2)}{(t_1 - s_2) (s_1 - t_2)} ,
\eeq
where $G_3(x)$ is defined in~\eqref{eq:G3}, and the normalisation factor in terms of the Bremsstrahlung function is motivated, as for the case of the 2-point function, by the results of~\cite{Correa:2012at}.

After plugging~\eqref{eqn:4ptG3} into~\eqref{eqn:int4pt2phperp2}, we should compute the double integral. The integral is log-divergent but such divergences will cancel after taking the derivatives. We will use a point-splitting regularisation where the integration region is restricted to $|s_1-s_2|>\delta$, $|s_i - t_j|>\delta$. 

It is convenient to split $G_3(x)$ into $G_3(x) = G_{3,\text{weak}}^{(0)}(x) + \delta G_3(x)$, where $G_{3,\text{weak}}^{(0)}(x)$ is the tree level contribution. For the tree level part, it is simplest to evaluate the integral directly and then take the derivatives in $t_i$. For the $\delta G_3(x)$ part, one can trade one integration variable (e.g. $s_2$) for the cross ratio $x$, and do the remaining integral explicitly. One has in particular, 
\begin{multline}
\int_{t_1+\delta <s_1<s_2-\delta<t_2-2 \delta} d s_1 d s_2 \frac{\delta G_3(x) }{(t_1-s_1)^2\; (s_2-t_2)^2}  \\ = -\frac{1}{t_{12}^2} \log\left(\frac{\delta^2}{t_{12}^2}\right)\; \int_0^1\frac{\delta G_3(x) }{x^2} dx + \frac{1}{t_{12}^2} \times\left( \texttt{$t_i$-independent} \right) , 
\end{multline}
where the omitted terms will drop out after dividing by $\langle W_{\theta=0}(t_1,t_2) \rangle$ and differentiating w.r.t. the endpoints. Subtracting the tree level part is crucial to make the $x$-integral above convergent. 

Including this subtraction, and taking the $t_i$-derivatives, once the dust settles from~\eqref{eqn:int4pt2phperp2} we find:
\beq
- \mathbb{B}_1 = 2\mathbb{B}\; \left(\int_0^1 \frac{\delta G_3(x) }{x^2} dx  - 1 \right) = -  2\mathbb{B} \; \left(-1 + \int_0^1 \partial_x\left( \frac{\delta f(x) }{x}\right) dx \right)\;,
\eeq
where we used~\eqref{eq:G3} in the second equality. Since we have a total derivative under the integral, using the limits of $\delta f(x)$ in \eqref{eq:limits2} we get
\beq
\frac{\mathbb{B}_1}{\mathbb{B}} = \mathbb{F} - 2  ,
\eeq
and since it is known~\cite{Liendo:2018ukf} how the constant $\mathbb{F}$ is related by superconformal symmetry to the OPE coefficient: $C^2_{BPS} = \mathbb{F} - 1$, we have proved from integrability that
\beq
C^2_{BPS} = 1 + \frac{\mathbb{B}_1}{\mathbb{B}},
\eeq
which reproduces the expression given in the main text.

\bibliographystyle{JHEP.bst}
\bibliography{references}

\providecommand{\href}[2]{#2}\begingroup\raggedright\begin{thebibliography}{100}

\bibitem{Lipatov:1993yb}
L.~N. Lipatov, {\it {Asymptotic behavior of multicolor QCD at high energies in
  connection with exactly solvable spin models}},  {\em JETP Lett.} {\bf 59}
  (1994) 596--599, [\href{http://arxiv.org/abs/hep-th/9311037}{{\tt
  hep-th/9311037}}]. [Pisma Zh. Eksp. Teor. Fiz.59,571(1994)].

\bibitem{Faddeev:1994zg}
L.~D. Faddeev and G.~P. Korchemsky, {\it {High-energy QCD as a completely
  integrable model}},  {\em Phys. Lett.} {\bf B342} (1995) 311--322,
  [\href{http://arxiv.org/abs/hep-th/9404173}{{\tt hep-th/9404173}}].

\bibitem{Minahan:2002ve}
J.~A. Minahan and K.~Zarembo, {\it {The Bethe ansatz for N=4 superYang-Mills}},
   {\em JHEP} {\bf 03} (2003) 013,
  [\href{http://arxiv.org/abs/hep-th/0212208}{{\tt hep-th/0212208}}].

\bibitem{Gromov:2013pga}
N.~Gromov, V.~Kazakov, S.~Leurent, and D.~Volin, {\it {Quantum Spectral Curve
  for Planar $\mathcal{N} = 4$ Super-Yang-Mills Theory}},  {\em Phys. Rev.
  Lett.} {\bf 112} (2014), no.~1 011602,
  [\href{http://arxiv.org/abs/1305.1939}{{\tt arXiv:1305.1939}}].

\bibitem{Gromov:2014caa}
N.~Gromov, V.~Kazakov, S.~Leurent, and D.~Volin, {\it {Quantum spectral curve
  for arbitrary state/operator in AdS$_{5}$/CFT$_{4}$}},  {\em JHEP} {\bf 09}
  (2015) 187, [\href{http://arxiv.org/abs/1405.4857}{{\tt arXiv:1405.4857}}].

\bibitem{Basso:2015zoa}
B.~Basso, S.~Komatsu, and P.~Vieira, ``{Structure Constants and Integrable
  Bootstrap in Planar N=4 SYM Theory}.'' 5, 2015.

\bibitem{Fleury:2016ykk}
T.~Fleury and S.~Komatsu, {\it {Hexagonalization of Correlation Functions}},
  {\em JHEP} {\bf 01} (2017) 130, [\href{http://arxiv.org/abs/1611.05577}{{\tt
  arXiv:1611.05577}}].

\bibitem{Eden:2016xvg}
B.~Eden and A.~Sfondrini, {\it {Tessellating cushions: four-point functions in
  $\mathcal{N} $ = 4 SYM}},  {\em JHEP} {\bf 10} (2017) 098,
  [\href{http://arxiv.org/abs/1611.05436}{{\tt arXiv:1611.05436}}].

\bibitem{Bargheer:2017nne}
T.~Bargheer, J.~Caetano, T.~Fleury, S.~Komatsu, and P.~Vieira, {\it {Handling
  Handles: Nonplanar Integrability in $\mathcal{N}=4$ Supersymmetric Yang-Mills
  Theory}},  {\em Phys. Rev. Lett.} {\bf 121} (2018), no.~23 231602,
  [\href{http://arxiv.org/abs/1711.05326}{{\tt arXiv:1711.05326}}].

\bibitem{Coronado:2018cxj}
F.~Coronado, {\it {Bootstrapping the Simplest Correlator in Planar $\mathcal N
  = 4$ Supersymmetric Yang-Mills Theory to All Loops}},  {\em Phys. Rev. Lett.}
  {\bf 124} (2020), no.~17 171601, [\href{http://arxiv.org/abs/1811.03282}{{\tt
  arXiv:1811.03282}}].

\bibitem{Kostov:2019stn}
I.~Kostov, V.~B. Petkova, and D.~Serban, {\it {Determinant Formula for the
  Octagon Form Factor in $N$=4 Supersymmetric Yang-Mills Theory}},  {\em Phys.
  Rev. Lett.} {\bf 122} (2019), no.~23 231601,
  [\href{http://arxiv.org/abs/1903.05038}{{\tt arXiv:1903.05038}}].

\bibitem{Bargheer:2019kxb}
T.~Bargheer, F.~Coronado, and P.~Vieira, {\it {Octagons I: Combinatorics and
  Non-Planar Resummations}},  {\em JHEP} {\bf 08} (2019) 162,
  [\href{http://arxiv.org/abs/1904.00965}{{\tt arXiv:1904.00965}}].

\bibitem{Bargheer:2019exp}
T.~Bargheer, F.~Coronado, and P.~Vieira, {\it {Octagons II: Strong Coupling}},
  \href{http://arxiv.org/abs/1909.04077}{{\tt arXiv:1909.04077}}.

\bibitem{Jiang:2015lda}
Y.~Jiang, S.~Komatsu, I.~Kostov, and D.~Serban, {\it {The hexagon in the
  mirror: the three-point function in the SoV representation}},  {\em J. Phys.
  A} {\bf 49} (2016), no.~17 174007,
  [\href{http://arxiv.org/abs/1506.09088}{{\tt arXiv:1506.09088}}].

\bibitem{Gromov:2016itr}
N.~Gromov, F.~Levkovich-Maslyuk, and G.~Sizov, {\it {New Construction of
  Eigenstates and Separation of Variables for SU(N) Quantum Spin Chains}},
  {\em JHEP} {\bf 09} (2017) 111, [\href{http://arxiv.org/abs/1610.08032}{{\tt
  arXiv:1610.08032}}].

\bibitem{Cavaglia:2018lxi}
A.~Cavagli{\`a}, N.~Gromov, and F.~Levkovich-Maslyuk, {\it {Quantum spectral
  curve and structure constants in $ \mathcal{N}=4 $ SYM: cusps in the ladder
  limit}},  {\em JHEP} {\bf 10} (2018) 060,
  [\href{http://arxiv.org/abs/1802.04237}{{\tt arXiv:1802.04237}}].

\bibitem{Giombi:2018hsx}
S.~Giombi and S.~Komatsu, {\it {More Exact Results in the Wilson Loop Defect
  CFT: Bulk-Defect OPE, Nonplanar Corrections and Quantum Spectral Curve}},
  {\em J. Phys.} {\bf A52} (2019), no.~12 125401,
  [\href{http://arxiv.org/abs/1811.02369}{{\tt arXiv:1811.02369}}].

\bibitem{Cavaglia:2019pow}
A.~Cavagli\`a, N.~Gromov, and F.~Levkovich-Maslyuk, {\it {Separation of
  variables and scalar products at any rank}},  {\em JHEP} {\bf 09} (2019) 052,
  [\href{http://arxiv.org/abs/1907.03788}{{\tt arXiv:1907.03788}}].

\bibitem{Gromov:2019wmz}
N.~Gromov, F.~Levkovich-Maslyuk, P.~Ryan, and D.~Volin, {\it {Dual Separated
  Variables and Scalar Products}},  {\em Phys. Lett. B} {\bf 806} (2020)
  135494, [\href{http://arxiv.org/abs/1910.13442}{{\tt arXiv:1910.13442}}].

\bibitem{Cavaglia:2021mft}
A.~Cavagli{\`a}, N.~Gromov, and F.~Levkovich-Maslyuk, {\it {Separation of
  variables in AdS/CFT: functional approach for the fishnet CFT}},  {\em JHEP}
  {\bf 06} (2021) 131, [\href{http://arxiv.org/abs/2103.15800}{{\tt
  arXiv:2103.15800}}].

\bibitem{Gromov:2022waj}
N.~Gromov, N.~Primi, and P.~Ryan, {\it {Form-factors and complete basis of
  observables via separation of variables for higher rank spin chains}},
  \href{http://arxiv.org/abs/2202.01591}{{\tt arXiv:2202.01591}}.

\bibitem{Rattazzi:2008pe}
R.~Rattazzi, V.~S. Rychkov, E.~Tonni, and A.~Vichi, {\it {Bounding scalar
  operator dimensions in 4D CFT}},  {\em JHEP} {\bf 12} (2008) 031,
  [\href{http://arxiv.org/abs/0807.0004}{{\tt arXiv:0807.0004}}].

\bibitem{El-Showk:2012cjh}
S.~El-Showk, M.~F. Paulos, D.~Poland, S.~Rychkov, D.~Simmons-Duffin, and
  A.~Vichi, {\it {Solving the 3D Ising Model with the Conformal Bootstrap}},
  {\em Phys. Rev. D} {\bf 86} (2012) 025022,
  [\href{http://arxiv.org/abs/1203.6064}{{\tt arXiv:1203.6064}}].

\bibitem{Cavaglia:2021bnz}
A.~Cavagli\`a, N.~Gromov, J.~Julius, and M.~Preti, {\it {Integrability and
  conformal bootstrap: One dimensional defect conformal field theory}},  {\em
  Phys. Rev. D} {\bf 105} (2022), no.~2 L021902,
  [\href{http://arxiv.org/abs/2107.08510}{{\tt arXiv:2107.08510}}].

\bibitem{Picco:2016ilr}
M.~Picco, S.~Ribault, and R.~Santachiara, {\it {A conformal bootstrap approach
  to critical percolation in two dimensions}},  {\em SciPost Phys.} {\bf 1}
  (2016), no.~1 009, [\href{http://arxiv.org/abs/1607.07224}{{\tt
  arXiv:1607.07224}}].

\bibitem{He:2020rfk}
Y.~He, J.~L. Jacobsen, and H.~Saleur, {\it {Geometrical four-point functions in
  the two-dimensional critical $Q$-state Potts model: The interchiral conformal
  bootstrap}},  {\em JHEP} {\bf 12} (2020) 019,
  [\href{http://arxiv.org/abs/2005.07258}{{\tt arXiv:2005.07258}}].

\bibitem{Nakayama:2016cim}
Y.~Nakayama, {\it {Bootstrapping critical Ising model on three-dimensional real
  projective space}},  {\em Phys. Rev. Lett.} {\bf 116} (2016), no.~14 141602,
  [\href{http://arxiv.org/abs/1601.06851}{{\tt arXiv:1601.06851}}].

\bibitem{Gliozzi:2015qsa}
F.~Gliozzi, P.~Liendo, M.~Meineri, and A.~Rago, {\it {Boundary and Interface
  CFTs from the Conformal Bootstrap}},  {\em JHEP} {\bf 05} (2015) 036,
  [\href{http://arxiv.org/abs/1502.07217}{{\tt arXiv:1502.07217}}].

\bibitem{Gliozzi:2016cmg}
F.~Gliozzi, {\it {Truncatable bootstrap equations in algebraic form and
  critical surface exponents}},  {\em JHEP} {\bf 10} (2016) 037,
  [\href{http://arxiv.org/abs/1605.04175}{{\tt arXiv:1605.04175}}].

\bibitem{Ferrero:2021bsb}
P.~Ferrero and C.~Meneghelli, {\it {Bootstrapping the half-BPS line defect CFT
  in N=4 supersymmetric Yang-Mills theory at strong coupling}},  {\em Phys.
  Rev. D} {\bf 104} (2021), no.~8 L081703,
  [\href{http://arxiv.org/abs/2103.10440}{{\tt arXiv:2103.10440}}].

\bibitem{Drukker:2006xg}
N.~Drukker and S.~Kawamoto, {\it {Small deformations of supersymmetric Wilson
  loops and open spin-chains}},  {\em JHEP} {\bf 07} (2006) 024,
  [\href{http://arxiv.org/abs/hep-th/0604124}{{\tt hep-th/0604124}}].

\bibitem{Correa:2012hh}
D.~Correa, J.~Maldacena, and A.~Sever, {\it {The quark anti-quark potential and
  the cusp anomalous dimension from a TBA equation}},  {\em JHEP} {\bf 08}
  (2012) 134, [\href{http://arxiv.org/abs/1203.1913}{{\tt arXiv:1203.1913}}].

\bibitem{Drukker:2012de}
N.~Drukker, {\it {Integrable Wilson loops}},  {\em JHEP} {\bf 10} (2013) 135,
  [\href{http://arxiv.org/abs/1203.1617}{{\tt arXiv:1203.1617}}].

\bibitem{Giombi:2017cqn}
S.~Giombi, R.~Roiban, and A.~A. Tseytlin, {\it {Half-BPS Wilson loop and
  AdS$_2$/CFT$_1$}},  {\em Nucl. Phys.} {\bf B922} (2017) 499--527,
  [\href{http://arxiv.org/abs/1706.00756}{{\tt arXiv:1706.00756}}].

\bibitem{Kim:2017sju}
M.~Kim, N.~Kiryu, S.~Komatsu, and T.~Nishimura, {\it {Structure Constants of
  Defect Changing Operators on the 1/2 BPS Wilson Loop}},  {\em JHEP} {\bf 12}
  (2017) 055, [\href{http://arxiv.org/abs/1710.07325}{{\tt arXiv:1710.07325}}].

\bibitem{Cooke:2017qgm}
M.~Cooke, A.~Dekel, and N.~Drukker, {\it {The Wilson loop CFT: Insertion
  dimensions and structure constants from wavy lines}},  {\em J. Phys.} {\bf
  A50} (2017), no.~33 335401, [\href{http://arxiv.org/abs/1703.03812}{{\tt
  arXiv:1703.03812}}].

\bibitem{Liendo:2018ukf}
P.~Liendo, C.~Meneghelli, and V.~Mitev, {\it {Bootstrapping the half-BPS line
  defect}},  {\em JHEP} {\bf 10} (2018) 077,
  [\href{http://arxiv.org/abs/1806.01862}{{\tt arXiv:1806.01862}}].

\bibitem{Giombi:2018qox}
S.~Giombi and S.~Komatsu, {\it {Exact Correlators on the Wilson Loop in
  $\mathcal{N}=4$ SYM: Localization, Defect CFT, and Integrability}},  {\em
  JHEP} {\bf 05} (2018) 109, [\href{http://arxiv.org/abs/1802.05201}{{\tt
  arXiv:1802.05201}}]. [Erratum: JHEP 11, 123 (2018)].

\bibitem{Giombi:2020amn}
S.~Giombi, J.~Jiang, and S.~Komatsu, {\it {Giant Wilson loops and
  AdS$_{2}$/dCFT$_{1}$}},  {\em JHEP} {\bf 11} (2020) 064,
  [\href{http://arxiv.org/abs/2005.08890}{{\tt arXiv:2005.08890}}].

\bibitem{Grabner:2020nis}
D.~Grabner, N.~Gromov, and J.~Julius, {\it {Excited States of One-Dimensional
  Defect CFTs from the Quantum Spectral Curve}},  {\em JHEP} {\bf 07} (2020)
  042, [\href{http://arxiv.org/abs/2001.11039}{{\tt arXiv:2001.11039}}].

\bibitem{Giombi:2021zfb}
S.~Giombi, S.~Komatsu, and B.~Offertaler, {\it {Large charges on the Wilson
  loop in $ \mathcal{N} $ = 4 SYM: matrix model and classical string}},  {\em
  JHEP} {\bf 03} (2022) 020, [\href{http://arxiv.org/abs/2110.13126}{{\tt
  arXiv:2110.13126}}].

\bibitem{Barrat:2021yvp}
J.~Barrat, A.~Gimenez-Grau, and P.~Liendo, {\it {Bootstrapping holographic
  defect correlators in $\mathcal{N}=4$ super Yang-Mills}},
  \href{http://arxiv.org/abs/2108.13432}{{\tt arXiv:2108.13432}}.

\bibitem{Barrat:2021tpn}
J.~Barrat, P.~Liendo, G.~Peveri, and J.~Plefka, {\it {Multipoint correlators on
  the supersymmetric Wilson line defect CFT}},
  \href{http://arxiv.org/abs/2112.10780}{{\tt arXiv:2112.10780}}.

\bibitem{Giombi:2022anm}
S.~Giombi, S.~Komatsu, and B.~Offertaler, {\it {Large Charges on the Wilson
  Line in $\mathcal{N}=4$ SYM: II. Quantum Fluctuations, OPE, and Spectral
  Curve}},  \href{http://arxiv.org/abs/2202.07627}{{\tt arXiv:2202.07627}}.

\bibitem{Polchinski:2011im}
J.~Polchinski and J.~Sully, {\it {Wilson Loop Renormalization Group Flows}},
  {\em JHEP} {\bf 10} (2011) 059, [\href{http://arxiv.org/abs/1104.5077}{{\tt
  arXiv:1104.5077}}].

\bibitem{Beccaria:2017rbe}
M.~Beccaria, S.~Giombi, and A.~Tseytlin, {\it {Non-supersymmetric Wilson loop
  in $ \mathcal{N} $ = 4 SYM and defect 1d CFT}},  {\em JHEP} {\bf 03} (2018)
  131, [\href{http://arxiv.org/abs/1712.06874}{{\tt arXiv:1712.06874}}].

\bibitem{Gimenez-Grau:2019hez}
A.~Gimenez-Grau and P.~Liendo, {\it {Bootstrapping line defects in
  $\mathcal{N}=2$ theories}},  {\em JHEP} {\bf 03} (2020) 121,
  [\href{http://arxiv.org/abs/1907.04345}{{\tt arXiv:1907.04345}}].

\bibitem{Julius:2021uka}
J.~Julius, {\em {Modern techniques for solvable models}}.
\newblock PhD thesis, King's Coll. London, 2021.

\bibitem{spec1DCFT}
N.~Gromov, J.~Julius, and N.~Sokolova. \textit{To appear}.

\bibitem{upcomingAJMNderivation}
A.~Cavagli{\`a}, N.~Gromov, J.~Julius, and M.~Preti. {\it To appear}.

\bibitem{Binder:2019jwn}
D.~J. Binder, S.~M. Chester, S.~S. Pufu, and Y.~Wang, {\it {$ \mathcal{N} $ = 4
  Super-Yang-Mills correlators at strong coupling from string theory and
  localization}},  {\em JHEP} {\bf 12} (2019) 119,
  [\href{http://arxiv.org/abs/1902.06263}{{\tt arXiv:1902.06263}}].

\bibitem{Chester:2021aun}
S.~M. Chester, R.~Dempsey, and S.~S. Pufu, {\it {Bootstrapping $\mathcal{N}=4$
  super-Yang-Mills on the conformal manifold}},
  \href{http://arxiv.org/abs/2111.07989}{{\tt arXiv:2111.07989}}.

\bibitem{Polyakov:1980ca}
A.~M. Polyakov, {\it {Gauge Fields as Rings of Glue}},  {\em Nucl. Phys. B}
  {\bf 164} (1980) 171--188.

\bibitem{Korchemsky:1987wg}
G.~P. Korchemsky and A.~V. Radyushkin, {\it {Renormalization of the Wilson
  Loops Beyond the Leading Order}},  {\em Nucl. Phys. B} {\bf 283} (1987)
  342--364.

\bibitem{Gromov:2015dfa}
N.~Gromov and F.~Levkovich-Maslyuk, {\it {Quantum Spectral Curve for a cusped
  Wilson line in $ \mathcal{N}=4 $ SYM}},  {\em JHEP} {\bf 04} (2016) 134,
  [\href{http://arxiv.org/abs/1510.02098}{{\tt arXiv:1510.02098}}].

\bibitem{Polyakov:2000ti}
A.~M. Polyakov and V.~S. Rychkov, {\it {Gauge field strings duality and the
  loop equation}},  {\em Nucl. Phys. B} {\bf 581} (2000) 116--134,
  [\href{http://arxiv.org/abs/hep-th/0002106}{{\tt hep-th/0002106}}].

\bibitem{Semenoff:2004qr}
G.~W. Semenoff and D.~Young, {\it {Wavy Wilson line and AdS / CFT}},  {\em Int.
  J. Mod. Phys. A} {\bf 20} (2005) 2833--2846,
  [\href{http://arxiv.org/abs/hep-th/0405288}{{\tt hep-th/0405288}}].

\bibitem{Zarembo:2016bbk}
K.~Zarembo, {\it {Localization and AdS/CFT Correspondence}},  {\em J. Phys. A}
  {\bf 50} (2017), no.~44 443011, [\href{http://arxiv.org/abs/1608.02963}{{\tt
  arXiv:1608.02963}}].

\bibitem{Drukker:1999zq}
N.~Drukker, D.~J. Gross, and H.~Ooguri, {\it {Wilson loops and minimal
  surfaces}},  {\em Phys. Rev. D} {\bf 60} (1999) 125006,
  [\href{http://arxiv.org/abs/hep-th/9904191}{{\tt hep-th/9904191}}].

\bibitem{Erickson:2000af}
J.~K. Erickson, G.~W. Semenoff, and K.~Zarembo, {\it {Wilson loops in N=4
  supersymmetric Yang-Mills theory}},  {\em Nucl. Phys.} {\bf B582} (2000)
  155--175, [\href{http://arxiv.org/abs/hep-th/0003055}{{\tt hep-th/0003055}}].

\bibitem{Zarembo:2002an}
K.~Zarembo, {\it {Supersymmetric Wilson loops}},  {\em Nucl. Phys. B} {\bf 643}
  (2002) 157--171, [\href{http://arxiv.org/abs/hep-th/0205160}{{\tt
  hep-th/0205160}}].

\bibitem{Drukker:2011za}
N.~Drukker and V.~Forini, {\it {Generalized quark-antiquark potential at weak
  and strong coupling}},  {\em JHEP} {\bf 06} (2011) 131,
  [\href{http://arxiv.org/abs/1105.5144}{{\tt arXiv:1105.5144}}].

\bibitem{Correa:2012at}
D.~Correa, J.~Henn, J.~Maldacena, and A.~Sever, {\it {An exact formula for the
  radiation of a moving quark in N=4 super Yang Mills}},  {\em JHEP} {\bf 06}
  (2012) 048, [\href{http://arxiv.org/abs/1202.4455}{{\tt arXiv:1202.4455}}].

\bibitem{Fiol:2012sg}
B.~Fiol, B.~Garolera, and A.~Lewkowycz, {\it {Exact results for static and
  radiative fields of a quark in N=4 super Yang-Mills}},  {\em JHEP} {\bf 05}
  (2012) 093, [\href{http://arxiv.org/abs/1202.5292}{{\tt arXiv:1202.5292}}].

\bibitem{Drukker:2000rr}
N.~Drukker and D.~J. Gross, {\it {An Exact prediction of N=4 SUSYM theory for
  string theory}},  {\em J. Math. Phys.} {\bf 42} (2001) 2896--2914,
  [\href{http://arxiv.org/abs/hep-th/0010274}{{\tt hep-th/0010274}}].

\bibitem{Drukker:2006ga}
N.~Drukker, {\it {1/4 BPS circular loops, unstable world-sheet instantons and
  the matrix model}},  {\em JHEP} {\bf 09} (2006) 004,
  [\href{http://arxiv.org/abs/hep-th/0605151}{{\tt hep-th/0605151}}].

\bibitem{Pestun:2009nn}
V.~Pestun, {\it {Localization of the four-dimensional N=4 SYM to a two-sphere
  and 1/8 BPS Wilson loops}},  {\em JHEP} {\bf 12} (2012) 067,
  [\href{http://arxiv.org/abs/0906.0638}{{\tt arXiv:0906.0638}}].

\bibitem{Gromov:2012eu}
N.~Gromov and A.~Sever, {\it {Analytic Solution of Bremsstrahlung TBA}},  {\em
  JHEP} {\bf 11} (2012) 075, [\href{http://arxiv.org/abs/1207.5489}{{\tt
  arXiv:1207.5489}}].

\bibitem{Gromov:2013qga}
N.~Gromov, F.~Levkovich-Maslyuk, and G.~Sizov, {\it {Analytic Solution of
  Bremsstrahlung TBA II: Turning on the Sphere Angle}},  {\em JHEP} {\bf 10}
  (2013) 036, [\href{http://arxiv.org/abs/1305.1944}{{\tt arXiv:1305.1944}}].

\bibitem{Sizov:2013joa}
G.~Sizov and S.~Valatka, {\it {Algebraic Curve for a Cusped Wilson Line}},
  {\em JHEP} {\bf 05} (2014) 149, [\href{http://arxiv.org/abs/1306.2527}{{\tt
  arXiv:1306.2527}}].

\bibitem{Bonini:2015fng}
M.~Bonini, L.~Griguolo, M.~Preti, and D.~Seminara, {\it {Bremsstrahlung
  function, leading L\"uscher correction at weak coupling and localization}},
  {\em JHEP} {\bf 02} (2016) 172, [\href{http://arxiv.org/abs/1511.05016}{{\tt
  arXiv:1511.05016}}].

\bibitem{Gunaydin:1990ag}
M.~Gunaydin and R.~J. Scalise, {\it {Unitary Lowest Weight Representations of
  the Noncompact Supergroup Osp(2m*/2n)}},  {\em J. Math. Phys.} {\bf 32}
  (1991) 599--606.

\bibitem{Liendo:2016ymz}
P.~Liendo and C.~Meneghelli, {\it {Bootstrap equations for $ \mathcal{N} $ = 4
  SYM with defects}},  {\em JHEP} {\bf 01} (2017) 122,
  [\href{http://arxiv.org/abs/1608.05126}{{\tt arXiv:1608.05126}}].

\bibitem{Marboe:2014gma}
C.~Marboe and D.~Volin, {\it {Quantum spectral curve as a tool for a
  perturbative quantum field theory}},  {\em Nucl. Phys.} {\bf B899} (2015)
  810--847, [\href{http://arxiv.org/abs/1411.4758}{{\tt arXiv:1411.4758}}].

\bibitem{Volin:2008kd}
D.~Volin, {\it {The 2-Loop generalized scaling function from the BES/FRS
  equation}},  \href{http://arxiv.org/abs/0812.4407}{{\tt arXiv:0812.4407}}.

\bibitem{ShotaTalk}
S.~Komatsu, ``Integrability in {{AdS/CFT}}.'' Talk at ICTP-SAIFR Strings 2021.

\bibitem{Kiryu:2018phb}
N.~Kiryu and S.~Komatsu, {\it {Correlation Functions on the Half-BPS Wilson
  Loop: Perturbation and Hexagonalization}},  {\em JHEP} {\bf 02} (2019) 090,
  [\href{http://arxiv.org/abs/1812.04593}{{\tt arXiv:1812.04593}}].

\bibitem{Chester:2019wfx}
S.~M. Chester, ``{Weizmann Lectures on the Numerical Conformal Bootstrap}.'' 7,
  2019.

\bibitem{Poland:2018epd}
D.~Poland, S.~Rychkov, and A.~Vichi, {\it {The Conformal Bootstrap: Theory,
  Numerical Techniques, and Applications}},  {\em Rev. Mod. Phys.} {\bf 91}
  (2019) 015002, [\href{http://arxiv.org/abs/1805.04405}{{\tt
  arXiv:1805.04405}}].

\bibitem{Simmons-Duffin:2015qma}
D.~Simmons-Duffin, {\it {A Semidefinite Program Solver for the Conformal
  Bootstrap}},  {\em JHEP} {\bf 06} (2015) 174,
  [\href{http://arxiv.org/abs/1502.02033}{{\tt arXiv:1502.02033}}].

\bibitem{Landry:2019qug}
W.~Landry and D.~Simmons-Duffin, ``{Scaling the semidefinite program solver
  SDPB}.'' 9, 2019.

\bibitem{Chester:2019ifh}
S.~M. Chester, W.~Landry, J.~Liu, D.~Poland, D.~Simmons-Duffin, N.~Su, and
  A.~Vichi, {\it {Carving out OPE space and precise $O(2)$ model critical
  exponents}},  {\em JHEP} {\bf 06} (2020) 142,
  [\href{http://arxiv.org/abs/1912.03324}{{\tt arXiv:1912.03324}}].

\bibitem{Remiddi:1999ew}
E.~Remiddi and J.~A.~M. Vermaseren, {\it {Harmonic polylogarithms}},  {\em Int.
  J. Mod. Phys. A} {\bf 15} (2000) 725--754,
  [\href{http://arxiv.org/abs/hep-ph/9905237}{{\tt hep-ph/9905237}}].

\bibitem{Maitre:2005uu}
D.~Maitre, {\it {HPL, a mathematica implementation of the harmonic
  polylogarithms}},  {\em Comput. Phys. Commun.} {\bf 174} (2006) 222--240,
  [\href{http://arxiv.org/abs/hep-ph/0507152}{{\tt hep-ph/0507152}}].

\bibitem{Maitre:2007kp}
D.~Maitre, {\it {Extension of HPL to complex arguments}},  {\em Comput. Phys.
  Commun.} {\bf 183} (2012) 846,
  [\href{http://arxiv.org/abs/hep-ph/0703052}{{\tt hep-ph/0703052}}].

\bibitem{Mazac:2018qmi}
D.~Maz\'a\v{c}, {\it {A Crossing-Symmetric OPE Inversion Formula}},  {\em JHEP}
  {\bf 06} (2019) 082, [\href{http://arxiv.org/abs/1812.02254}{{\tt
  arXiv:1812.02254}}].

\bibitem{Dorigoni:2021bvj}
D.~Dorigoni, M.~B. Green, and C.~Wen, {\it {Novel Representation of an
  Integrated Correlator in $\mathcal N$ = 4 Supersymmetric Yang-Mills Theory}},
   {\em Phys. Rev. Lett.} {\bf 126} (2021), no.~16 161601,
  [\href{http://arxiv.org/abs/2102.08305}{{\tt arXiv:2102.08305}}].

\bibitem{Dorigoni:2021guq}
D.~Dorigoni, M.~B. Green, and C.~Wen, {\it {Exact properties of an integrated
  correlator in $ \mathcal{N} $ = 4 SU(N) SYM}},  {\em JHEP} {\bf 05} (2021)
  089, [\href{http://arxiv.org/abs/2102.09537}{{\tt arXiv:2102.09537}}].

\bibitem{Mazac:2016qev}
D.~Mazac, {\it {Analytic bounds and emergence of AdS$_{2}$ physics from the
  conformal bootstrap}},  {\em JHEP} {\bf 04} (2017) 146,
  [\href{http://arxiv.org/abs/1611.10060}{{\tt arXiv:1611.10060}}].

\bibitem{Mazac:2018mdx}
D.~Mazac and M.~F. Paulos, {\it {The analytic functional bootstrap. Part I: 1D
  CFTs and 2D S-matrices}},  {\em JHEP} {\bf 02} (2019) 162,
  [\href{http://arxiv.org/abs/1803.10233}{{\tt arXiv:1803.10233}}].

\bibitem{Mazac:2018ycv}
D.~Mazac and M.~F. Paulos, {\it {The analytic functional bootstrap. Part II.
  Natural bases for the crossing equation}},  {\em JHEP} {\bf 02} (2019) 163,
  [\href{http://arxiv.org/abs/1811.10646}{{\tt arXiv:1811.10646}}].

\bibitem{Paulos:2019fkw}
M.~F. Paulos and B.~Zan, {\it {A functional approach to the numerical conformal
  bootstrap}},  {\em JHEP} {\bf 09} (2020) 006,
  [\href{http://arxiv.org/abs/1904.03193}{{\tt arXiv:1904.03193}}].

\bibitem{Ferrero:2019luz}
P.~Ferrero, K.~Ghosh, A.~Sinha, and A.~Zahed, {\it {Crossing symmetry,
  transcendentality and the Regge behaviour of 1d CFTs}},  {\em JHEP} {\bf 07}
  (2020) 170, [\href{http://arxiv.org/abs/1911.12388}{{\tt arXiv:1911.12388}}].

\bibitem{Bianchi:2021piu}
L.~Bianchi, G.~Bliard, V.~Forini, and G.~Peveri, {\it {Mellin amplitudes for 1d
  CFT}},  {\em JHEP} {\bf 10} (2021) 095,
  [\href{http://arxiv.org/abs/2106.00689}{{\tt arXiv:2106.00689}}].

\bibitem{Kos:2016ysd}
F.~Kos, D.~Poland, D.~Simmons-Duffin, and A.~Vichi, {\it {Precision Islands in
  the Ising and $O(N)$ Models}},  {\em JHEP} {\bf 08} (2016) 036,
  [\href{http://arxiv.org/abs/1603.04436}{{\tt arXiv:1603.04436}}].

\bibitem{Caron-Huot:2017vep}
S.~Caron-Huot, {\it {Analyticity in Spin in Conformal Theories}},  {\em JHEP}
  {\bf 09} (2017) 078, [\href{http://arxiv.org/abs/1703.00278}{{\tt
  arXiv:1703.00278}}].

\bibitem{Alday:2017vkk}
L.~F. Alday and S.~Caron-Huot, {\it {Gravitational S-matrix from CFT dispersion
  relations}},  {\em JHEP} {\bf 12} (2018) 017,
  [\href{http://arxiv.org/abs/1711.02031}{{\tt arXiv:1711.02031}}].

\bibitem{Caron-Huot:2020adz}
S.~Caron-Huot, D.~Mazac, L.~Rastelli, and D.~Simmons-Duffin, {\it {Dispersive
  CFT Sum Rules}},  {\em JHEP} {\bf 05} (2021) 243,
  [\href{http://arxiv.org/abs/2008.04931}{{\tt arXiv:2008.04931}}].

\bibitem{Bissi:2022mrs}
A.~Bissi, A.~Sinha, and X.~Zhou, {\it {Selected Topics in Analytic Conformal
  Bootstrap: A Guided Journey}},  \href{http://arxiv.org/abs/2202.08475}{{\tt
  arXiv:2202.08475}}.

\bibitem{Bissi:2021spj}
A.~Bissi, P.~Dey, and G.~Fardelli, {\it {Two Applications of the Analytic
  Conformal Bootstrap: A Quick Tour Guide}},  {\em Universe} {\bf 7} (2021),
  no.~7 247, [\href{http://arxiv.org/abs/2107.10097}{{\tt arXiv:2107.10097}}].

\bibitem{Aharony:2016dwx}
O.~Aharony, L.~F. Alday, A.~Bissi, and E.~Perlmutter, {\it {Loops in AdS from
  Conformal Field Theory}},  {\em JHEP} {\bf 07} (2017) 036,
  [\href{http://arxiv.org/abs/1612.03891}{{\tt arXiv:1612.03891}}].

\bibitem{Alday:2017xua}
L.~F. Alday and A.~Bissi, {\it {Loop Corrections to Supergravity on $AdS_5
  \times S^5$}},  {\em Phys. Rev. Lett.} {\bf 119} (2017), no.~17 171601,
  [\href{http://arxiv.org/abs/1706.02388}{{\tt arXiv:1706.02388}}].

\bibitem{Bianchi:2017ozk}
L.~Bianchi, L.~Griguolo, M.~Preti, and D.~Seminara, {\it {Wilson lines as
  superconformal defects in ABJM theory: a formula for the emitted radiation}},
   {\em JHEP} {\bf 10} (2017) 050, [\href{http://arxiv.org/abs/1706.06590}{{\tt
  arXiv:1706.06590}}].

\bibitem{Bianchi:2018scb}
L.~Bianchi, M.~Preti, and E.~Vescovi, {\it {Exact Bremsstrahlung functions in
  ABJM theory}},  {\em JHEP} {\bf 07} (2018) 060,
  [\href{http://arxiv.org/abs/1802.07726}{{\tt arXiv:1802.07726}}].

\bibitem{Bianchi:2020hsz}
L.~Bianchi, G.~Bliard, V.~Forini, L.~Griguolo, and D.~Seminara, {\it {Analytic
  bootstrap and Witten diagrams for the ABJM Wilson line as defect CFT$_{1}$}},
   {\em JHEP} {\bf 08} (2020) 143, [\href{http://arxiv.org/abs/2004.07849}{{\tt
  arXiv:2004.07849}}].

\bibitem{Griguolo:2012iq}
L.~Griguolo, D.~Marmiroli, G.~Martelloni, and D.~Seminara, {\it {The
  generalized cusp in ABJ(M) N = 6 Super Chern-Simons theories}},  {\em JHEP}
  {\bf 05} (2013) 113, [\href{http://arxiv.org/abs/1208.5766}{{\tt
  arXiv:1208.5766}}].

\bibitem{Bonini:2016fnc}
M.~Bonini, L.~Griguolo, M.~Preti, and D.~Seminara, {\it {Surprises from the
  resummation of ladders in the ABJ(M) cusp anomalous dimension}},  {\em JHEP}
  {\bf 05} (2016) 180, [\href{http://arxiv.org/abs/1603.00541}{{\tt
  arXiv:1603.00541}}].

\bibitem{Bianchi:2014laa}
M.~S. Bianchi, L.~Griguolo, M.~Leoni, S.~Penati, and D.~Seminara, {\it {BPS
  Wilson loops and Bremsstrahlung function in ABJ(M): a two loop analysis}},
  {\em JHEP} {\bf 06} (2014) 123, [\href{http://arxiv.org/abs/1402.4128}{{\tt
  arXiv:1402.4128}}].

\bibitem{Correa:2014aga}
D.~H. Correa, J.~Aguilera-Damia, and G.~A. Silva, {\it {Strings in $AdS_4
  \times \mathbb{CP}^{3}$ Wilson loops in $\mathcal N=$6 super
  Chern-Simons-matter and bremsstrahlung functions}},  {\em JHEP} {\bf 06}
  (2014) 139, [\href{http://arxiv.org/abs/1405.1396}{{\tt arXiv:1405.1396}}].

\bibitem{Bianchi:2017svd}
M.~S. Bianchi, L.~Griguolo, A.~Mauri, S.~Penati, M.~Preti, and D.~Seminara,
  {\it {Towards the exact Bremsstrahlung function of ABJM theory}},  {\em JHEP}
  {\bf 08} (2017) 022, [\href{http://arxiv.org/abs/1705.10780}{{\tt
  arXiv:1705.10780}}].

\bibitem{Drukker:2019bev}
N.~Drukker et~al., {\it {Roadmap on Wilson loops in 3d
  Chern\textendash{}Simons-matter theories}},  {\em J. Phys. A} {\bf 53}
  (2020), no.~17 173001, [\href{http://arxiv.org/abs/1910.00588}{{\tt
  arXiv:1910.00588}}].

\bibitem{Cavaglia:2014exa}
A.~Cavagli\`a, D.~Fioravanti, N.~Gromov, and R.~Tateo, {\it {Quantum Spectral
  Curve of the $\mathcal N=$ 6 Supersymmetric Chern-Simons Theory}},  {\em
  Phys. Rev. Lett.} {\bf 113} (2014), no.~2 021601,
  [\href{http://arxiv.org/abs/1403.1859}{{\tt arXiv:1403.1859}}].

\bibitem{Bombardelli:2017vhk}
D.~Bombardelli, A.~Cavagli\`a, D.~Fioravanti, N.~Gromov, and R.~Tateo, {\it
  {The full Quantum Spectral Curve for $AdS_4/CFT_3$}},  {\em JHEP} {\bf 09}
  (2017) 140, [\href{http://arxiv.org/abs/1701.00473}{{\tt arXiv:1701.00473}}].

\bibitem{Bombardelli:2018bqz}
D.~Bombardelli, A.~Cavagli\`a, R.~Conti, and R.~Tateo, {\it {Exploring the
  spectrum of planar AdS$_{4}$/CFT$_{3}$ at finite coupling}},  {\em JHEP} {\bf
  04} (2018) 117, [\href{http://arxiv.org/abs/1803.04748}{{\tt
  arXiv:1803.04748}}].

\bibitem{Cavaglia:2021eqr}
A.~Cavagli\`a, N.~Gromov, B.~Stefa\'nski, Jr., Jr., and A.~Torrielli, {\it
  {Quantum Spectral Curve for AdS$_{3}$/CFT$_{2}$: a proposal}},  {\em JHEP}
  {\bf 12} (2021) 048, [\href{http://arxiv.org/abs/2109.05500}{{\tt
  arXiv:2109.05500}}].

\bibitem{Ekhammar:2021pys}
S.~Ekhammar and D.~Volin, {\it {Monodromy Bootstrap for SU(2|2) Quantum
  Spectral Curves: From Hubbard model to AdS3/CFT2}},
  \href{http://arxiv.org/abs/2109.06164}{{\tt arXiv:2109.06164}}.

\bibitem{Gurdogan:2015csr}
O.~G\"urdo\u{g}an and V.~Kazakov, {\it {New Integrable 4D Quantum Field
  Theories from Strongly Deformed Planar $\mathcal N = $ 4 Supersymmetric
  Yang-Mills Theory}},  {\em Phys. Rev. Lett.} {\bf 117} (2016), no.~20 201602,
  [\href{http://arxiv.org/abs/1512.06704}{{\tt arXiv:1512.06704}}]. [Addendum:
  Phys.Rev.Lett. 117, 259903 (2016)].

\bibitem{Caetano:2016ydc}
J.~a. Caetano, O.~G\"urdo\u{g}an, and V.~Kazakov, {\it {Chiral limit of $
  \mathcal{N} $ = 4 SYM and ABJM and integrable Feynman graphs}},  {\em JHEP}
  {\bf 03} (2018) 077, [\href{http://arxiv.org/abs/1612.05895}{{\tt
  arXiv:1612.05895}}].

\bibitem{Grabner:2017pgm}
D.~Grabner, N.~Gromov, V.~Kazakov, and G.~Korchemsky, {\it {Strongly
  $\gamma$-Deformed $\mathcal{N}=4$ Supersymmetric Yang-Mills Theory as an
  Integrable Conformal Field Theory}},  {\em Phys. Rev. Lett.} {\bf 120}
  (2018), no.~11 111601, [\href{http://arxiv.org/abs/1711.04786}{{\tt
  arXiv:1711.04786}}].

\bibitem{Basso:2017jwq}
B.~Basso and L.~J. Dixon, {\it {Gluing Ladder Feynman Diagrams into Fishnets}},
   {\em Phys. Rev. Lett.} {\bf 119} (2017), no.~7 071601,
  [\href{http://arxiv.org/abs/1705.03545}{{\tt arXiv:1705.03545}}].

\bibitem{Gromov:2018hut}
N.~Gromov, V.~Kazakov, and G.~Korchemsky, {\it {Exact Correlation Functions in
  Conformal Fishnet Theory}},  {\em JHEP} {\bf 08} (2019) 123,
  [\href{http://arxiv.org/abs/1808.02688}{{\tt arXiv:1808.02688}}].

\bibitem{Kazakov:2018gcy}
V.~Kazakov, E.~Olivucci, and M.~Preti, {\it {Generalized fishnets and exact
  four-point correlators in chiral CFT$_{4}$}},  {\em JHEP} {\bf 06} (2019)
  078, [\href{http://arxiv.org/abs/1901.00011}{{\tt arXiv:1901.00011}}].

\bibitem{Pittelli:2019ceq}
A.~Pittelli and M.~Preti, {\it {Integrable fishnet from $\gamma$-deformed
  $\mathcal{N}=2$ quivers}},  {\em Phys. Lett. B} {\bf 798} (2019) 134971,
  [\href{http://arxiv.org/abs/1906.03680}{{\tt arXiv:1906.03680}}].

\bibitem{Derkachov:2019tzo}
S.~Derkachov and E.~Olivucci, {\it {Exactly solvable magnet of conformal spins
  in four dimensions}},  {\em Phys. Rev. Lett.} {\bf 125} (2020), no.~3 031603,
  [\href{http://arxiv.org/abs/1912.07588}{{\tt arXiv:1912.07588}}].

\bibitem{Derkachov:2020zvv}
S.~Derkachov and E.~Olivucci, {\it {Exactly solvable single-trace four point
  correlators in $\chi$CFT$_4$}},  {\em JHEP} {\bf 02} (2021) 146,
  [\href{http://arxiv.org/abs/2007.15049}{{\tt arXiv:2007.15049}}].

\bibitem{Shahpo:2021xax}
O.~Shahpo and E.~Vescovi, {\it {Correlation functions of determinant operators
  in conformal fishnet theory}},  \href{http://arxiv.org/abs/2110.09458}{{\tt
  arXiv:2110.09458}}.

\bibitem{Gromov:2017cja}
N.~Gromov, V.~Kazakov, G.~Korchemsky, S.~Negro, and G.~Sizov, {\it
  {Integrability of Conformal Fishnet Theory}},  {\em JHEP} {\bf 01} (2018)
  095, [\href{http://arxiv.org/abs/1706.04167}{{\tt arXiv:1706.04167}}].

\bibitem{Gromov:2019jfh}
N.~Gromov and A.~Sever, {\it {The Holographic Dual of Strongly
  $\gamma$-deformed N=4 SYM Theory: Derivation, Generalization, Integrability
  and Discrete Reparametrization Symmetry}},
  \href{http://arxiv.org/abs/1908.10379}{{\tt arXiv:1908.10379}}.

\bibitem{Cavaglia:2020hdb}
A.~Cavagli{\`a}, D.~Grabner, N.~Gromov, and A.~Sever, {\it {Colour-Twist
  Operators I: Spectrum and Wave Functions}},
  \href{http://arxiv.org/abs/2001.07259}{{\tt arXiv:2001.07259}}.

\bibitem{Levkovich-Maslyuk:2020rlp}
F.~Levkovich-Maslyuk and M.~Preti, {\it {Exploring the ground state spectrum of
  $\gamma$-deformed $N=4$ SYM}},  \href{http://arxiv.org/abs/2003.05811}{{\tt
  arXiv:2003.05811}}.

\bibitem{Gromov:2021ahm}
N.~Gromov, J.~Julius, and N.~Primi, {\it {Open Fishchain in N=4 Supersymmetric
  Yang-Mills Theory}},  \href{http://arxiv.org/abs/2101.01232}{{\tt
  arXiv:2101.01232}}.

\bibitem{Correa:2018fgz}
D.~Correa, M.~Leoni, and S.~Luque, {\it {Spin chain integrability in
  non-supersymmetric Wilson loops}},  {\em JHEP} {\bf 12} (2018) 050,
  [\href{http://arxiv.org/abs/1810.04643}{{\tt arXiv:1810.04643}}].

\end{thebibliography}\endgroup

\end{document}